\documentclass[twocolumn,showpacs,preprintnumbers,amsmath,amssymb]{revtex4-2}
\usepackage{graphicx}
\usepackage{dcolumn}
\usepackage{color}
\usepackage{bm}
\usepackage{epsfig}
\usepackage{enumerate}
\usepackage{array}
\usepackage{multirow}
\usepackage{hyperref}
\hypersetup{
    colorlinks=true,
    linkcolor=blue,
    urlcolor=blue,
	citecolor=blue
}
\usepackage{upgreek}
\begin{document}
\title{Thermoelectric performance of a minimally nonlinear voltage probe and voltage-temperature probe heat engine with broken time-reversal symmetry}

\author{Jayasmita Behera,$^1$ Salil Bedkihal,$^2$ Bijay Kumar Agarwalla,$^3$ Malay Bandyopadhyay$^1$}
\affiliation{$^{1}$School of Basic Sciences, Indian Institute of Technology Bhubaneswar, Odisha, India 752050,\\
$^{2}$Dartmouth Engineering Thayer School, 15 Thayer Drive, Hanover, NH 03755, USA,\\
$^{3}$Department of Physics, Indian Institute of Science Education and Research, Pune 411008, India.}

\vskip-2.8cm
\date{\today}
\vskip-0.9cm

\begin{abstract}
We investigate the thermoelectric performance of minimally nonlinear irreversible heat engines with broken time-reversal symmetry (TRS), realized through voltage and voltage-temperature probe configurations. Our framework extends Onsager relations by including a nonlinear power dissipation term in the heat current. We derive and analyze analytical expressions for the efficiency at a given power and the efficiency at maximum power (EMP), expressed in terms of asymmetry parameters and generalized figures of merit. Our analysis reveals that the combined effects of broken TRS and nonlinear dissipation give rise to two universal bounds on the EMP that can surpass the Curzon-Ahlborn (CA) limit. Although these bounds share a similar analytical form, differences in Carnot efficiency and asymmetry parameters lead to distinct operational characteristics, as shown through numerical simulations. We consider a triple-quantum-dot Aharonov-Bohm heat engine incorporating either a voltage probe or a voltage-temperature probe. In both cases, TRS is broken by the magnetic flux. However, the voltage-temperature probe requires an additional anisotropy in the system for its TRS-breaking effects to significantly influence transport. We examine the role of this anisotropy in enhancing performance. Our results show that the EMP and efficiency at a given power can be enhanced by increasing the strength of nonlinear power dissipation, even though the output power remains unchanged. The voltage probe configuration generally yields higher power, while the voltage-temperature probe is more efficient, except in certain regimes where large asymmetries and high figures of merit allow the voltage probe setup to outperform.
\end{abstract}

%below there are some pacs that we would typically use, please search every time to find the most suitable
\pacs{85.25.Cp%Josephson devices
, 42.50.Dv} %Quantum state engineering and measurements
\maketitle

%==========================
\section{Introduction}\label{intro}
\begin{figure}[t!]
    \centering
    \includegraphics[scale=0.35]{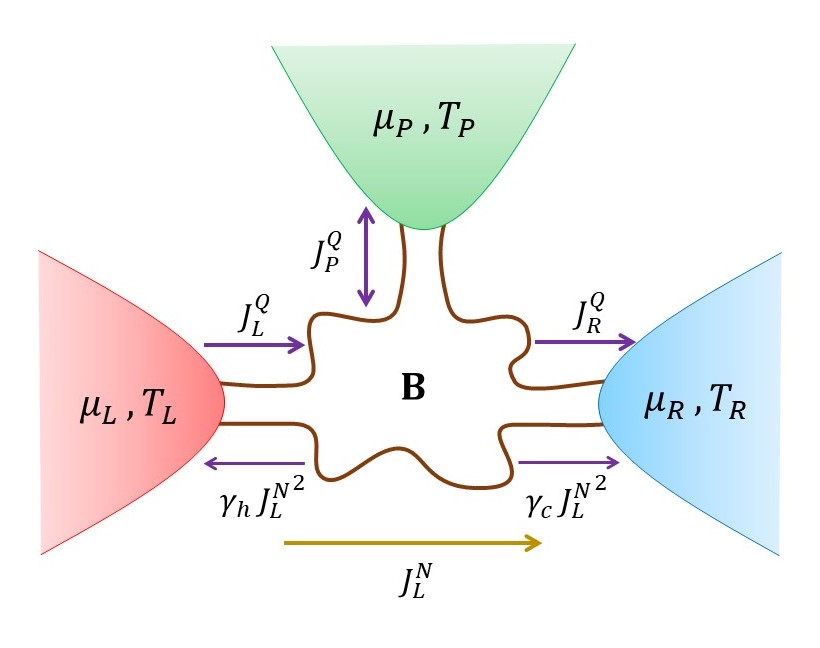}
    \caption{Schematic illustration of a minimally nonlinear three-terminal heat engine with broken time-reversal symmetry. The arrows show the direction of the particle current ($J_L^N$), heat currents ($J_L^Q$, $J_P^Q$, and $J_R^Q$), and the dissipation terms $\gamma_h{J_L^N}^2$ and $\gamma_c{J_L^N}^2$ included in the heat fluxes $J_L^Q$ and $J_R^Q$, respectively.}
    \label{fig:three_terminal}
\end{figure}

Recent advancements in energy harvesting technologies have spurred extensive research into optimizing real-world thermodynamic cycles and designing efficient nanoscale thermoelectric heat engines \cite{energy1,energy2,energy3,energy4}. Unlike bulk thermoelectrics, nanoscale heat engines perform heat-to-work conversion in steady-state without any macroscopic moving parts through steady-state flows of microscopic particles like electrons, photons, phonons, etc. Hence, these thermal devices exhibit quantum interference effects which can induce and modify the thermoelectric response of a system \cite{BENENTI20171, engine1, Whitney2014, Whitney2015, Whitney2018, Taniguchi2020}. The discrete energy levels of nanostructures based on a few quantum dots (QDs) or single molecular junctions provide energy filtering effects crucial for enhanced thermoelectric performance of quantum heat engines \cite{Whitney2015, Whitney2018, Mahan, thermopower, interf1, interf2, interf3, interf4, interf5, interf6, quantumphase1, quantumphase2, quantumphase3, quantumphase4, quantumphase5}.\\
\indent
Improvement in the performance of a nanoscale heat engine demands a proper understanding of the fundamental thermodynamic constraints on the efficiency \cite{Hershfield, Apertet, Yamamoto1, Yamamoto2, Jiang, Whitney2013}. The efficiency of a thermoelectric reversible heat engine is bounded from above by the Carnot efficiency, $\eta_C=1-\frac{T_c}{T_h}$ \cite{Sadi}, working between two heat reservoirs at temperatures $T_h$ and $T_c$ ($T_h>T_c$).
%However, an ideal reversible heat engine operates infinitely slowly to satisfy reversibility resulting in zero output power.
With the development of finite-time thermodynamics \cite{Curzon, Broeck, DeVos, Hernandez, Anderson, Esposito}, Curzon and Ahlborn \cite{Curzon} proposed the efficiency at maximum power (EMP) of a finite-time Carnot model widely known as Curzon-Ahlborn (CA) efficiency: $\eta_{CA}=1-\sqrt{T_c/T_h}$. The EMP, a key performance measure, is extensively used in the study of quantum and stochastic heat engines \cite{Whitney2014, Whitney2015, Ryabov1, Ryabov2, Ryabov3}.\\
\indent
Usually, the presence of a magnetic field breaks the time-reversal symmetry (TRS) of a system \cite{Benenti1}, however, this is not the case for noninteracting two-terminal systems, where the thermopower is an even function of the magnetic field due to the symmetry properties of the scattering matrix. A third terminal, specifically a voltage probe or a voltage-temperature probe, can introduce inelastic scattering effects in the noninteracting system, creating asymmetric thermopower concerning the magnetic field \cite{Buttiker1, Buttiker2, Buttiker3, Bedkihal1}. Within the linear irreversible framework, Brandner {\it et al.} proposed a strong bound on Onsager coefficients as a consequence of the unitarity of the scattering matrix for a multiterminal thermoelectric heat engine with broken TRS \cite{Brandner1, Brandner2, Brandner3}. Within the context of broken TRS, the EMP exceeds the CA limit for three-terminal and multi-terminal systems in the linear-response regime \cite{Benenti2, Benenti3, Balachandran, Brandner1, Brandner2, Brandner3, Sanchenz2, Zhang}. The study of three-terminal QDs-based heat engines shows enhanced thermoelectric performance in the linear-response regime \cite{Mazza, Lu2019}. Recently, Sartipi {\it et al.} showed that the voltage probe heat engine performs better than the voltage-temperature probe heat engine in the linear response regime with broken TRS \cite{Zahra}.\\
\indent
%The CA limit is no longer an upper bound on the efficiency at maximum power with broken time-reversal symmetry \cite{Benenti2, Balachandran, Brandner2, Brandner3, Zhang, Zahra} and it can be exceeded in the nonlinear regime \cite{Seifert, Esposito2, Esposito3, Izumida1, Izumida2}.
In the present manuscript, we extend the study of Sartipi {\it et al.} \cite{Zahra} for the minimally nonlinear (MNL) irreversible heat engine. Traditional heat engines are often analyzed under linear assumptions. The MNL heat engines help explore the boundaries of efficiency, revealing how small nonlinearities can impact the performance, potentially leading to more efficient designs. The idealized linear-response regime results are recoverable under the condition of the Joule dissipation parameter $\gamma_h\rightarrow 0$. Besides, the performance of nanoscale thermoelectric devices is severely affected by heat dissipation, which necessitates a detailed analysis of optimizing heat engines with broken TRS at maximum power and arbitrary power, considering a leading-order nonlinearity such as Joule dissipation. While our analysis operates in the MNL regime, where the thermodynamic flux–force relations and the transport coefficients can be computed using the Landauer-B\"{u}ttiker scattering formalism. In this framework, the voltage probe and voltage-temperature probe parameters are determined self-consistently, ensuring that the resulting expressions for currents remain gauge invariant (see Appendix \ref{app:entropy_gauge}), as established in Refs. \cite{TChristen_1996,DSanchez_2013,Meair_2013}.\\
\indent
With this motivation for introducing the MNL term, we move to the literature of MNL heat engines. Izumida {\it et al.} first proposed a MNL irreversible heat engine model that incorporates Onsager relations with an additional nonlinear dissipation term and obtained a new upper bound on the EMP, $\eta_+=\eta_C/(2-\eta_C)$, which exceeds the CA limit \cite{Izumida3, Izumida4}. The generality of the extended Onsager relations is validated by comparing them with the low-dissipation Carnot engine \cite{Esposito, Johal1}. Further studies on two-terminal MNL heat engines can be found in Refs. \cite{Long1, Long2, Iyyappan1, Ponmurugan, Bai, Zhang2, entropy_Liu, Johal2}. Despite the studies mentioned above, a three-terminal heat engine with broken TRS working in the MNL regime has not been investigated yet. This paper aims to optimize the thermoelectric performance and determine the upper bound on the EMP for an MNL three-terminal heat engine under broken TRS, incorporating a B\"uttiker voltage probe or a voltage-temperature probe. Our analysis shows that the characterization of EMP and efficiency at a given power require two dimensionless parameters, an asymmetry parameter $x_m$ (for voltage probe) or $x$ (for voltage-temperature probe) and a generalized figure of merit parameter $y_m$ (for voltage probe) or $y$ (voltage-temperature probe) which reduces to $ZT$ \cite{BENENTI20171, Benenti1, Benenti3} in the symmetric limit $x_m (x) \rightarrow 1$. In the case $|x_m|>1$ or $|x|>1$, it is possible to overcome the CA limit. The presence of an additional parameter is significant since it allows us additional freedom in designing high-performance thermoelectric devices.\\
\indent
We begin by analyzing a generalized three-terminal system subjected to an external magnetic field and connected to three fermionic reservoirs ($L, P, R$), each maintained at different temperatures and chemical potentials, as illustrated in Fig. \ref{fig:three_terminal}. The third reservoir ($P$) could be a (i) voltage probe that blocks the net particle current but allows the heat current to the probe \cite{Bedkihal1, BANDYOPADHYAY, Saha, Rafael1, Rafael2, Zahra} or a (ii) voltage-temperature probe that blocks both particle and heat currents to the probe to introduce inelastic scattering effects into the system \cite{Bedkihal1, Brandner2, Brandner3, Benenti2, Balachandran, Zhang, Zahra}. A nonlinear term ($-\gamma_h{J_L^N}^2$), meaning the power dissipation, is added to the linear Onsager relation for heat current with $\gamma_h$ denoting the strength of dissipation to the hot reservoir, and $J_L^N$ is the particle current flowing from the left reservoir. One can recover the linear-response results from the MNL results by setting $\gamma_h=0$ \cite{Zhang, Zahra}. With broken time-reversal symmetry, we obtain two universal upper bounds on the EMP for an MNL voltage probe and voltage-temperature probe heat engines, respectively. We consider many-body interactions modeled through the probe at the mean-field level, extending beyond the non-interacting regime, incorporating leading order nonlinearity by including the Joule heating term.\\
\indent
To explicitly demonstrate our derived results for a generalized three-terminal system, we perform a numerical simulation for a triple-dot Aharonov-Bohm (AB) heat engine comprised of three quantum dots arranged in a triangular geometry with dot 2 either connected to a voltage probe or a voltage-temperature probe (see Fig. \ref{fig:TQD}). The AB interferometer is widely investigated as a phase-tunable quantum device and a broken TRS system \cite{ABring1, ABring2, ABring3, Haack2019, Haack2021, Wohlman, Bedkihal1, Bedkihal2, BANDYOPADHYAY, JB}, exhibiting higher-order quantum interference effects \cite{JB, Coish}, and experimentally demonstrated in \cite{Gaudreau, Rogge_2009, Amaha}. Our recent investigation \cite{JB} of a two-terminal triple-dot AB heat engine shows a significantly high value of $0.8\eta_C$ for EMP in the nonlinear regime, which motivates us to investigate the system for the three-terminal setup in the MNL regime.\\ %We demonstrate the tunability of the device by controlling the $t/\gamma$ ratio in three regimes i.e., (a) $t/\gamma<1$, (b) $t/\gamma=1$, and (c) $t/\gamma>1$ with $t$ and $\gamma$ denoting the inter-dot tunneling strength and dot-reservoir coupling strength, respectively. Note that the coupling strength $\gamma$ differs from the dissipation strength $\gamma_h$.\\
\indent
The rest of the paper is organized as follows: We introduce a general three-terminal model and develop linear Onsager relations between the currents and the thermodynamic biases in Sec. \ref{sec:linear}. Section \ref{sec:mnl} elaborates on the MNL voltage probe and voltage-temperature probe heat engines utilizing extended Onsager relations. A triple-dot AB heat engine is investigated in Sec. \ref{sec:TQD} and the numerical results are discussed in Sec. \ref{sec:results}. We conclude in Sec. \ref{sec:conclusion}. Additional details are provided in appendix \ref{coefficients}, \ref{sec:inelastic}, \ref{Carnot}, \ref{additional}, \ref{fully_nonlinear}, and \ref{sec:LP}.

\section{Model and Formulation}\label{sec:linear}
This section is dedicated to a brief theoretical background on linear irreversible thermodynamics and the phenomenological equation connecting the fluxes with the thermodynamic forces described as $\textbf{J}=\mathcal{L}\textbf{X}$. We use the B\"{u}ttiker probe technique, i.e., (i) voltage probe and (ii) voltage-temperature probe, to simulate inelastic scattering effects into the system \cite{Buttiker1, Buttiker2, Buttiker3, Bedkihal1, BANDYOPADHYAY, Saha, Rafael1, Rafael2}. Further, strong bounds on the Onsager coefficients are obtained in the subsequent section.
\subsection{Linear irreversible three-terminal heat engine}
A three-terminal thermoelectric heat engine with broken time-reversal symmetry is depicted in Fig. \ref{fig:three_terminal}. A scattering region subjected to an external magnetic field $\textbf{B}$ is connected to three different fermionic reservoirs termed as left ($L$), right ($R$), and probe ($P$) that can exchange both heat and particles with the system. The temperature and chemical potential of the left and right reservoirs are set as $T_L>T_R$ and $\mu_L<\mu_R$ to ensure our system works as a thermoelectric heat engine. The probe parameters ($T_P$, $\mu_P$) are adjusted to block particle flow or both particle and heat flow. We assume the right reservoir as a reference by setting $(\mu_R, T_R)=(\mu, T)$ and write $\mu_{\nu}=\mu+\Delta\mu_{\nu}$, $T_{\nu}=T+\Delta T_{\nu}$ for $\nu=L,P$. Our system operates in the linear response regime if $|\Delta\mu_{\nu}/T|\ll 1$ and $|\Delta T_{\nu}|/T\ll 1$. The particle currents ($J_{\nu}^N$) and the energy currents ($J_{\nu}^U$) flowing from the reservoir $\nu$ satisfy the current conservation laws i.e., $\sum_{\nu}J_{\nu}^{N(U)}=0$. The heat current ($J_{\nu}^Q$) flowing from the reservoir $\nu$ can be expressed in terms of $J_{\nu}^N$ and $J_{\nu}^U$ as $J_{\nu}^Q=J_{\nu}^U-\mu_{\nu}J_{\nu}^N$. Considering the above conditions, one can linearly expand the particle and heat currents defined in Eq. (\ref{eq:particle}) and Eq. (\ref{eq:heat}), respectively, to obtain the linear Onsager relation between the thermodynamic fluxes (particle and heat currents) and forces (chemical potential and temperature biases) as ${\bf J}={\bf\mathcal{L}X}$ \cite{Mazza, Brandner2, Benenti2, Balachandran, Zhang, Zahra}:
\begin{equation}\label{onsager}
    \renewcommand{\arraystretch}{1.4}
        \begin{pmatrix}
            J_L^N\\
            J_L^Q\\
            J_P^N\\
            J_P^Q
        \end{pmatrix}
        =
        \begin{pmatrix}
            \mathcal{L}_{11} & \mathcal{L}_{12} & \mathcal{L}_{13} & \mathcal{L}_{14}\\
            \mathcal{L}_{21} & \mathcal{L}_{22} & \mathcal{L}_{23} & \mathcal{L}_{24}\\
            \mathcal{L}_{31} & \mathcal{L}_{32} & \mathcal{L}_{33} & \mathcal{L}_{34}\\
            \mathcal{L}_{41} & \mathcal{L}_{42} & \mathcal{L}_{43} & \mathcal{L}_{44}
        \end{pmatrix}
        \begin{pmatrix}
            X_L^{\mu}\\
            X_L^T\\
            X_P^{\mu}\\
            X_P^T
        \end{pmatrix},
\end{equation}
where $X_{\nu}^{\mu}=(\mu_{\nu}-\mu)/T$ and $X_{\nu}^T=1/T-1/T_{\nu}$ ($X_{\nu}^T\approx\Delta T_{\nu}/T^2$) for $\nu=L, P$ are the generalized forces. Here ${\bf\mathcal{L}}$ is $4\times4$ Onsager matrix, and its elements $\mathcal{L}_{ij}$ are called the Onsager coefficients, and we will suppress $\textbf{B}$ in the Onsager coefficients to keep the notation simple, unless necessary. The explicit expressions of the Onsager coefficients are given in Appendix \ref{coefficients}.
\subsection{Voltage probe and voltage-temperature probe heat engine}
The voltage probe introduces dissipative inelastic scattering processes into the system by demanding that the net particle current flowing to the probe vanishes (i.e, $J_P^N=0$) but allowing a nonzero heat current ($J_P^Q\ne 0$) from the probe \cite{Bedkihal1, BANDYOPADHYAY, Saha, Rafael1, Rafael2, Zahra}. Thus, from Eq. (\ref{onsager}), we can derive $X_P^{\mu}$ as follows:
\begin{equation}\label{xpv}
    X_P^{\mu}=-\frac{(\mathcal{L}_{31}X_L^{\mu}+\mathcal{L}_{32}X_L^T+\mathcal{L}_{34}X_P^T)}{\mathcal{L}_{33}}.
\end{equation}
Substituting Eq. (\ref{xpv}) in Eq. (\ref{onsager}) and after some algebra we can express the remaining finite currents in terms of a ($3\times 3$) reduced Onsager matrix $\mathcal{L}^{\prime}$ as follows \cite{Zahra}:
\begin{equation}\label{vprobe}
    \renewcommand{\arraystretch}{1.4}
        \begin{pmatrix}
            J_L^N\\
            J_L^Q\\
            J_P^Q
        \end{pmatrix}=
        \begin{pmatrix}
            \mathcal{L}_{11}^{\prime} & \mathcal{L}_{12}^{\prime} & \mathcal{L}_{13}^{\prime}\\
            \mathcal{L}_{21}^{\prime} & \mathcal{L}_{22}^{\prime} & \mathcal{L}_{23}^{\prime}\\
            \mathcal{L}_{31}^{\prime} & \mathcal{L}_{32}^{\prime} & \mathcal{L}_{33}^{\prime}
        \end{pmatrix}
        \begin{pmatrix}
            X_L^{\mu}\\
            X_L^T\\
            X_P^T
        \end{pmatrix},
\end{equation}
where the reduced Onsager coefficients $\mathcal{L}^{\prime}_{ij}$ are expressed as follows \cite{Zahra}:
  \begin{equation}\label{eq:coeff_V}
      \begin{aligned}
          \mathcal{L}_{11}^{\prime}=\frac{\mathcal{L}_{33}\mathcal{L}_{11}-\mathcal{L}_{13}\mathcal{L}_{31}}{\mathcal{L}_{33}},\,\
          \mathcal{L}_{12}^{\prime}=\frac{\mathcal{L}_{33}\mathcal{L}_{12}-\mathcal{L}_{13}\mathcal{L}_{32}}{\mathcal{L}_{33}},\\
          \mathcal{L}_{13}^{\prime}=\frac{\mathcal{L}_{33}\mathcal{L}_{14}-\mathcal{L}_{13}\mathcal{L}_{34}}{\mathcal{L}_{33}},\,\
          \mathcal{L}_{21}^{\prime}=\frac{\mathcal{L}_{33}\mathcal{L}_{21}-\mathcal{L}_{23}\mathcal{L}_{31}}{\mathcal{L}_{33}},\\
          \mathcal{L}_{22}^{\prime}=\frac{\mathcal{L}_{33}\mathcal{L}_{22}-\mathcal{L}_{23}\mathcal{L}_{32}}{\mathcal{L}_{33}},\,\
          \mathcal{L}_{23}^{\prime}=\frac{\mathcal{L}_{33}\mathcal{L}_{24}-\mathcal{L}_{23}\mathcal{L}_{34}}{\mathcal{L}_{33}},\\
          \mathcal{L}_{31}^{\prime}=\frac{\mathcal{L}_{33}\mathcal{L}_{41}-\mathcal{L}_{43}\mathcal{L}_{31}}{\mathcal{L}_{33}},\,\
          \mathcal{L}_{32}^{\prime}=\frac{\mathcal{L}_{33}\mathcal{L}_{42}-\mathcal{L}_{43}\mathcal{L}_{32}}{\mathcal{L}_{33}},\\
          \mathcal{L}_{33}^{\prime}=\frac{\mathcal{L}_{33}\mathcal{L}_{44}-\mathcal{L}_{43}\mathcal{L}_{34}}{\mathcal{L}_{33}}.
      \end{aligned}
  \end{equation}
\indent Likewise, for a voltage-temperature probe, the chemical potential and the probe's temperature are adjusted so that both the net particle current and heat current from the probe vanishes i.e., $J_P^N=0$ and $J_P^Q=0$ \cite{Bedkihal1, Brandner2, Benenti2, Balachandran, Zhang, Zahra}. We can obtain a ($2\times 2$) reduced Onsager matrix $\mathcal{L}^{\prime\prime}$ for the voltage-temperature probe by setting $J_P^Q=0$ in Eq. (\ref{vprobe}) and it is expressed as follows \cite{Zahra}:
\begin{equation}\label{vtprobe}
     \renewcommand{\arraystretch}{1.4}
        \begin{pmatrix}
            J_L^N\\
            J_L^Q
        \end{pmatrix}
        =
        \begin{pmatrix}
            \mathcal{L}^{\prime\prime}_{11} & \mathcal{L}^{\prime\prime}_{12}\\
            \mathcal{L}^{\prime\prime}_{21} & \mathcal{L}^{\prime\prime}_{11}
        \end{pmatrix}
        \begin{pmatrix}
            X_L^{\mu}\\
            X_L^T
        \end{pmatrix},
    \end{equation}
where the reduced Onsager coefficients for the voltage-temperature probe are
\begin{equation}\label{eq:coeff_VT}
    \begin{aligned}
        \mathcal{L}^{\prime\prime}_{11}= \frac{\mathcal{L}_{33}^{\prime}\mathcal{L}_{11}^{\prime}-\mathcal{L}_{13}^{\prime}\mathcal{L}_{31}^{\prime}}{\mathcal{L}_{33}^{\prime}},\,\
        \mathcal{L}^{\prime\prime}_{12}= \frac{\mathcal{L}_{33}^{\prime}\mathcal{L}_{12}^{\prime}-\mathcal{L}_{13}^{\prime}\mathcal{L}_{32}^{\prime}}{\mathcal{L}_{33}^{\prime}},\\
        \mathcal{L}^{\prime\prime}_{21}= \frac{\mathcal{L}_{33}^{\prime}\mathcal{L}_{21}^{\prime}-\mathcal{L}_{23}^{\prime}\mathcal{L}_{31}^{\prime}}{\mathcal{L}_{33}^{\prime}},\,\
        \mathcal{L}^{\prime\prime}_{22}= \frac{\mathcal{L}_{33}^{\prime}\mathcal{L}_{22}^{\prime}-\mathcal{L}_{23}^{\prime}\mathcal{L}_{32}^{\prime}}{\mathcal{L}_{33}^{\prime}}.
    \end{aligned}
\end{equation}
\indent The time-reversal symmetry is broken in the presence of a nonzero magnetic field, and the Onsager coefficients follow the Onsager-Casimir relation \cite{OnsagerReciprocal}
\begin{equation}
    \mathcal{L}^{\prime}_{ij}({\bf B})=\mathcal{L}^{\prime}_{ji}(-{\bf B})\,\,\ \mathrm{and}\,\,\ \mathcal{L}^{\prime\prime}_{ij}({\bf B})=\mathcal{L}^{\prime\prime}_{ji}(-{\bf B}).
\end{equation}
\subsection{Bounds on the Onsager coefficients}
Onsager coefficients are subjected to some constraints due to the second law of thermodynamics, which can be derived from the positivity of the entropy production rate for a two-terminal heat engine \cite{Benenti1}. However, for a multi-terminal system ($n\ge3$) with broken time-reversal symmetry, the unitarity of the scattering matrix imposes stronger bounds on Onsager coefficients than the positivity of the entropy production rate \cite{Brandner1, Brandner2}. So, the reduced Onsager coefficients for the voltage probe heat engine satisfy the following inequalities \cite{Zahra}:
\begin{equation}\label{boundv}
    \begin{aligned}
        \mathfrak{L}_{11}\ge 0,\,\,\ \mathfrak{L}_{22}\ge 0,\\
        \mathfrak{L}_{11}\mathfrak{L}_{22}+\mathfrak{L}_{12}\mathfrak{L}_{21}-\mathfrak{L}_{12}^2-\mathfrak{L}_{21}^2\ge 0,
    \end{aligned}
\end{equation}
where
\begin{equation}
    \begin{aligned}
        \mathfrak{L}_{11} &=\mathcal{L}_{11}^{\prime},\\
        \mathfrak{L}_{12} &=\mathcal{L}_{12}^{\prime}+\mathcal{L}_{13}^{\prime}\xi,\\
        \mathfrak{L}_{21} &=\mathcal{L}_{21}^{\prime}+\mathcal{L}_{31}^{\prime}\xi,\\
        \mathfrak{L}_{22} &=\mathcal{L}_{22}^{\prime}+\mathcal{L}_{33}^{\prime}\xi^2+(\mathcal{L}_{23}^{\prime}+\mathcal{L}_{32}^{\prime})\xi,
    \end{aligned}
\end{equation}
with $\xi=X_P^T/X_L^T$. Similarly, the Onsager bounds for the voltage-temperature probe heat engine \cite{Brandner1, Brandner2, Zahra} are
\begin{equation}\label{boundvt}
    \begin{aligned}
        \mathcal{L}^{\prime\prime}_{11}\ge0,\,\,\ \mathcal{L}^{\prime\prime}_{22}\ge0,\\
        \mathcal{L}^{\prime\prime}_{11}\mathcal{L}^{\prime\prime}_{22}+ \mathcal{L}^{\prime\prime}_{12}\mathcal{L}^{\prime\prime}_{21}-{\mathcal{L}^{\prime\prime}_{12}}^2-{\mathcal{L}^{\prime\prime}_{21}}^2\ge0.
    \end{aligned}
\end{equation}
\section{Minimally nonlinear three-terminal heat engine: New bounds on efficiency at maximum power}\label{sec:mnl}
We consider an MNL irreversible heat engine model which accounts for possible thermal dissipation effects due to the interaction between the system and the reservoir \cite{Izumida3, Izumida4, Long1, Long2, Iyyappan1, Ponmurugan, Bai, Zhang2, entropy_Liu}. A nonlinear term $-\gamma_h{J_L^N}^2$ meaning power dissipation, is added to the linear Onsager relation for the heat current from the left reservoir ($J_L^Q$) described in Eqs. (\ref{vprobe}) and (\ref{vtprobe}). Here $-\gamma_h{J_L^N}^2$ denotes the dissipation to the hot (left) reservoir and $\gamma_h (>0)$ is the strength of dissipation \cite{Izumida3, Izumida4}. The addition of this nonlinear term can be treated as a natural extension of the linear irreversible heat engine. We assume that no other higher-order nonlinear terms contribute to the currents. The dissipation is assumed to be weak, so the bounds on the Onsager coefficients discussed in Eq. (\ref{boundv}) and Eq. (\ref{boundvt}) still hold for the MNL heat engine model. Unlike the linear response regime, the generalized forces are not restricted to small values for the MNL heat engine. Considering the above assumptions, we discuss the thermoelectric performance of the MNL heat engine with voltage probe and voltage-temperature probe in the subsequent sections.

\subsection{Minimally nonlinear voltage probe heat engine}
The extended Onsager relations for the MNL voltage probe heat engine are expressed as follows:
\begin{equation}\label{eq:jlnv}
    J_L^N= \mathcal{L}_{11}^{\prime}X_L^{\mu}+(\mathcal{L}_{12}^{\prime}+\mathcal{L}_{13}\xi)X_L^T,
\end{equation}
\begin{equation}\label{eq:jlqv}
    J_L^Q = \mathcal{L}_{21}^{\prime}X_L^{\mu}+(\mathcal{L}_{22}^{\prime}+\mathcal{L}_{23}^{\prime}\xi)X_L^T-\gamma_h{J_L^N}^2,
\end{equation}
\begin{equation}\label{eq:jpqv}
    J_P^Q = \mathcal{L}_{31}^{\prime}X_L^{\mu}+(\mathcal{L}_{32}^{\prime}+\mathcal{L}_{33}^{\prime}\xi)X_L^T,
\end{equation}
where $\xi=X_P^T/X_L^T$. The linear Onsager relations are recovered by putting $\gamma_h=0$. The nonlinear term ($-\gamma_h{J_L^N}^2$) represents the inevitable power loss due to the finite time operation of the heat engine \cite{Izumida1, Izumida2}.\\
\indent Since $J_L^N$ and $X_L^{\mu}$ are uniquely related at a fixed values of $X_L^T$ and $X_P^T$, we can choose $J_L^N$ as our control parameter instead of $X_L^{\mu}$ by using Eq. (\ref{eq:jlnv}) and express the output power, $\mathcal{P}=-TX_L^{\mu}J_L^N>0$ as
\begin{equation}\label{power}
    \mathcal{P}=\frac{T}{\mathcal{L}_{11}^{\prime}}(\mathcal{L}_{12}^{\prime}+\mathcal{L}_{13}^{\prime}\xi)X_L^TJ_L^N-\frac{T}{\mathcal{L}_{11}^{\prime}}{J_L^N}^2.
\end{equation}
We notice that the nonlinear term in Eq. (\ref{eq:jlqv}) does not contribute to the output power and it only depends on $J_L^N$ (or $X_L^{\mu}$) at fixed $X_L^T$ and $X_P^T$.\\
\indent The efficiency of a three-terminal heat engine operating between the temperatures $T=T_R<T_P<T_L$ at a given power is defined as
\begin{equation}
    \eta=\frac{\mathcal{P}}{\sum_{\nu+}J_{\nu}^Q}=\frac{-TX_L^{\mu}J_L^N}{\sum_{\nu+}J_{\nu}^Q},
\end{equation}
where $\sum_{\nu+}J_{\nu}^Q$ represents only positive heat currents ($J_{\nu}^Q>0$) absorbed by the system from the reservoir. We consider $J_R^Q<0$ and $J_L^Q>0$ for our system to work as a heat engine. Depending on the sign of $J_P^Q$, the efficiency of the voltage probe heat engine when only $J_L^Q>0$ is given by
\begin{equation}\label{etaL}
    \eta_L=\frac{\mathcal{P}}{J_L^Q},
\end{equation}
 or when both $J_L^Q>0$ and $J_P^Q>0$,
\begin{equation}\label{etaLP}
    \eta_{LP}=\frac{\mathcal{P}}{J_L^Q+J_P^Q}.
\end{equation}
Combinedly, we can represent these two cases of the voltage probe heat engine as $\eta_m$ with index $m = L, LP$.
\subsubsection{Efficiency at maximum power (EMP)}
We now derive the maximum power by setting $\partial\mathcal{P}/\partial J_L^N=0$ which leads to the following:
\begin{equation}\label{jln}
    {J_L^N}^*=(\mathcal{L}_{12}^{\prime}+\mathcal{L}_{13}^{\prime}\xi)\frac{X_L^T}{2},\,\,\, {X_L^{\mu}}^*=\frac{-(\mathcal{L}_{12}^{\prime}+\mathcal{L}_{13}^{\prime}\xi)X_L^T}{2\mathcal{L}_{11}^{\prime}}.
\end{equation}
Substituting the value of ${J_L^N}^*$ in Eq. (\ref{power}), the maximum power can be expressed as
\begin{equation}\label{pmax}
    \mathcal{P}_{max}=\frac{T}{4\mathcal{L}_{11}^{\prime}}(\mathcal{L}_{12}^{\prime}+\mathcal{L}_{13}^{\prime}\xi)^2{X_L^T}^2.
\end{equation}
\indent Using Eq. (\ref{jln}) and Eq. (\ref{pmax}) in the efficiency expression defined in Eq. (\ref{etaL}) and Eq. (\ref{etaLP}), the EMP for the MNL voltage probe heat engine is derived as follows:
\begin{equation}\label{etaPmax}
    \eta_{m}(\mathcal{P}_{max})=\eta_{C,m}(\mathcal{P}_{max})\frac{x_my_m}{2(y_m+2)-\alpha TX_L^T x_my_m},
\end{equation}
where $\eta_{C,m}(\mathcal{P}_{max})$ is the value of the Carnot efficiency derived by substituting $X_L^{\mu}={X_L^{\mu}}^*$ (details in Appendix \ref{Carnot}). \\
%The Carnot efficiency for a three-terminal heat engine is discussed in Appendix \ref{Carnot}.
\indent We introduce the asymmetry parameter $x_m$ for the voltage probe heat engine defined as \cite{Zahra}
\begin{equation}
    x_m=\frac{\mathfrak{L}_{12}}{\mathfrak{L}_{21}}=\frac{(\mathcal{L}_{12}^{\prime}+\mathcal{L}_{13}^{\prime}\xi)}{(\mathcal{L}_{21}^{\prime}+\mathcal{L}_{31}^{\prime}\xi)}.
\end{equation}
%where
%\begin{equation}
    %r_m=(2\delta\mathcal{Z}_m^A+\mathcal{Z}_m^B+\delta^2\mathcal{Z}_m^C)T.
%\end{equation}
The generalized figure of merit $y_m$ for the voltage probe heat engine is expressed as \cite{Zahra}
%\begin{equation}
    %y_m=[\delta(\mathcal{Z}_m^{A^{\prime}}+\mathcal{Z}_m^{A^{\prime\prime}})+\mathcal{Z}_m^{B^{\prime}}+\delta^2\mathcal{Z}_m^{C^{\prime}}],
%\end{equation}
\begin{equation}
    y_m=\frac{\mathfrak{L}_{12}\mathfrak{L}_{21}}{\mathfrak{L}_{11}\mathfrak{L}_{22}-\mathfrak{L}_{12}\mathfrak{L}_{21}},
\end{equation}
where $\mathfrak{L}_{ij}$ is defined in Eq. (9).
%where $\delta=1/\xi$ and the term $\mathcal{Z}_m^{\theta}T$ with $\theta=A,A^{\prime},A^{\prime\prime},B,B^{\prime},C,C^{\prime}$ are given in Eq. (\ref{ZT}) of Appendix \ref{transport}. The time-reversal symmetry case refers to $x_m=1$.\\
%\indent The parameter $d_m$ in Eq. (\ref{etaPmax}) is related to the thermal conductance  and temperature bias ratio of the system \cite{Zahra} and it is defined as
%\begin{equation}
    %\begin{aligned}
        %d_L &=\delta\frac{K_{PL}+K_{LP}}{K_{LL}}+\frac{K_{PP}}{K_{LL}}+\delta^2\,\,\,\,\,  \mathrm{if}\,\, J_L^Q>0,\\
        %d_{LP} &=\frac{\delta K_{PL}+K_{PP}}{K_{LP}}+\delta^2\frac{K_{LL}}{K_{LP}}+\delta\,\,\,\,\,  \mathrm{if}\,\, J_L^Q, J_P^Q>0.
    %\end{aligned}
%\end{equation}
%Here $K_{ij}$ are thermal conductance terms defined in Eq. (\ref{Kv}) of Appendix \ref{transport}.\\
\indent From Eq. (\ref{etaPmax}), the dimensionless parameter $\alpha$ representing dissipation strength ratio is defined as
\begin{equation}
    \alpha=\frac{1}{\big(1+\frac{\gamma_c}{\gamma_h}\big)},
\end{equation}
where $\gamma_c=\frac{T}{\mathcal{L}^{\prime}_{11}}-\gamma_h>0$ denotes the strength of dissipation to the cold (right) reservoir for the MNL voltage probe heat engine (detailed in Appendix \ref{additional}). The asymmetry dissipation limits $\gamma_c/\gamma_h\to\infty$ (i.e., $\gamma_h\to 0$) and $\gamma_c/\gamma_h\to 0$ (i.e., $\gamma_c\to 0$) corresponds to $\alpha=0$ and $\alpha=1$, respectively and by setting $\alpha=0$ we will recover the linear-response results.

\subsubsection{Efficiency at a given power}
The optimal performance of the MNL voltage probe heat engine can be examined by defining the relative power as
\begin{equation}\label{delP}
    \frac{\mathcal{P}}{\mathcal{P}_{max}}=\frac{J_L^NX_L^{\mu}}{{J_L^N}^*{X_L^{\mu}}^*}=\chi(2-\chi),
\end{equation}
where $\chi=X_L^{\mu}/{X_L^{\mu}}^*$. Further, by solving Eq. (\ref{delP}), we can obtain
\begin{equation}\label{chi}
    \chi_{\pm}=1\pm\sqrt{1-\mathcal{P}/\mathcal{P}_{max}}\,.
\end{equation}
The $\chi_+$ refers to the favorable branch where $X_L^{\mu}/{X_L^{\mu}}^*>1$ and the efficiency goes beyond the EMP and reaches a maximum value for a given power output. Likewise, $\chi_-$ is called the unfavorable branch with $X_L^{\mu}/{X_L^{\mu}}^*<1$, where the efficiency at the given power is lower than the EMP.\\
\indent The efficiency at a given power for the MNL voltage probe heat engine can be expressed in terms of the relative bias $\chi$ and $\mathcal{P}/\mathcal{P}_{max}$, respectively, as follows:
\begin{widetext}
    \begin{equation}\label{v_effc_chi}
        \eta_m=\frac{\eta_{C,m}\,\chi(2-\chi)x_my_m}{4(y_m+1)-2y_m\chi-\alpha TX_L^T(2-\chi)^2x_my_m},
    \end{equation}
    \begin{equation}\label{v_effc_delP}
        \eta_m=\frac{\eta_{C,m}\big(\mathcal{P}/\mathcal{P}_{max}\big)x_my_m}{4(y_m+1)-2y_m\Big(1\pm\sqrt{1-\mathcal{P}/\mathcal{P}_{max}}\Big)-\alpha TX_L^T\Big[2-\Big(1\pm\sqrt{1-\mathcal{P}/\mathcal{P}_{max}}\Big)\Big]^2x_my_m}.
    \end{equation}
\end{widetext}
\subsubsection{Upper bound on EMP}
The asymmetry parameter $x_m$ measures the asymmetry of the off-diagonal elements of the Onsager matrix by breaking the time-reversal symmetry in the presence of a nonzero magnetic field. Although there is no constraint on $x_m$, the generalized figure of merit $y_m$ is bounded from above due to the bounds on the Onsager coefficients defined in Eq. (\ref{boundv}). For the voltage probe heat engine, the third inequality in Eq. (\ref{boundv}) can be expressed in terms $x_m$ and $y_m$ as follows:
\begin{equation}\label{hmymxm}
    \begin{aligned}
        H_m\le y_m\le 0\,\,\ \mathrm{for}\,\,\ x_m\le 0,\\
        H_m\ge y_m\ge 0\,\,\ \mathrm{for}\,\,\ x_m\ge 0,
    \end{aligned}
\end{equation}
where the bound function $H_m$ is defined as \cite{Zahra}
\begin{equation}\label{Hv}
    H_m=\frac{x_m}{(x_m-1)^2}.
\end{equation}
Equation (\ref{Hv}) shows that the sign of $y_m$ is associated with the sign of both $x_m$.\\
\indent By setting $y_m=H_m$ in Eq. (\ref{etaPmax}), we can obtain the upper bound on the EMP for an MNL voltage probe heat engine as follows:
\begin{equation}\label{npmaxb}
    \eta^*_{m}(\mathcal{P}_{max})=\frac{\eta_{C,m}(\mathcal{P}_{max})x_m^2}{2(2x_m^2-3x_m+2)-\alpha TX_L^Tx_m^2}.
\end{equation}
Equation (\ref{npmaxb}) is a universal upper bound on the EMP for an MNL voltage probe heat engine with broken time-reversal symmetry.\\
\indent For the time-reversal symmetry case, i.e., $x_m=1$ with $0\le\alpha\le 1$, $\eta^*_{m}(\mathcal{P}_{max})$ can take values between
\begin{equation}\label{vbound}
    \frac{\eta_{C,m}(\mathcal{P}_{max})}{2}\le\eta^*_{m}(\mathcal{P}_{max})\le\frac{\eta_{C,m}(\mathcal{P}_{max})}{2-TX_L^T}.
\end{equation}
We observe from Eq. (\ref{vbound}) that the upper bound on the EMP is enhanced and can exceed the CA limit by introducing nonlinearity into the system.
\subsection{Minimally nonlinear voltage-temperature probe heat engine}
For an MNL voltage-temperature probe heat engine, the extended Onsager relations for the particle current and the heat current from the left reservoir are expressed as follows:
\begin{equation}\label{eq:jlnvt}
    J_L^N =\mathcal{L}^{\prime\prime}_{11}X_L^{\mu}+\mathcal{L}^{\prime\prime}_{12}X_L^T,
\end{equation}
\begin{equation}\label{eq:jlqvt}
    J_L^Q =\mathcal{L}^{\prime\prime}_{21}X_L^{\mu}+\mathcal{L}^{\prime\prime}_{22}X_L^T-\gamma_h{J_L^N}^2,
\end{equation}
where $-\gamma_h{J_L^N}^2$ denotes the dissipation to the hot reservoir and $\gamma_h>0$ indicates the dissipation strength.\\
\indent By changing the variable from $X_L^{\mu}$ to $J_L^N$ using Eq. (\ref{eq:jlnvt}) at a fixed value of $X_L^T$, we can write the output power $\mathcal{P}=-TX_L^{\mu}J_L^N>0$ for the voltage-temperature probe heat engine is given by
\begin{equation}\label{powervt}
    \mathcal{P}=\frac{TX_L^T\mathcal{L}^{\prime\prime}_{12}}{\mathcal{L}^{\prime\prime}_{11}}J_L^N-\frac{T}{\mathcal{L}^{\prime\prime}_{11}}{J_L^N}^2,
\end{equation}

\subsubsection{Efficiency at maximum power (EMP)}
The maximum output power for a voltage-temperature probe heat engine can be obtained by taking the derivative of output power in Eq. (\ref{powervt}) with $J_L^N$ to be zero i.e., $\partial\mathcal{P}/\partial J_L^N=0$. This implies
\begin{equation}
    {J_L^N}^*=\frac{\mathcal{L}^{\prime\prime}_{12}X_L^T}{2},\,\ {X_L^{\mu}}^*=-\frac{\mathcal{L}^{\prime\prime}_{12}X_L^T}{2\mathcal{L}^{\prime\prime}_{11}}.
\end{equation}
The maximum output power is obtained by substituting ${J_L^N}^*$ in Eq. (\ref{powervt}) and it is expressed as follows:
\begin{equation}
    \mathcal{P}_{max}=\frac{T}{4\mathcal{L}_{11}^{\prime\prime}}{\mathcal{L}_{12}^{\prime\prime}}^2{X_L^T}^2.
\end{equation}
\indent The EMP for the MNL voltage-temperature probe heat engine is derived as follows:
\begin{equation}\label{etapmaxvt}
    \eta(\mathcal{P}_{max})=\eta_C\frac{xy}{2(y+2)-\alpha TX_L^T xy},
\end{equation}
where $\eta_C=1-T_R/T_L$ ($=TX_L^T$) is the Carnot efficiency for the voltage-temperature probe heat engine.\\
\indent The asymmetry parameter $x$ and the generalized figure of merit $y$ for the voltage-temperature probe heat engine are defined as follows:
\begin{equation}
    x=\frac{\mathcal{L}^{\prime\prime}_{12}}{\mathcal{L}^{\prime\prime}_{21}},
\end{equation}
\begin{equation}
    y=\frac{\mathcal{L}^{\prime\prime}_{12}\mathcal{L}^{\prime\prime}_{21}}{\mathcal{L}^{\prime\prime}_{11}\mathcal{L}^{\prime\prime}_{22}-\mathcal{L}^{\prime\prime}_{12}\mathcal{L}^{\prime\prime}_{21}}.
\end{equation}
\indent Likewise, the voltage probe heat engine, from Eq. (\ref{etapmaxvt}), we define the dimensionless parameter denoting dissipation strength ratio as $\alpha=1/(1+\gamma_c/\gamma_h)$. Here $\gamma_c=T/\mathcal{L}^{\prime\prime}_{11}-\gamma_h>0$ is the dissipation strength to the cold reservoir (detailed in Appendix \ref{additional}). It takes values from $0\le\alpha\le 1$ for asymmetric dissipation limits $\gamma_c/\gamma_h\to\infty$ and $\gamma_c/\gamma_h\to 0$. Linear Onsager relations are obtained by taking $\alpha=0$.
\subsubsection{Efficiency at a given power}
The performance of the MNL voltage-temperature heat engine can be optimized by defining the relation
\begin{equation}\label{chivt}
    \chi_{\pm}=1\pm\sqrt{1-\mathcal{P}/\mathcal{P}_{max}}\,,
\end{equation}
where $\chi=X_L^{\mu}/{X_L^{\mu}}^*$ and the plus (minus) sign corresponds to the favorable (unfavorable) branch.\\
\indent The efficiency at a given power for the MNL voltage-temperature probe heat engine can be expressed in terms of $\chi$ and $\mathcal{P}/\mathcal{P}_{max}$ as follows:
\begin{equation}\label{vt_effc_chi}
    \eta=\frac{\eta_C\,\chi(2-\chi)xy}{4(y+1)-2y\chi-\alpha TX_L^T(2-\chi)^2xy},
\end{equation}
\begin{widetext}
    \begin{equation}\label{vt_effc_delP}
        \eta=\frac{\eta_{C}\big(\mathcal{P}/\mathcal{P}_{max}\big)xy}{4(y+1)-2y\Big(1\pm\sqrt{1-\mathcal{P}/\mathcal{P}_{max}}\Big)-\alpha TX_L^T\Big[2-\Big(1\pm\sqrt{1-\mathcal{P}/\mathcal{P}_{max}}\Big)\Big]^2xy}.
    \end{equation}
\end{widetext}
\subsubsection{Upper bound on EMP}
The generalized figure of merit $y$ for the voltage-temperature probe heat engine is bounded due to the constraints on the Onsager coefficients as discussed in Eq. (\ref{boundvt}), and it is expressed as follows:
\begin{equation}\label{hyx}
    \begin{aligned}
        H\le y\le 0\,\,\ \mathrm{for}\,\,\ x\le 0,\\
        H\ge y\ge 0\,\,\ \mathrm{for}\,\,\ x\ge 0,
    \end{aligned}
\end{equation}
where the bound function $H$ for the voltage-temperature probe heat engine is defined as
\begin{equation}
    H=\frac{x}{(x-1)^2}.
\end{equation}
By setting $y=H(x)$ in Eq. (\ref{etapmaxvt}), we can obtain the upper bound on the EMP for an MNL voltage-temperature probe heat engine as follows:
\begin{equation}\label{npmaxb_vtprobe}
    \eta^*(\mathcal{P}_{max})=\eta_C\frac{x^2}{2(2x^2-3x+2)-\alpha TX_L^T x^2}.
\end{equation}
Equation (\ref{npmaxb_vtprobe}) is a universal upper bound on the EMP for an MNL voltage-temperature probe heat engine with broken time-reversal symmetry.\\
\indent
For time-reversal symmetric case ($x=1$) with $0\le\alpha\le1$, the above upper bound reduces to
\begin{equation}
    \frac{\eta_C}{2}\le\eta^*(\mathcal{P}_{max})\le\frac{\eta_C}{2-\eta_C},
\end{equation}
where $\eta_C=TX_L^T$. Similar upper bounds on the EMP are also obtained for a two-terminal system with time-reversal symmetry \cite{Izumida1, Izumida2, Iyyappan1}.
We find that introducing nonlinearity and the broken time-reversal symmetry helps enhance the upper bound on the EMP, allowing one to overcome the CA limit described in the linear response regime.
\section{Triple-dot AB Heat Engine}\label{sec:TQD}
\begin{figure}[t!]
    \centering
    \includegraphics[scale=0.29]{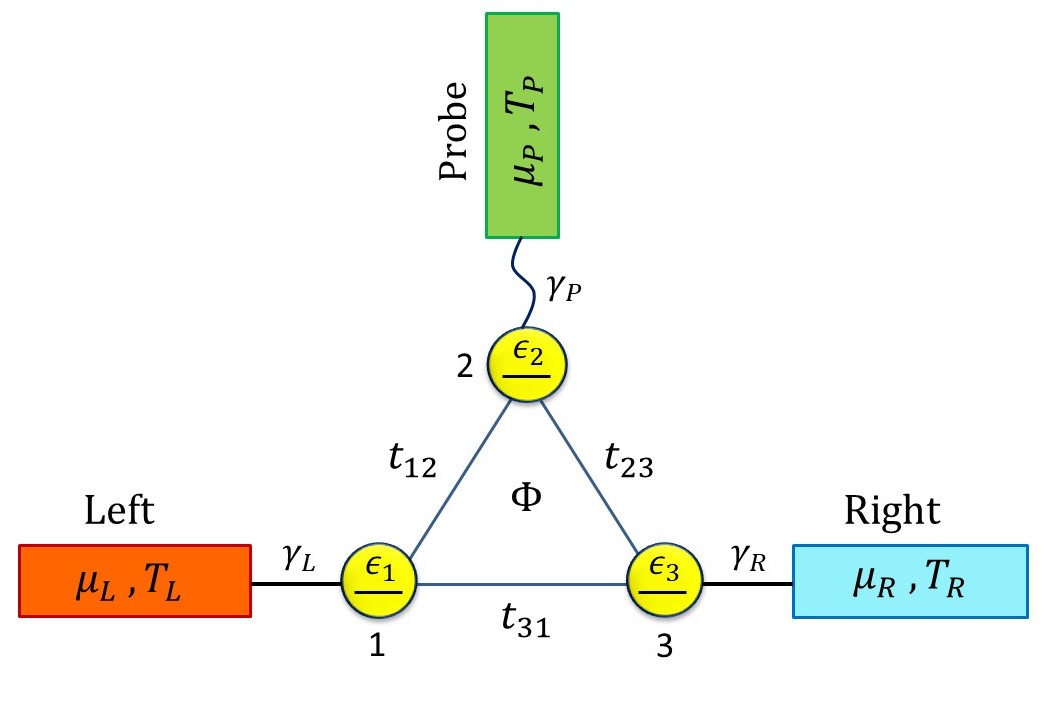}
    \caption{Model of a three-terminal triple-dot AB heat engine consisting of three quantum dots arranged in a triangular geometry and a magnetic flux $\Phi$ pierces the triangular AB ring perpendicularly. Dots 1 and 3 are connected to two metallic leads (reservoirs) termed left ($L$) and right ($R$), respectively, which are maintained at different temperatures and chemical potentials. Dot 2 is connected to a probe ($P$) that introduces inelastic scattering processes into the system. We set $T_L>T_R$ and $\mu_L<\mu_R$ to ensure our system works as a thermoelectric heat engine. The probe parameters ($T_P, \mu_P$) are adjusted to satisfy the condition for a voltage probe and voltage-temperature probe.}
    \label{fig:TQD}
\end{figure}
We consider a three-terminal setup of triple-quantum-dot AB interferometer \cite{TQD1, TQD2, TQD3, BANDYOPADHYAY}, as illustrated in Fig. \ref{fig:TQD}, to numerically investigate the theory developed in the previous sections. Transport in such a triangular triple-quantum-dot geometry has been experimentally demonstrated in Refs. \cite{Gaudreau, Rogge_2009, Amaha}. Here, three QDs are arranged in a triangular geometry with each dot located at the vertex of an equilateral triangle and interconnected through quantum tunneling. A magnetic flux $\Phi$ threads the triangular ring perpendicularly. Dots 1 and 3 are connected to two reservoirs termed left ($L$) and right ($R$), respectively, maintained at different temperatures ($T_L>T_R$) and chemical potentials ($\mu_L<\mu_R$). A probe ($P$) is connected to dot 2 to introduce inelastic scattering mechanisms into the system. The probe parameters ($T_P, \mu_P$) are adjusted to examine a voltage probe and a voltage-temperature probe for our setup. Here, we do not consider the electron-electron interactions and the spin degrees of freedom for simplicity. Thus, we can describe each QD by a spin-less electronic level and disregard the Zeeman effect. The total Hamiltonian $\hat{H}$ of the whole system is expressed as,
\begin{equation}\label{eq1}
    \hat{H} =\hat{H}_{TQD}+\hat{H}_B+\hat{H}_{TQD,B}+\hat{H}_P+\hat{H}_{TQD,P},
\end{equation}
where, $\hat{H}_{TQD}$ is the subsystem Hamiltonian for the triple QD system, $\hat{H}_B$ and $\hat{H}_P$ represent the bath (reservoir) Hamiltonian, and the probe Hamiltonian, respectively. $\hat{H}_{TQD, B}$ and $\hat{H}_{TQD, P}$ are the Hamiltonian for the interaction of the triple-dot system with the bath and the probe, respectively. We represent the individual Hamiltonian in the second quantization form. The subsystem Hamiltonian consisting of three interconnected dots is given as \cite{BANDYOPADHYAY}
\begin{equation}\label{eq2}
    \begin{split}
        \hat{H}_{TQD} =\sum_{i=1,2,3}\epsilon_i\hat{d}_i^{\dagger}\hat{d}_i+(t_{12}d_1^\dagger d_2e^{i\phi_{12}}
        +t_{23}d_2^\dagger d_3e^{i\phi_{23}}\\
        +t_{31}d_3^\dagger d_1e^{i\phi_{31}}+h.c.),
    \end{split}
\end{equation}
where, $\epsilon_i$ represents the energy of the $i^{th}$ dot, $\hat{d}_i^{\dagger}$ ($\hat{d}_i$) is the creation (annihilation) operator of the electrons in the $i^{th}$ dot. The three quantum dots are connected by the inter-dot tunneling strength $t_{ij}$ and $\phi_{ij}$ is the AB phase factor acquired by the electron wave while tunneling through the dots.\\
\indent The Hamiltonian for the two reservoirs (leads) is given by
\begin{equation}\label{eq3}
    \hat{H}_B =\sum_{k,\nu\in{L,R}}\epsilon_{\nu,k}\hat{c}_{\nu,k}^{\dagger}
    \hat{c}_{\nu,k}.
\end{equation}
Here $\epsilon_{\nu,k}$ is the energy of the $k^{th}$ state in the reservoir $\nu$ and $\hat{c}_{\nu,k}^{\dagger}$ and $\hat{c}_{\nu,k}$ are the creation and annihilation operators of the electrons for the reservoirs.\\
The system-bath interaction Hamiltonian is represented as
\begin{equation}\label{eq4}
    \hat{H}_{TQD,B}=V_{1,L,k}\hat{d}_1^{\dagger}\hat{c}_{L,k}+
    V_{3,R,k}\hat{d}_3^{\dagger}\hat{c}_{R,k}+h.c..
\end{equation}
Here, $V_{1,L,k}$ and $V_{3,R,k}$ denote the interaction strength of the dot with the left ($L$) and the right ($R$) bath, respectively.
Similarly, the probe Hamiltonian is given by
\begin{equation}
    \hat{H}_P=\sum_{k}\epsilon_{P,k}\hat{c}_{P,k}^{\dagger}
    \hat{c}_{P,k},
\end{equation}
and the system-probe interaction Hamiltonian is given by
\begin{equation}
    \hat{H}_{TQD,P}=V_{2,P,k}\hat{d}_2^{\dagger}\hat{c}_{P,k}+h.c..
\end{equation}
Here, $\hat{c}_{P,k}^{\dagger}$ and $\hat{c}_{P,k}$ are the creation and annihilation operators of the electron in the $k^{th}$ state of the probe, respectively. $V_{2,P,k}$ is the interaction strength between dot and the probe.\\
\indent The AB phases $\phi_{ij}$ acquired by the electron's wave while tunneling through the dots satisfy the following relation \cite{BANDYOPADHYAY, JB}:
\begin{equation}\label{phi}
    \phi_{12}+\phi_{23}+\phi_{31}=\phi=2\pi\frac{\Phi}{\Phi_0},
\end{equation}
where $\Phi$ is the total magnetic flux enclosed by the triangular AB ring and $\Phi_0=h/e$ is the flux quantum. As the dots are at the vertices of an equilateral triangular loop and physical observables are gauge invariant in the steady-state, we may choose the gauge as $\phi_{12}=\phi_{23}=\phi_{31}=\phi/3$. We use the natural units convention $\hbar=c=e=k_B=1$ for simplicity.\\
\begin{figure*}[t!]
    \centering
    \includegraphics[scale=0.12]{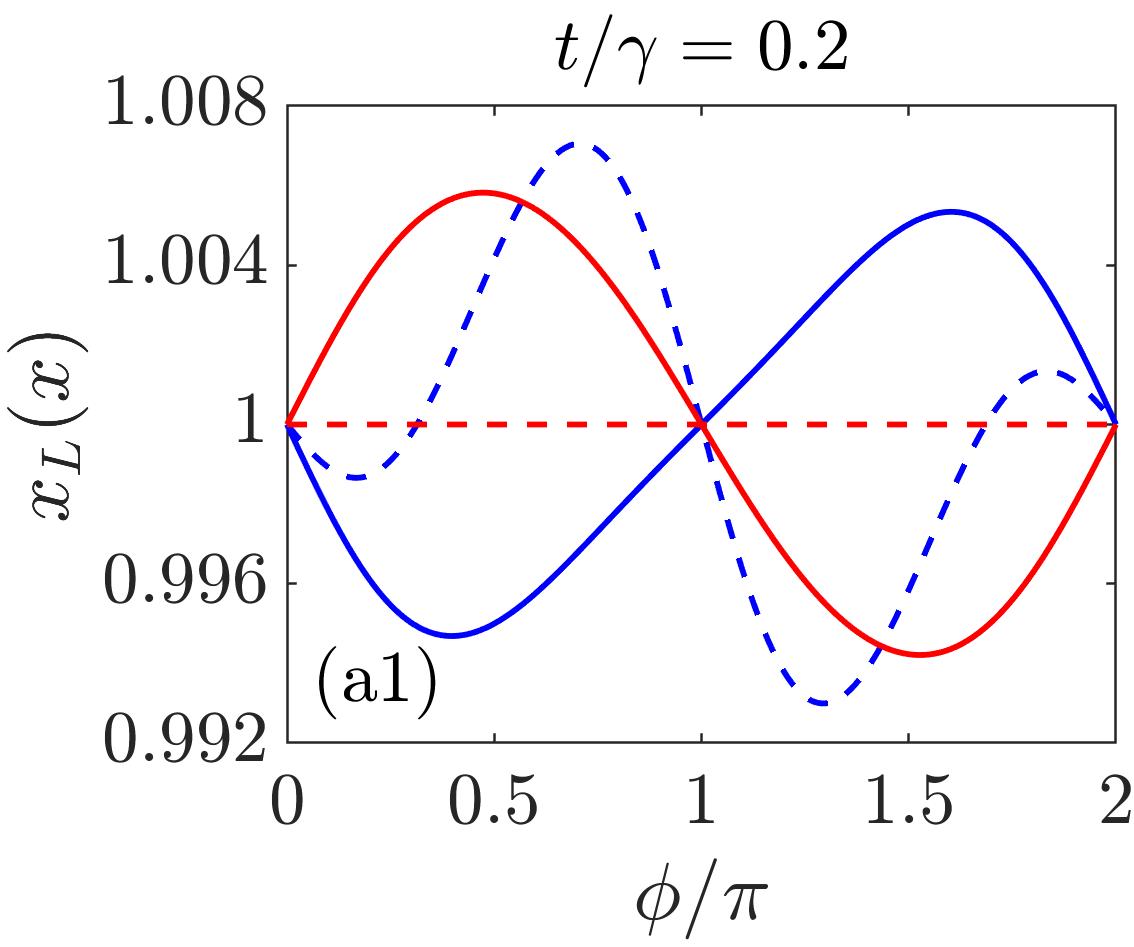}
    \includegraphics[scale=0.12]{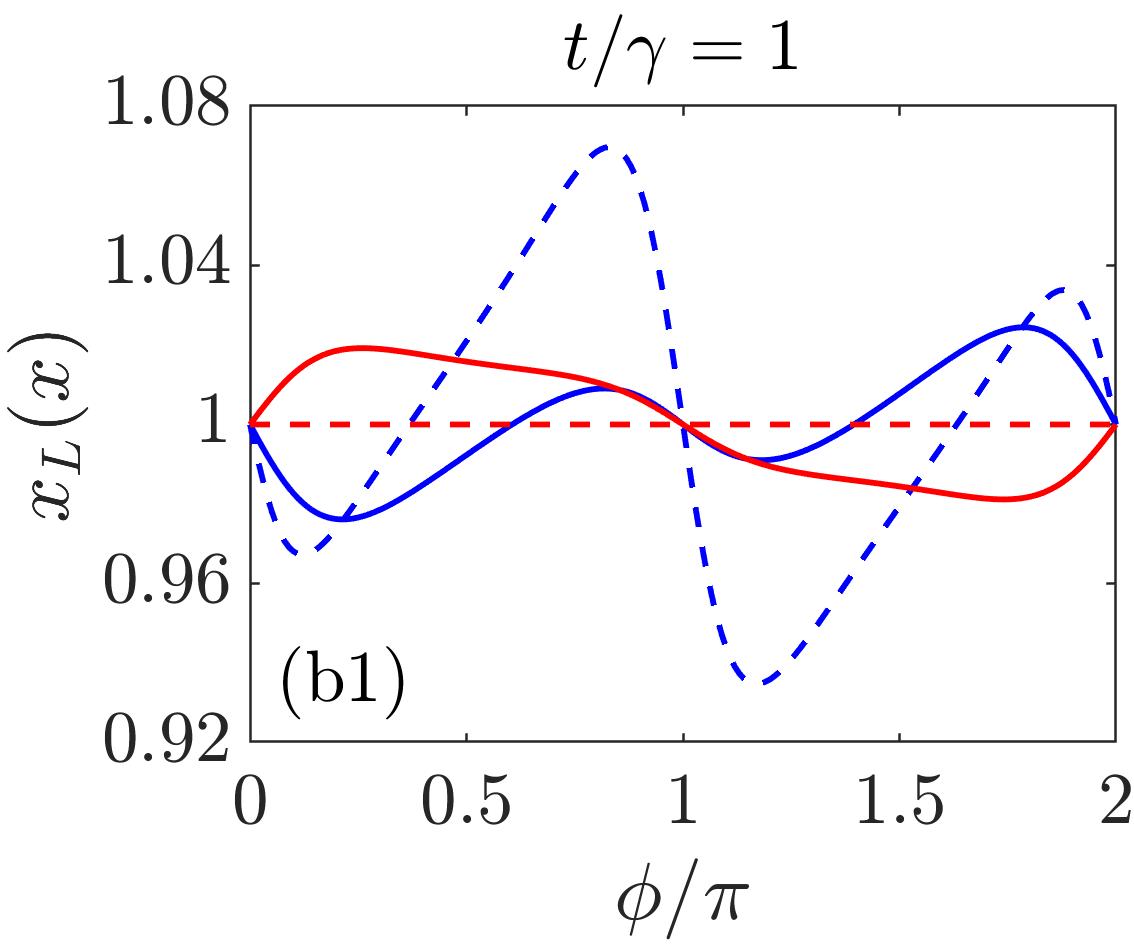}
    \includegraphics[scale=0.12]{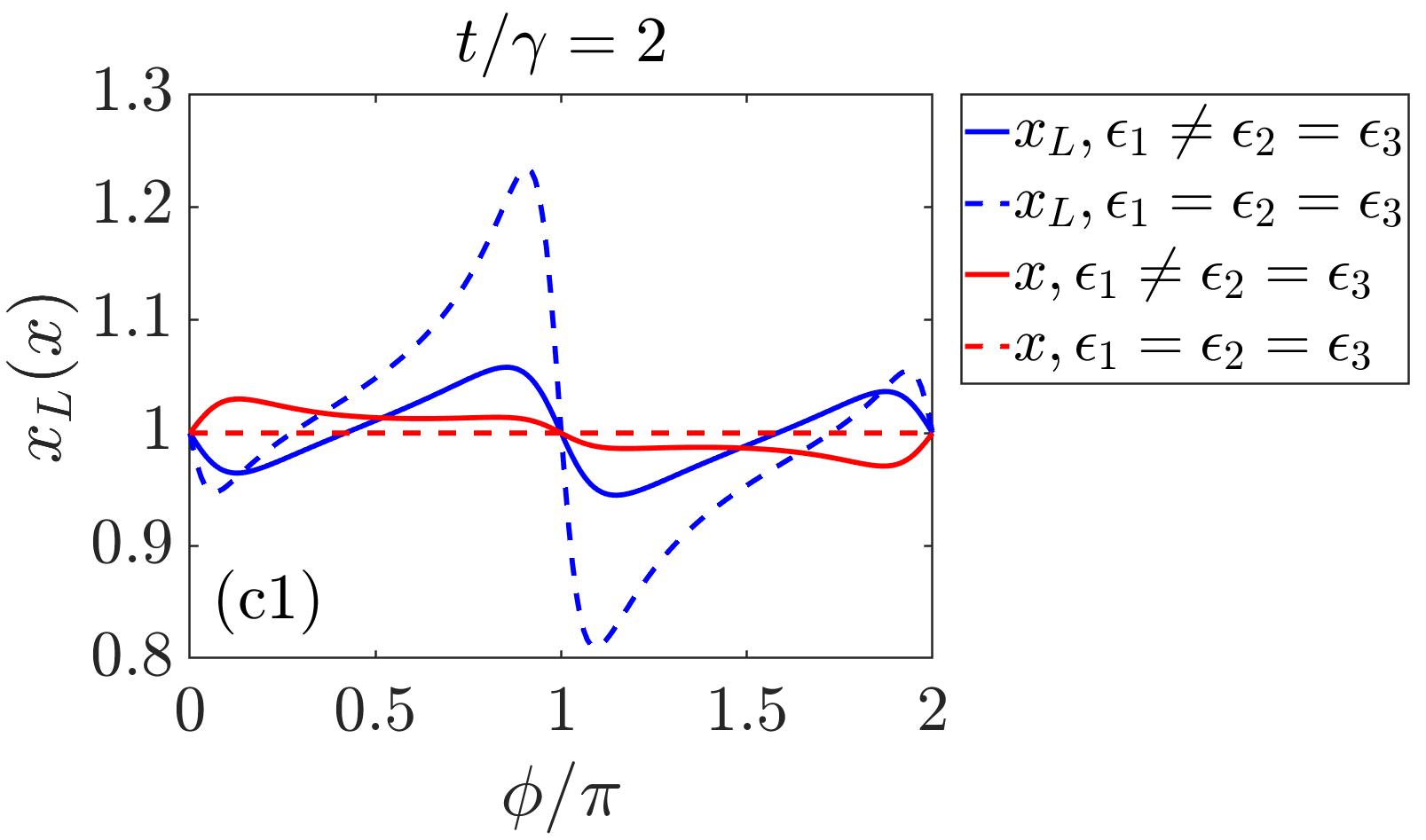}
    \includegraphics[scale=0.12]{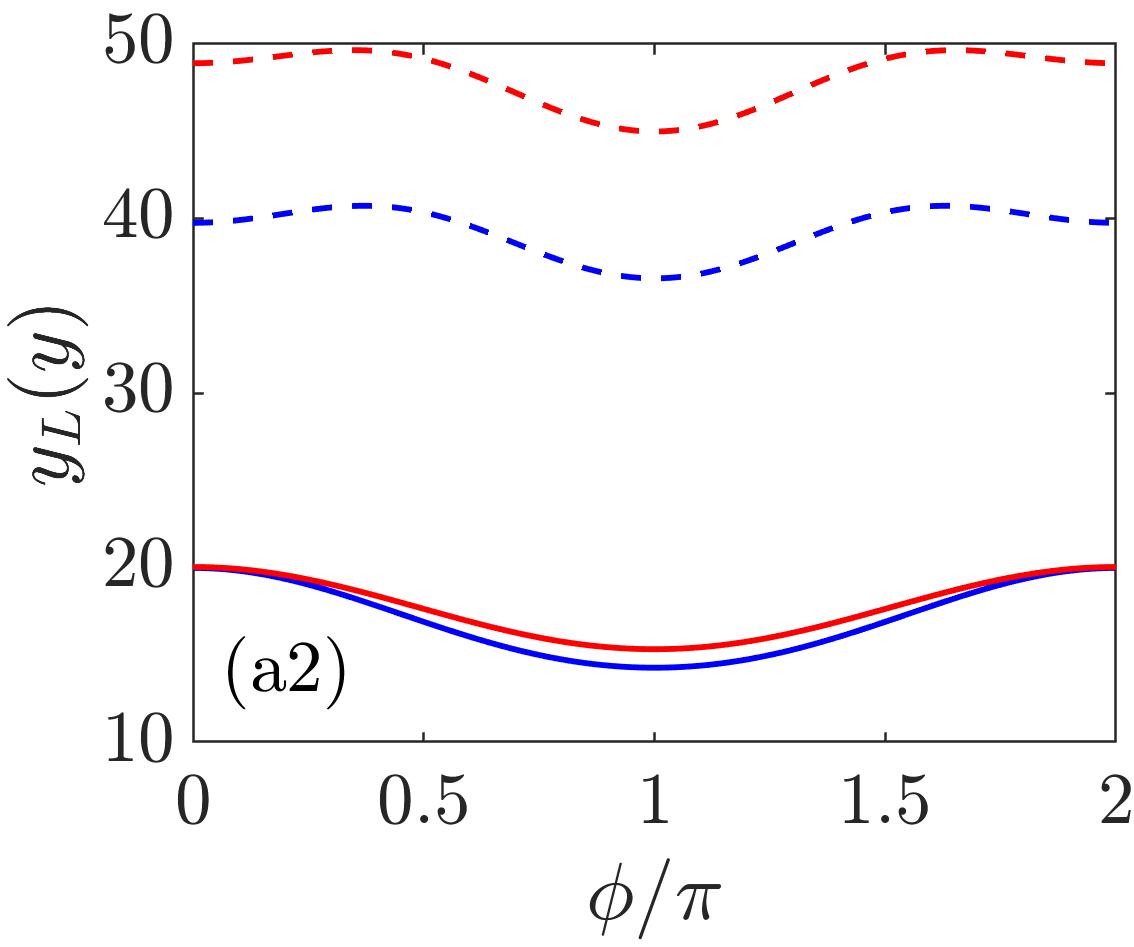}
    \includegraphics[scale=0.12]{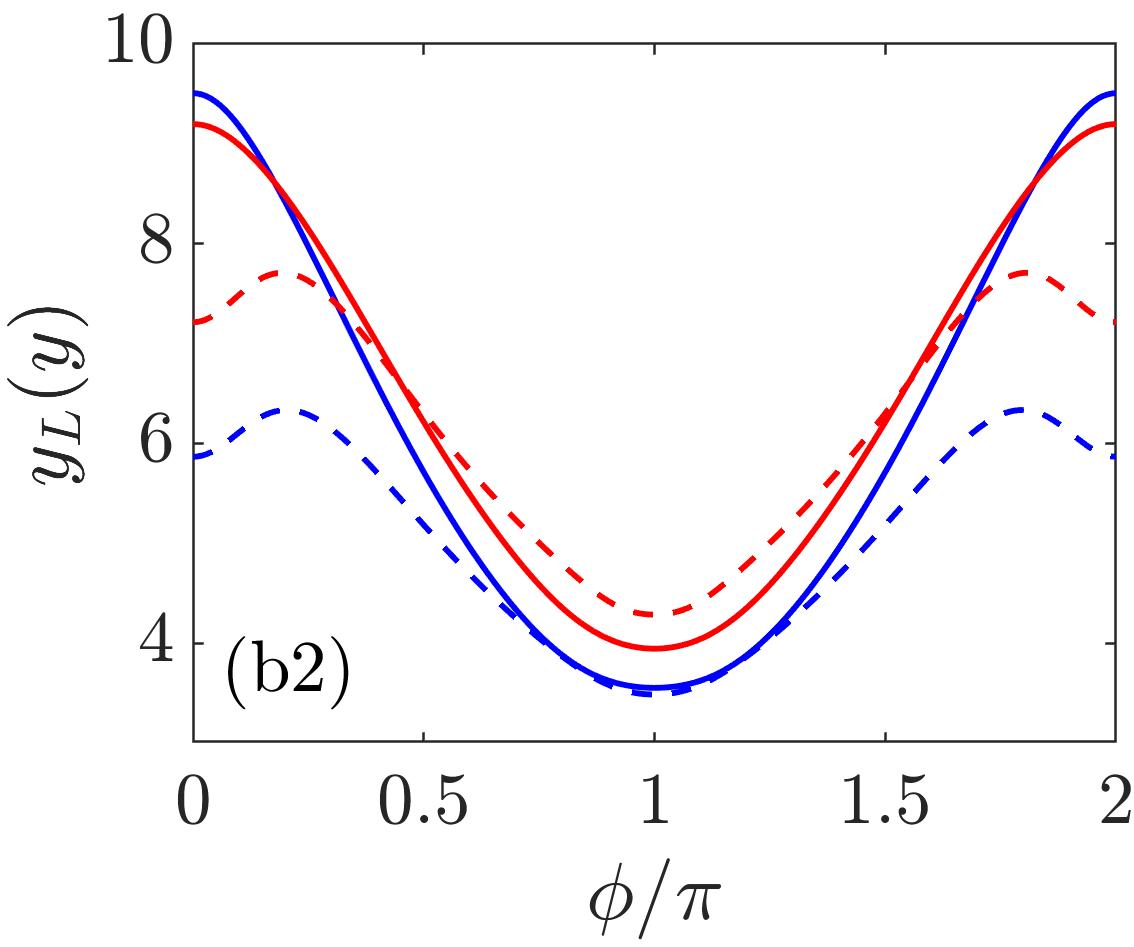}
    \includegraphics[scale=0.12]{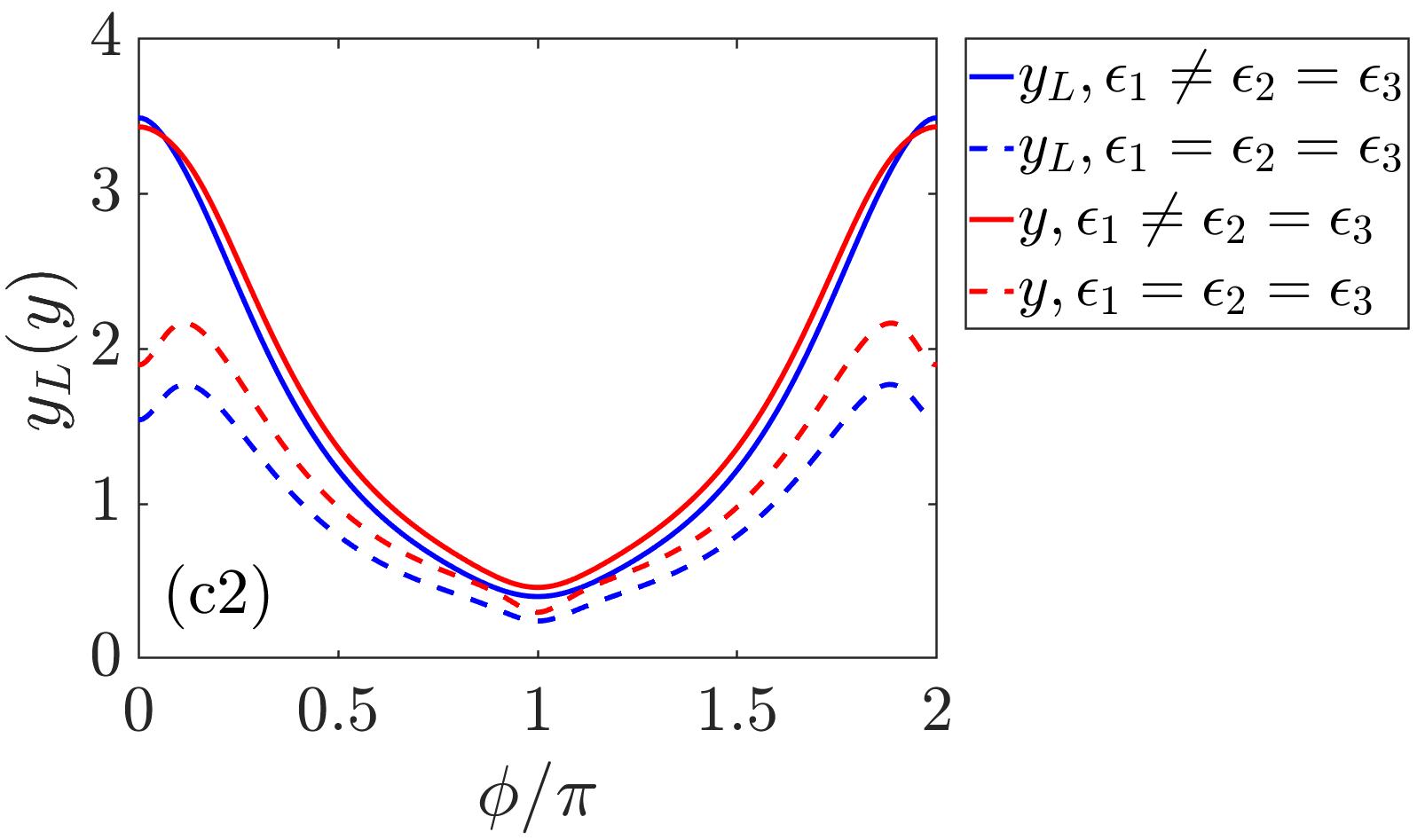}
    \caption{Asymmetry parameter $x_L (x)$ and the generalized figure of merit $y_L (y)$ as a function of $\phi$ for the voltage probe (blue) and voltage-temperature probe (red) heat engines in three different regimes of $t/\gamma$ ratio: (a1-a2) $t/\gamma=0.2$, (b1-b2) $t/\gamma=1$, and (c1-c2) $t/\gamma=2$. Here, the solid lines denote the non-degenerate dot setup ($\epsilon_1\ne\epsilon_2=\epsilon_3$) and the dashed lines denote the degenerate dot setup ($\epsilon_1=\epsilon_2=\epsilon_3$). Parameters used are: $\epsilon_1=0.6$ (solid line), $\epsilon_1=0.5$ (dashed line), $\epsilon_2=\epsilon_3=0.5$, $\gamma=\gamma_L=\gamma_R=\gamma_P=0.05$, $T_R=T=0.1$, $T_L=0.13$, $T_P=0.102$ (for the voltage probe), $\mu_R=\mu=0.3$.}
    \label{fig:xLyL}
\end{figure*}
\indent
We use the Nonequilibrium Green's Function (NEGF) approach \cite{meir1992, wangNEGF, textbook} to solve the model Hamiltonian and compute the observables. We solve the Heisenberg equation of motion for both the bath and the subsystem variables (see Ref. \cite{BANDYOPADHYAY, JB, Dhar2006} for detailed calculations) and obtain the retarded [$G^+(\omega)$] and advanced [$G^-(\omega)$] Green's function for our system by using the quantum Langevin equation \cite{Dhar2006}.
\begin{figure*}
    \centering
    \includegraphics[scale=0.12]{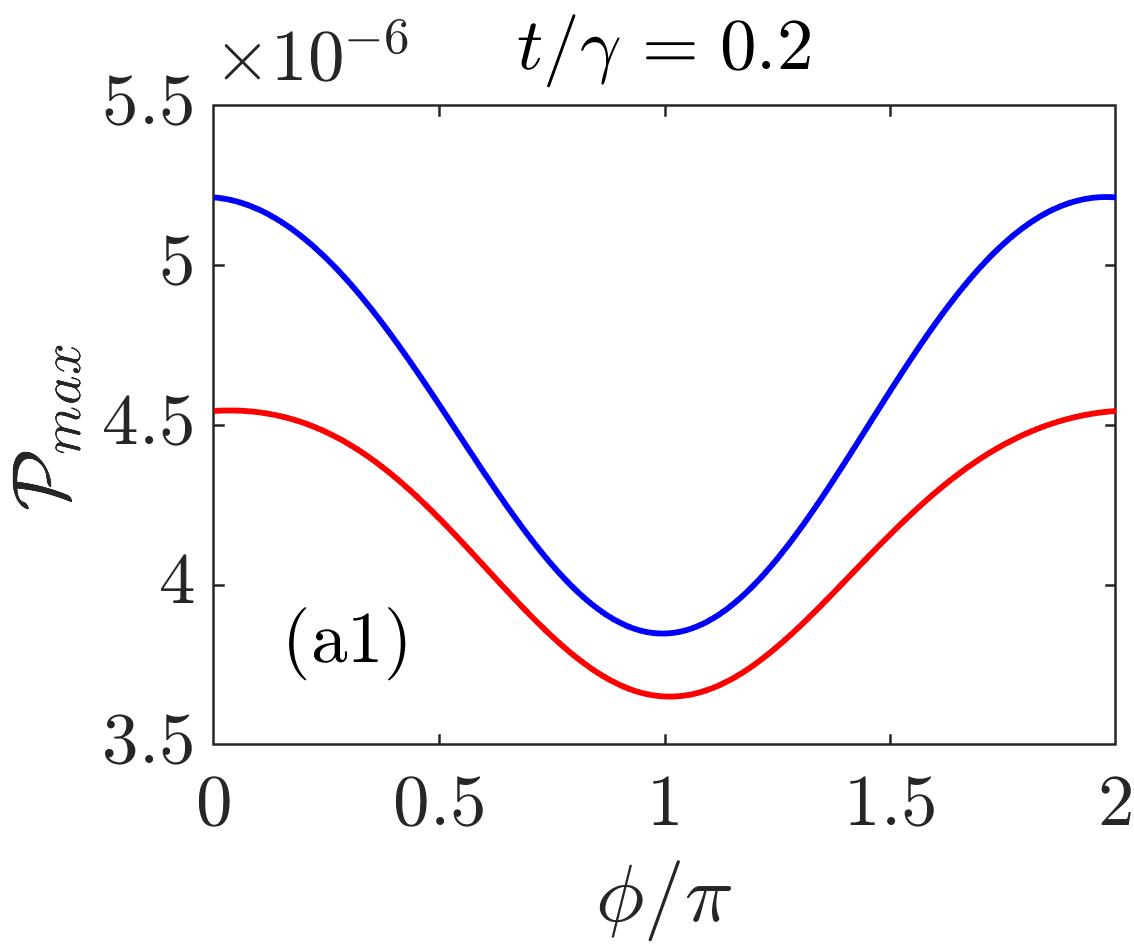}
    \includegraphics[scale=0.12]{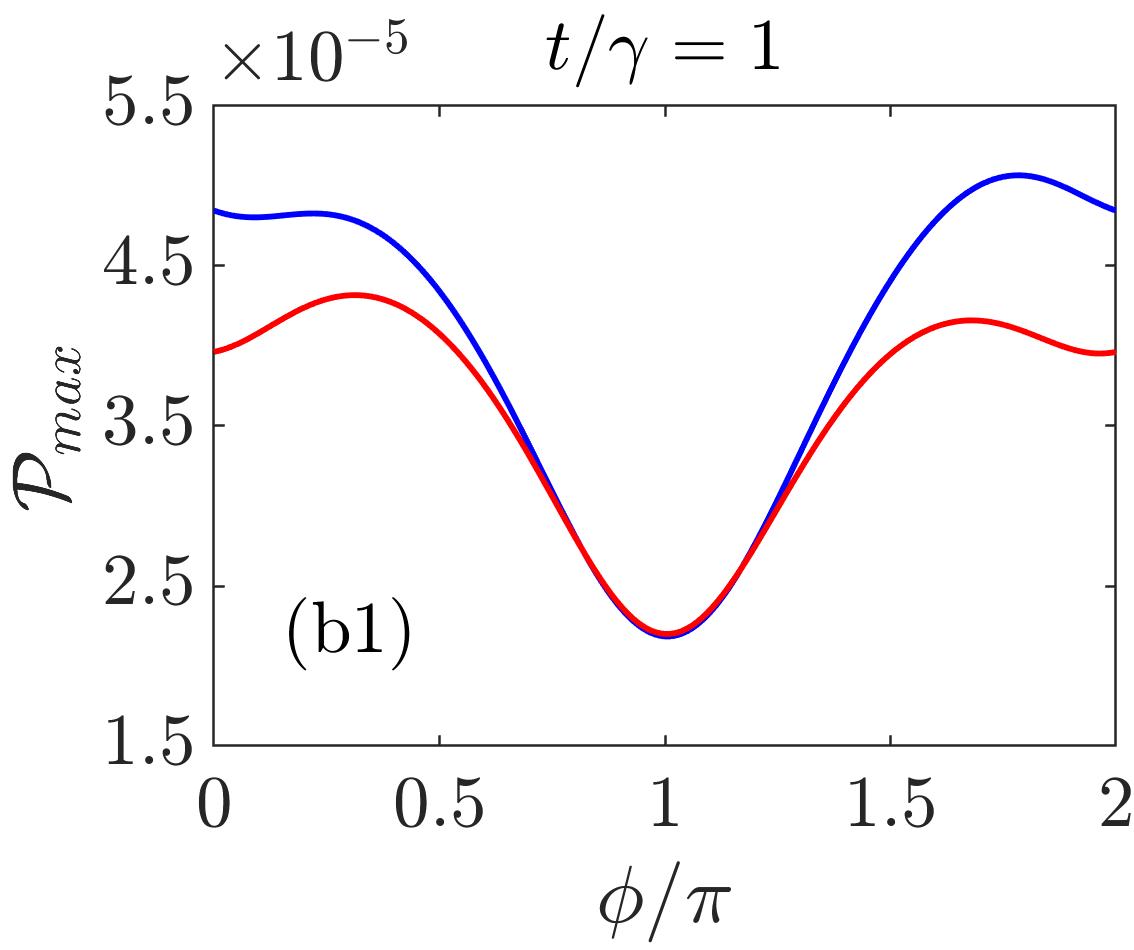}
    \includegraphics[scale=0.12]{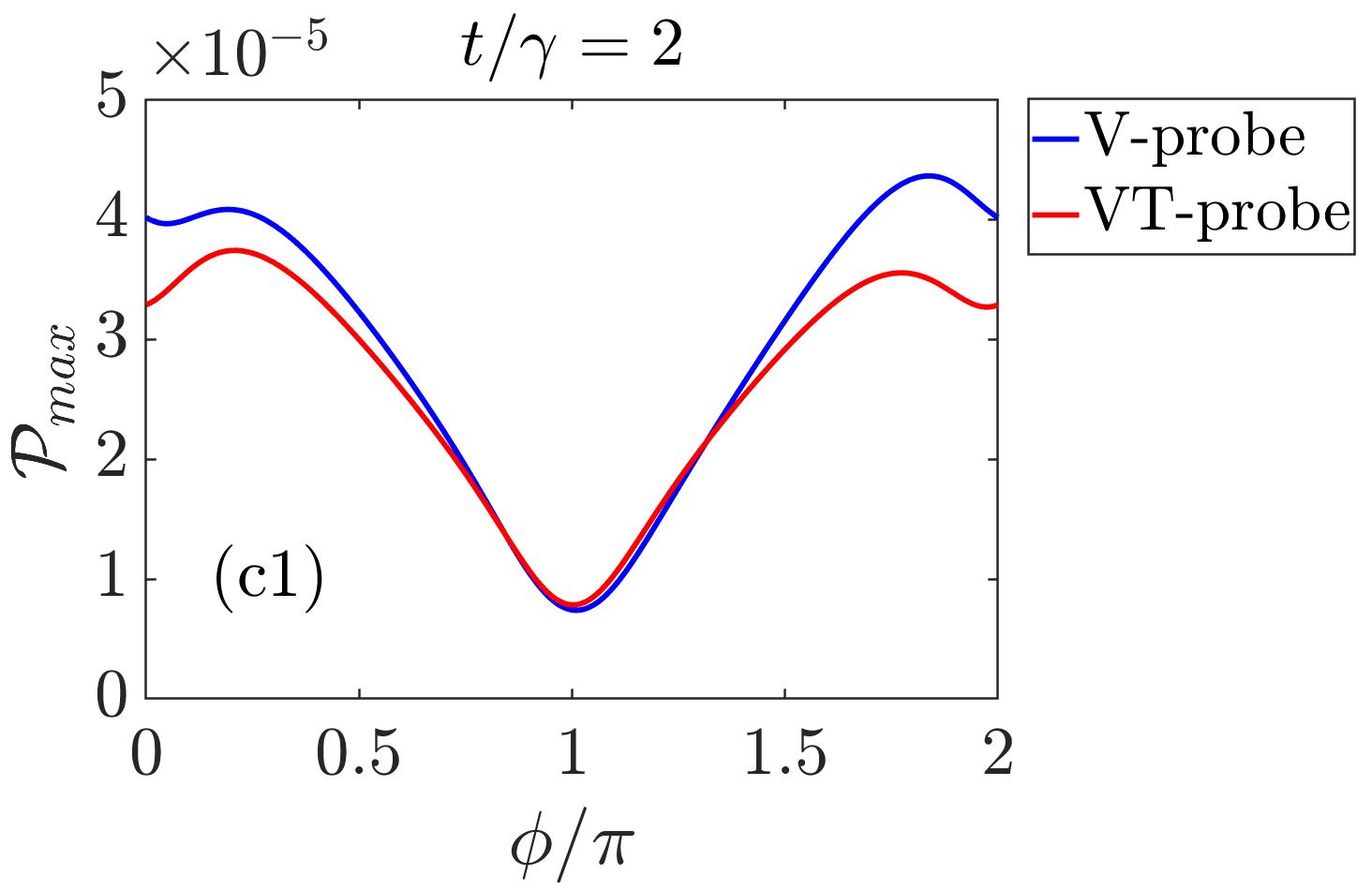}
    \includegraphics[scale=0.12]{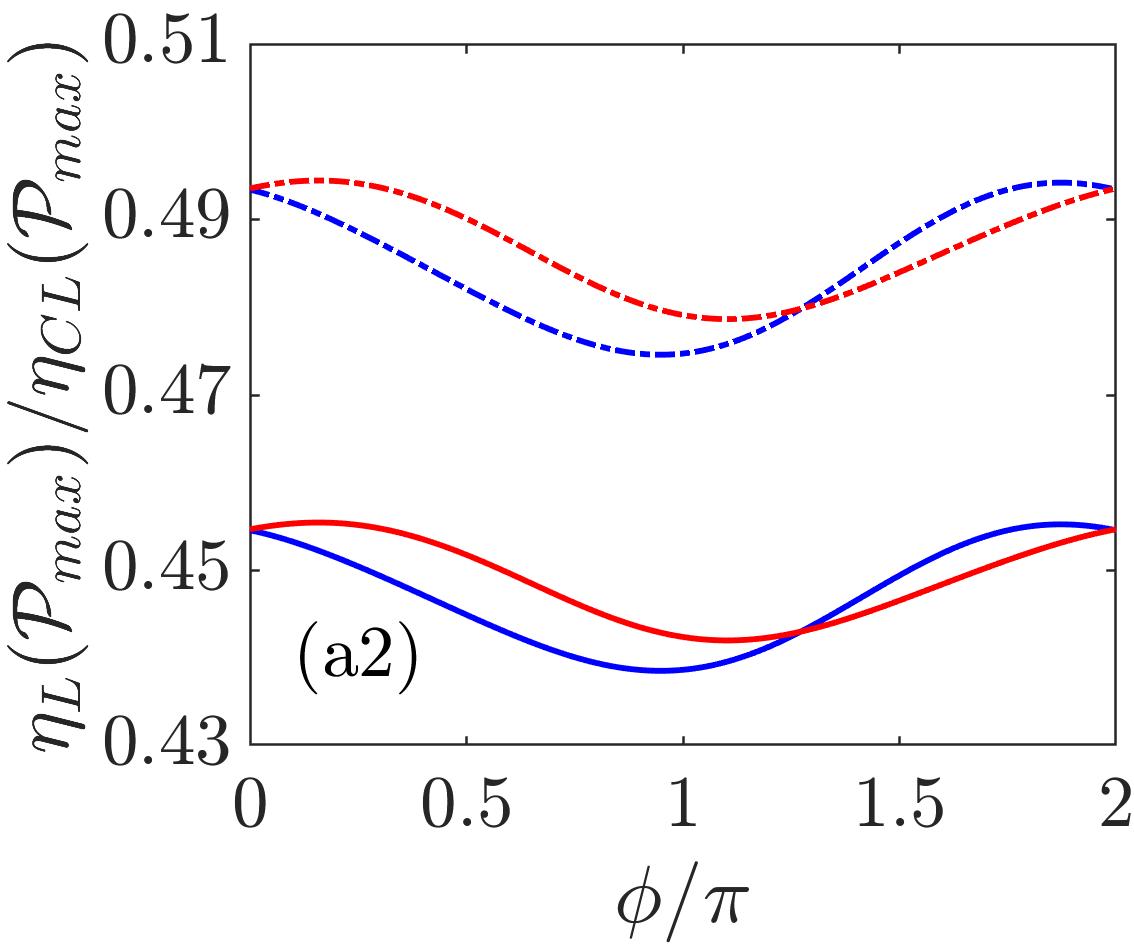}
    \includegraphics[scale=0.12]{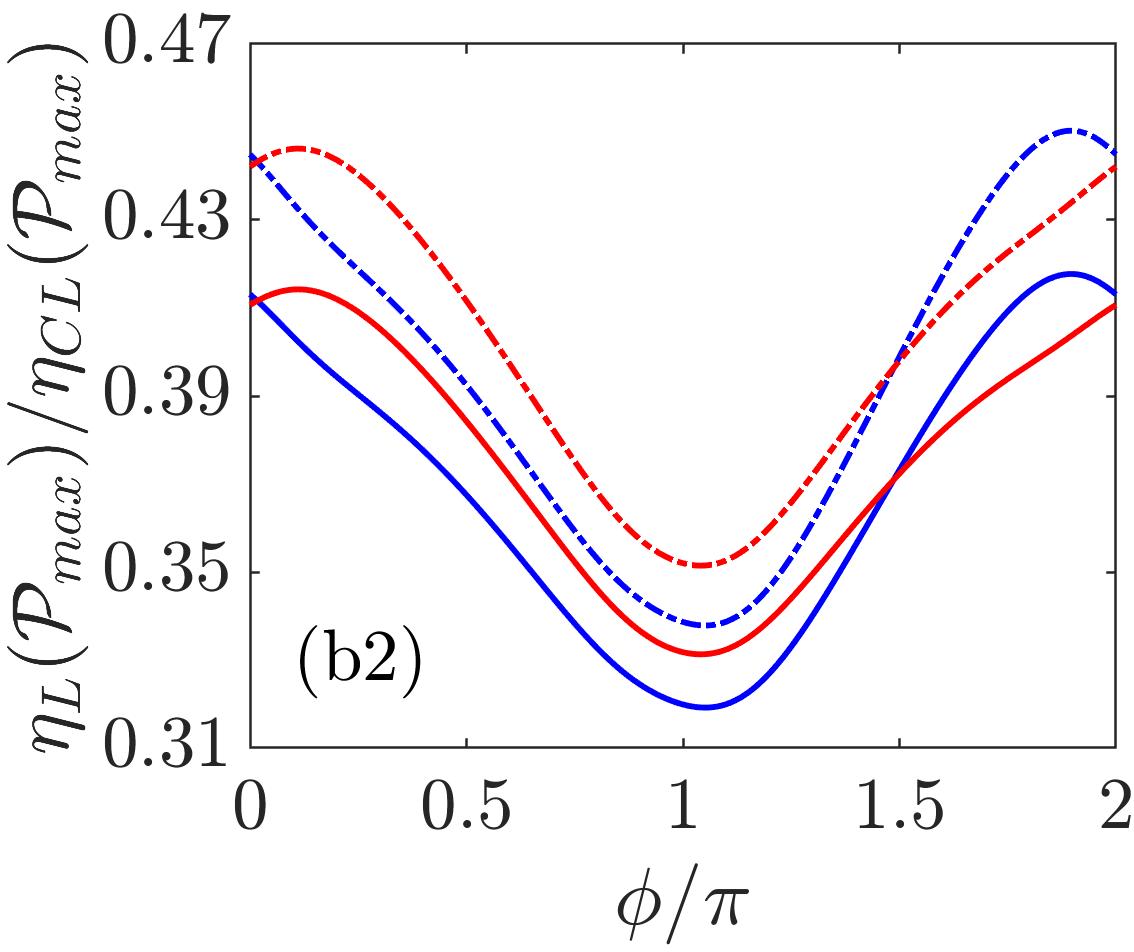}
    \includegraphics[scale=0.12]{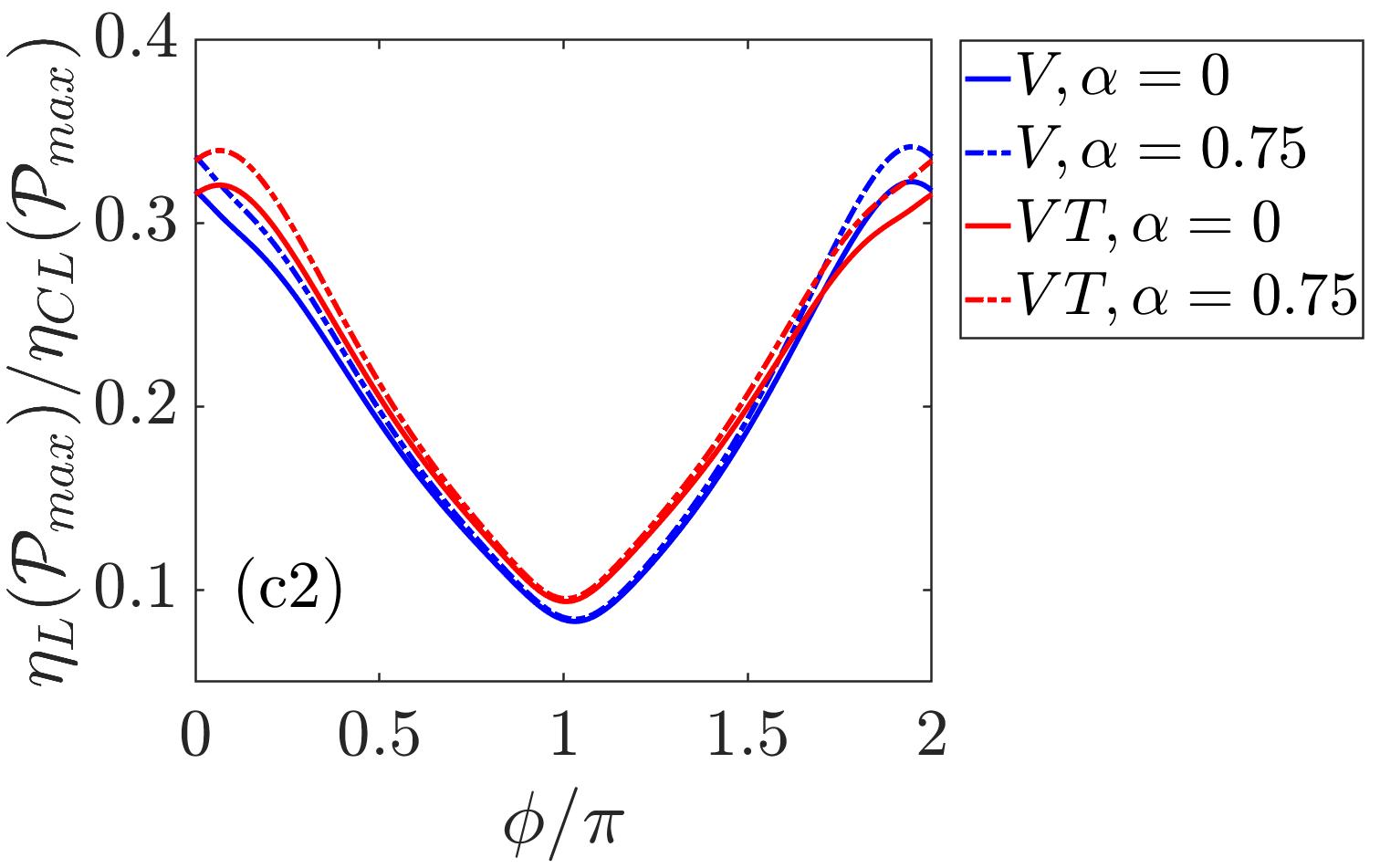}
    \caption{Comparison of thermoelectric performance between the voltage probe (blue) and voltage-temperature probe (red) heat engines in the MNL regime: maximum power, $\mathcal{P}_{max}$ and EMP, $\eta_L(\mathcal{P}_{max})/\eta_{CL}(\mathcal{P}_{max})$ as a function of $\phi$ in three different regimes of $t/\gamma$ ratio: (a1-a2) $t/\gamma=0.2$, (b1-b2) $t/\gamma=1$, and (c1-c2) $t/\gamma=2$. Note that the EMP for the voltage-temperature probe heat engine is denoted by $\eta(\mathcal{P}_{max})/\eta_{C}$ in the main text. Parameters used are: $\epsilon_1=0.6$, $\epsilon_2=\epsilon_3=0.5$, $\gamma=\gamma_L=\gamma_R=\gamma_P=0.05$, $T_R=T=0.1$, $T_L=0.13$, $\mu_R=\mu=0.3$, $T_P=0.102$ (for the voltage probe).}
    \label{fig:vvsvt}
\end{figure*}
The retarded Green's function which gives the dynamics of the electrons inside the system is defined as \cite{BANDYOPADHYAY}
\begin{equation}
    G^{+}(\omega)=\big[\omega I-H_{TQD}-\Sigma^+_L(\omega)-\Sigma^+_R(\omega)-\Sigma^+_P(\omega)\big]^{-1},
\end{equation}
where $I$ is a ($3\times3$) identity matrix and $\Sigma_{\nu}^+(\omega)$ is the self-energy of the reservoir $\nu=L, P, R$. For simplicity, we use the wide-band limit approximation where the real part of the self-energy vanishes when the density of states of the leads are energy independent and only the imaginary part of the self-energy exists which is given in terms of the hybridization matrix $\Gamma_{\nu}$ as $\Sigma^{\pm}_{\nu}=\mp i\Gamma^{\nu}/2$, where
\begin{equation}
    \Gamma_{i,i^{\prime}}^{\nu} =2\pi\sum_{k}V_{i^{\prime},k}^{\nu^*}V_{i,k}^{\nu}\delta(\omega-\omega_k).
\end{equation}
By setting $t_{ij}=t$, the retarded Green's function for the whole system can be expressed in the matrix form as
    \begin{equation}\label{eq15}
\renewcommand{\arraystretch}{2}
    G^{+}(\omega)=
    \begin{pmatrix}
    \omega-\epsilon_1+i\frac{\gamma_L}{2} & -te^{i\phi/3} & -te^{-i\phi/3}\\
    -te^{-i\phi/3} & \omega-\epsilon_2+i\frac{\gamma_P}{2} & -te^{i\phi/3}\\
    -te^{i\phi/3} & -te^{-i\phi/3} & \omega-\epsilon_3+i\frac{\gamma_R}{2}
    \end{pmatrix}^{-1}.
\end{equation}
The hybridization matrices which account for the subsystem-bath coupling are given by
\begin{equation}\label{eq16}
    \Gamma^L=
    \begin{pmatrix}
    \gamma_L & 0 & 0\\
    0 & 0 & 0\\
    0 & 0 & 0
    \end{pmatrix},
    \Gamma^R=
    \begin{pmatrix}
    0 & 0 & 0\\
    0 & 0 & 0\\
    0 & 0 & \gamma_R
    \end{pmatrix},
    \Gamma^P=
    \begin{pmatrix}
    0 & 0 & 0\\
    0 & \gamma_P & 0\\
    0 & 0 & 0
    \end{pmatrix},
\end{equation}
where $\gamma_{\nu}$ represents the dot-reservoir coupling strength.\\
\indent
The transmission of electrons from reservoir $\nu$ to $\nu^{\prime}$ terminal is defined by the transmission function as \cite{datta1997electronic}:
\begin{equation}
T_{\nu\nu^{\prime}}={\rm Tr}[\Gamma^{\nu}G^+\Gamma^{\nu^{\prime}}G^-].
\end{equation}
Here the advanced Green's function $G^-(\omega)$ is the conjugate transpose of the retarded Green's function $G^+(\omega)$. \\
\begin{figure*}
    \centering
    \includegraphics[scale=0.12]{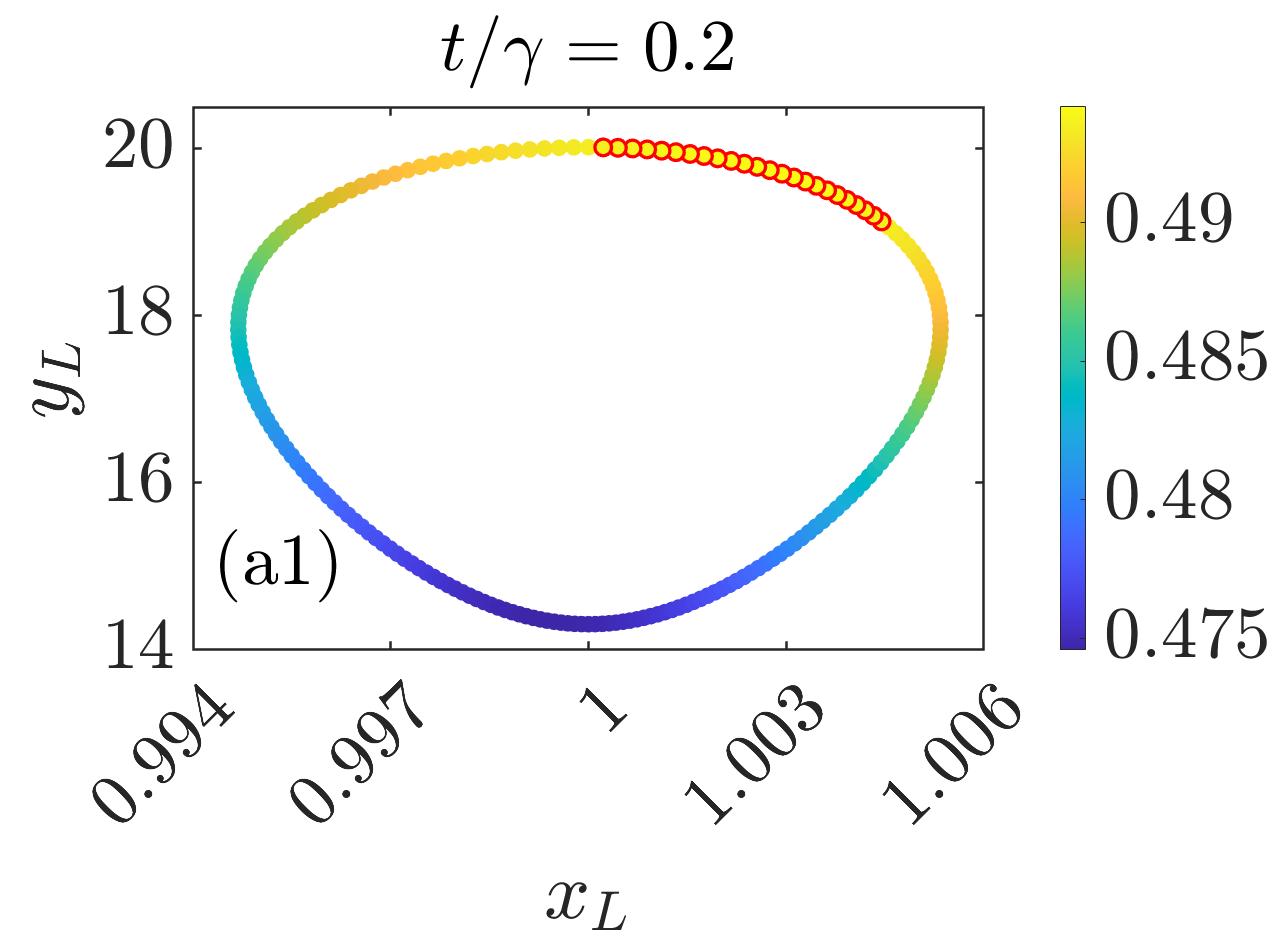}
    \includegraphics[scale=0.12]{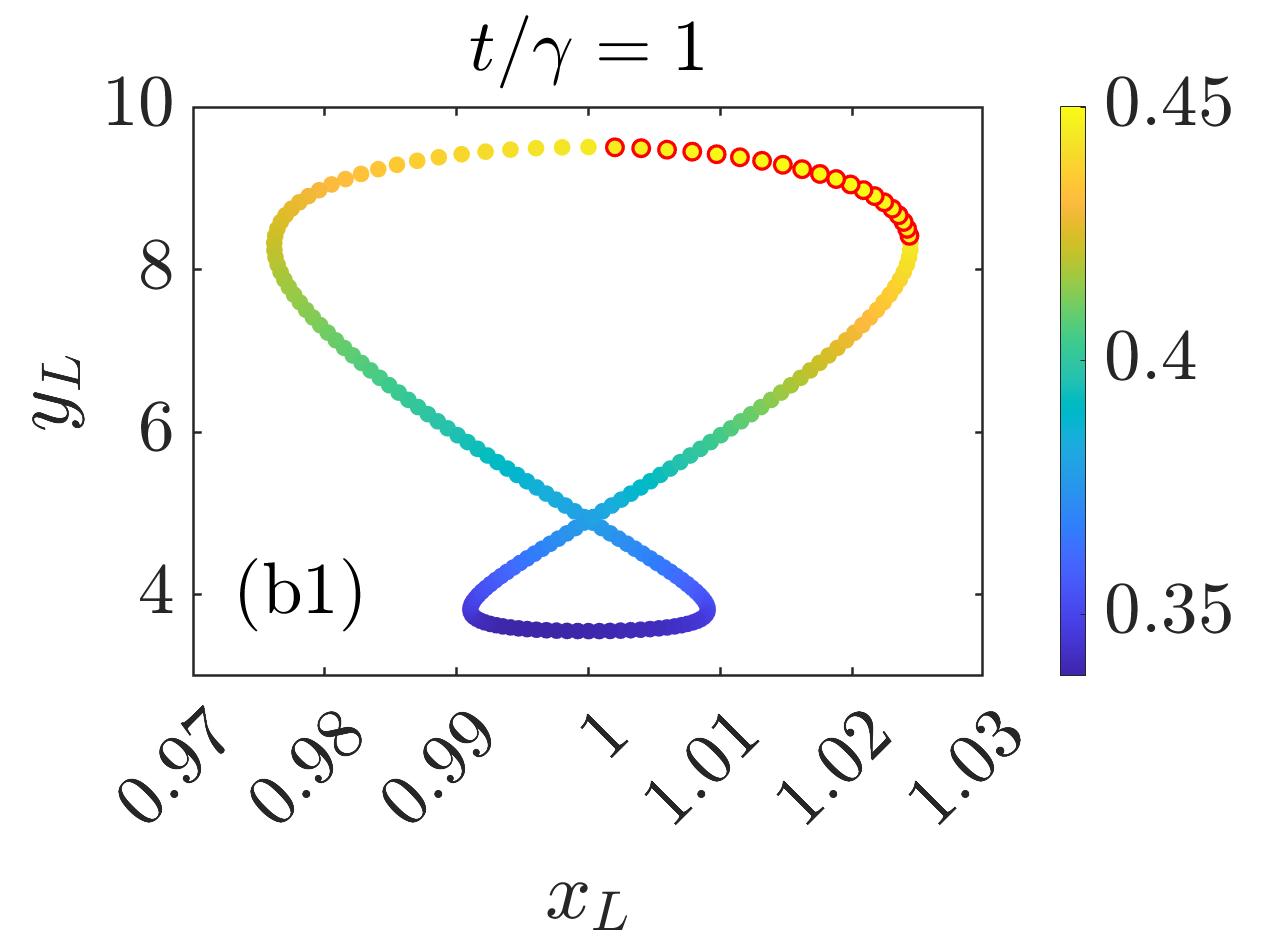}
    \includegraphics[scale=0.12]{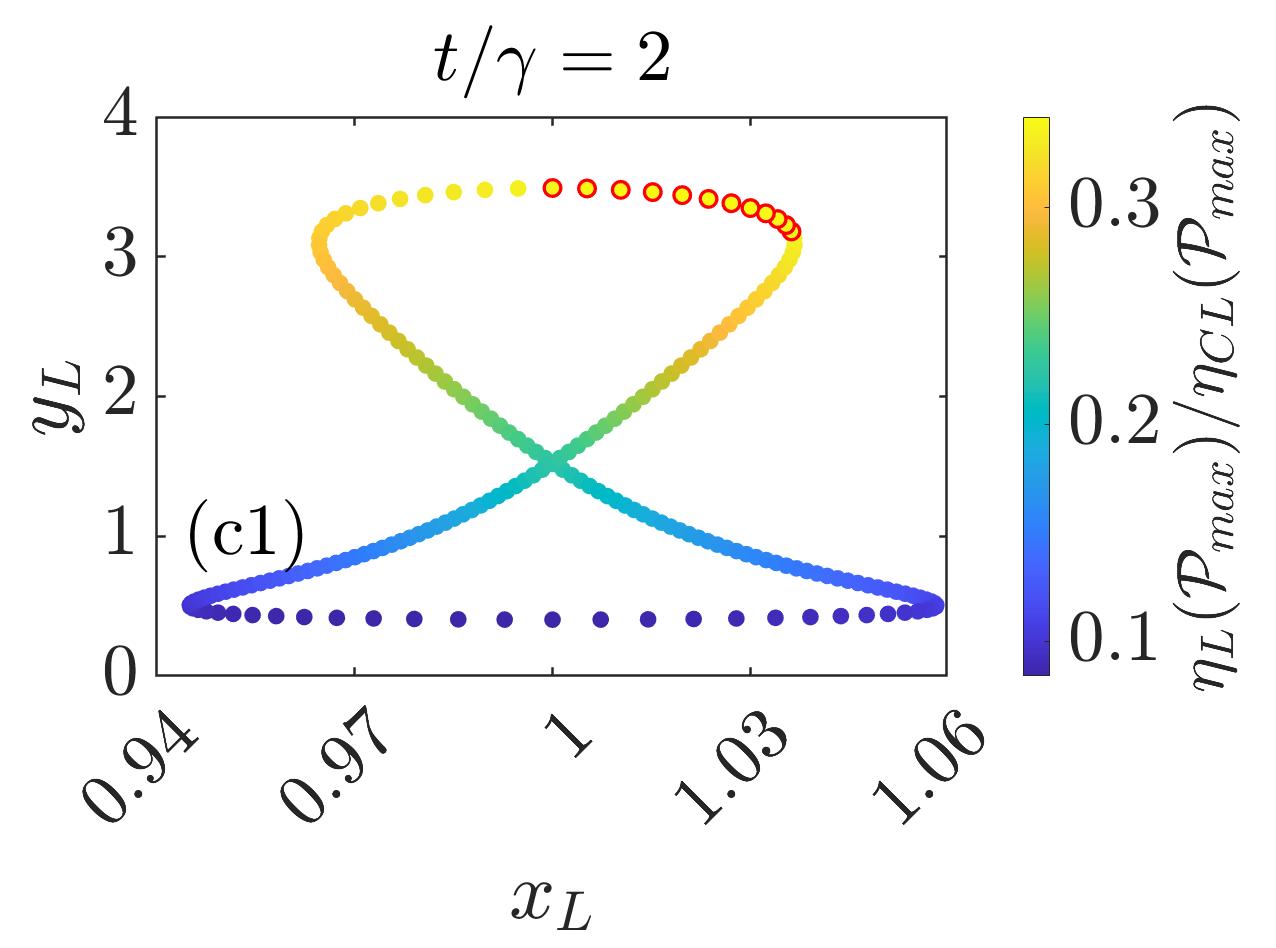}
    \includegraphics[scale=0.12]{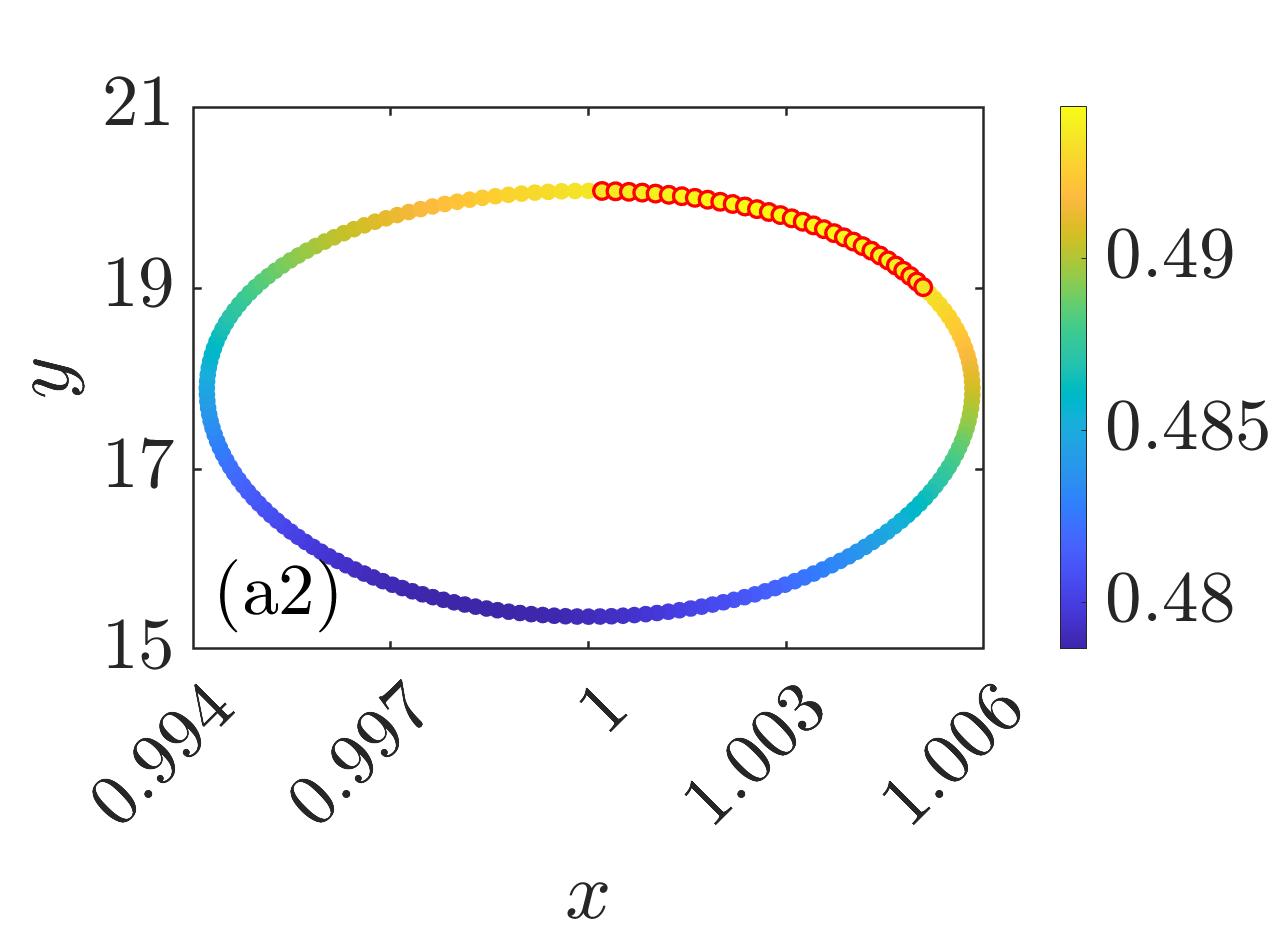}
    \includegraphics[scale=0.12]{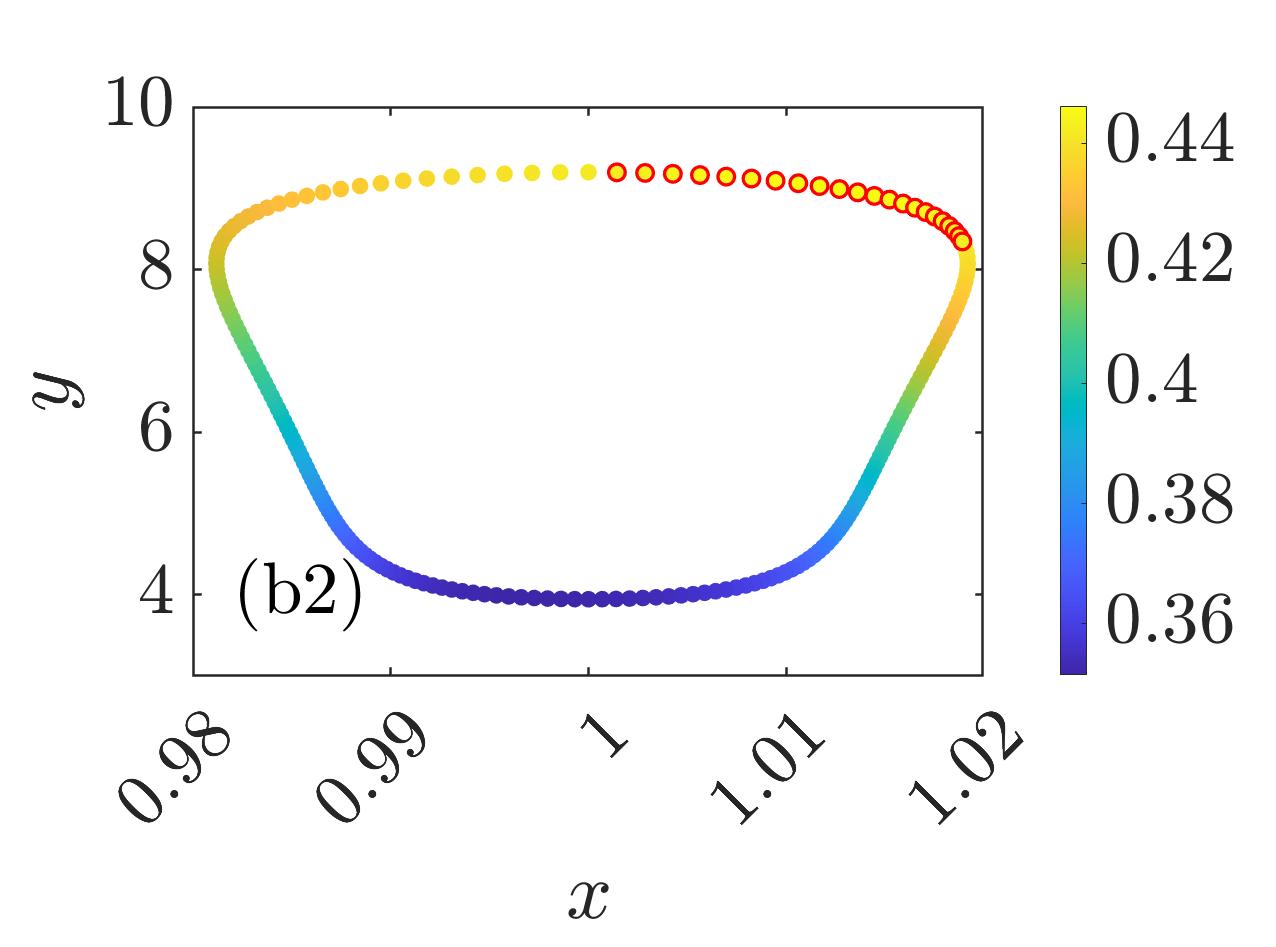}
    \includegraphics[scale=0.12]{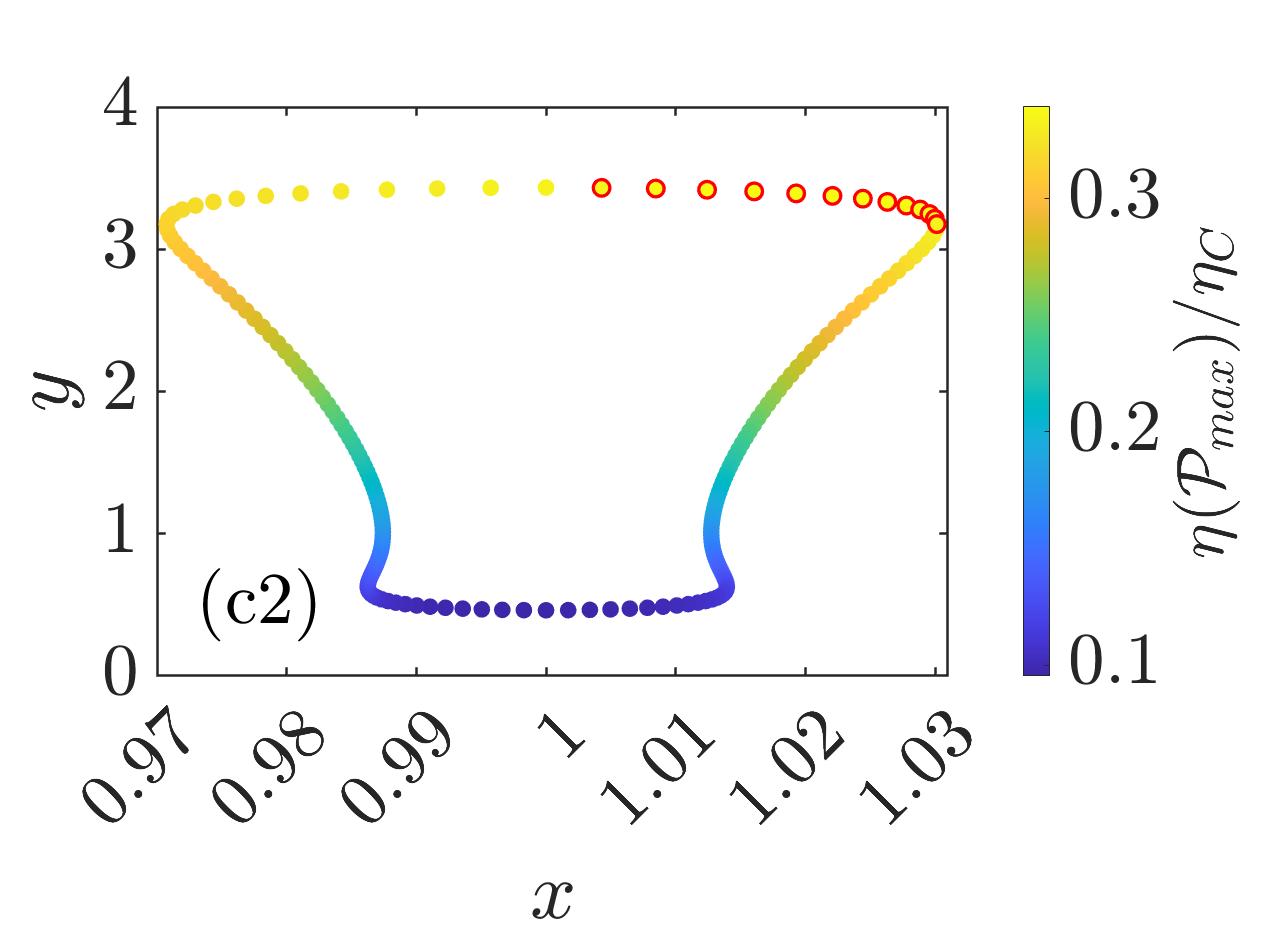}
    \caption{(Scatter plot for the EMP in the MNL regime: (Top) $\eta_L(P_{max})/\eta_{CL}(P_{max})$ as a function of $x_L$ and $y_L$ for the voltage probe heat engine and (Bottom) $\eta(P_{max})/\eta_{C}$ as a function of $x$ and $y$ for the voltage-temperature probe heat engine in three different regimes of $t/\gamma$ ratio: (a1-a2) $t/\gamma=0.2$, (b1-b2) $t/\gamma=1$, and (c1-c2) $t/\gamma=2$. The red-edged circles highlight
 points where the EMP exceeds that of the symmetric case. Parameters used are: $\epsilon_1=0.6$, $\epsilon_2=\epsilon_3=0.5$, $\gamma=\gamma_L=\gamma_R=\gamma_P=0.05$, $T_R=T=0.1$, $T_L=0.13$, $\mu_R=\mu=0.3$, $T_P=0.102$ (for the voltage probe), $\alpha=0.75$.}
    \label{fig:scatter}
\end{figure*}
\indent
The Landauer formula for the particle current ($J_{\nu}^N$) flowing from reservoir $\nu$ to the central system (with $\nu,\nu^{\prime}=L,P,R$) is defined as \cite{datta1997electronic, Sivan, Butcher_1990}
\begin{equation}\label{eq:particle}
\begin{split}
    J_{\nu}^N=\int_{-\infty}^{\infty} d\omega \Big[\sum_{\nu^{\prime}\ne\nu}T_{\nu\nu^{\prime}}(\omega,\phi)f_{\nu}(\omega)\\
    -\sum_{\nu^{\prime}\ne\nu}T_{\nu^{\prime}\nu}(\omega,\phi)f_{\nu^{\prime}}(\omega)\Big].
\end{split}
\end{equation}
\indent Similarly, the heat current ($J_{\nu}^Q$) flowing from reservoir $\nu$ to the central system is defined as \cite{datta1997electronic, Sivan, Butcher_1990, Bergfield, Galperin1, Galperin2, Galperin3, Galperin4, Topp_2015}
    \begin{equation}\label{eq:heat}
        \begin{split}
            J_{\nu}^Q=\int_{-\infty}^{\infty} d\omega(\omega-\mu_{\nu}) \Big[\sum_{\nu^{\prime}\ne\nu}T_{\nu\nu^{\prime}}(\omega,\phi)f_{\nu}(\omega)\\
            -\sum_{\nu^{\prime}\ne\nu}T_{\nu^{\prime}\nu}(\omega,\phi)f_{\nu^{\prime}}(\omega)\Big],
        \end{split}
    \end{equation}
where $f_{\nu}(\omega,\mu_{\nu},T_{\nu})=[\mathrm{exp}(\frac{\omega-\mu_{\nu}}{T_{\nu}})+1]^{-1}$ for $\nu=L,P,R$ is Fermi-Dirac distribution function. \\
\indent In the linear-response regime, we can expand the Fermi-Dirac distribution function around the reference point ($\mu,T$) as
\begin{equation}\label{fermiexpand1}
    \begin{split}
        f_{\nu}(\omega,\mu_{\nu},T_{\nu})=f_a(\omega,\mu,T)-(\mu_{\nu}-\mu)\frac{\partial f_a}{\partial\omega}\\
        -(T_{\nu}-T)\Big(\frac{\omega-\mu}{T}\Big)\frac{\partial f_a}{\partial\omega},
    \end{split}
\end{equation}
where $f_a(\omega,\mu,T)=[\mathrm{exp}(\frac{\omega-\mu}{T})+1]^{-1}$ is equilibrium Fermi distribution at the reference point $(\mu,T)$. We consider the right reservoir as the reference point by setting $(\mu_R,T_R)=(\mu,T)$.\\
\indent By substituting Eq. (\ref{fermiexpand1}) in Eqs. (\ref{eq:particle}) and (\ref{eq:heat}), we can obtain the particle currents ($J_{\nu}^N$) and heat currents ($J_{\nu}^Q$) in the linear-response regime in terms of the Onsager relations as discussed in Eq. (\ref{onsager}). The detailed derivations for the Onsager relations using Landauer-B\"{u}ttiker formalism are discussed in Appendix \ref{coefficients}. Further, the introduction of inelastic effects through the voltage probe and voltage-temperature probe in the linear-response and MNL regimes is elaborated in Appendix \ref{sec:inelastic}.
\section{Results and Discussion}\label{sec:results}
This section discusses the numerical results of the three-terminal triple-dot AB heat engine with (i) a voltage probe and (ii) a voltage-temperature probe. The thermoelectric performance of these two heat engines is analyzed by considering the effects of broken time-reversal symmetry in both the linear-response and MNL regimes. We highlight the impact of breaking time-reversal symmetry and introducing nonlinearity on enhancing the thermoelectric performance of these two heat engines. Additionally, we compare our results across three distinct regimes of $t/\gamma$ ratio: (a) $t/\gamma<1$, (b) $t/\gamma=1$, and (c) $t/\gamma>1$. The regimes (b) and (c) correspond to the cases where the interdot tunneling strength is comparable to the dot-lead tunneling rate. In this case, we expect the localization behavior, where the electron winds around the loop multiple times before exiting it. We generally observe an order of magnitude increase in the maximum power compared to the case (a).\\
\indent The efficiency at a given power and the EMP of the voltage probe and voltage-temperature probe heat engine are determined by the asymmetry parameters ($x_L$ and $x$), which reflect TRS breaking by the external magnetic field, while the generalized figures of merit ($y_L$ and $y$) optimize the interplay between thermopower, electrical conductance, and thermal conductance. These dependencies are provided in Eqs. (\ref{etaPmax}), (\ref{v_effc_chi}), (\ref{etapmaxvt}) and (\ref{vt_effc_chi}) of Sec. \ref{sec:mnl}. Under broken TRS, the asymmetry parameters $x_m$ and $x$ can deviate from 1 ($x_L(x)=1$ is the TRS case), while the generalized figures of merit $y_m$ and $y$ are constrained by the upper bound given by the functions $H_m$ and $H$, respectively, as discussed in Eq. (\ref{hmymxm}) and Eq. (\ref{hyx}). The optimization of the performance of these two heat engines requires a thorough analysis of these parameters under broken TRS with a nonzero magnetic field. To demonstrate TRS breaking in the triple-dot AB heat engines, we examine both (i) non-degenerate dot setup with $\epsilon_1\ne\epsilon_2=\epsilon_3$ and (ii) degenerate dot setup with $\epsilon_1=\epsilon_2=\epsilon_3$. Also, we consider the heat current from the left reservoir $J_L^Q>0$ for both the heat engines and the heat current from the probe $J_P^Q<0$ for the voltage probe heat engine for all numerical simulations.\\
\indent
We analyze the thermoelectric performance of a minimally nonlinear voltage probe and voltage-temperature probe heat engine, where a nonlinear dissipative term $-\gamma_h {J_L^{N}}^2$ is introduced to the heat current $J_L^Q$, as defined in Eq. (\ref{eq:jlqv}) and Eq. (\ref{eq:jlqvt}), respectively. We use the dissipation strength ratio, $\alpha=1/(1+\gamma_c/\gamma_h)$ with $0\le\alpha\le1$ to demonstrate the effect of nonlinearity in all our calculations and simulations with $\alpha=0$ recovering the linear-response results.\\
\indent
Figure \ref{fig:xLyL} demonstrates the asymmetry parameters ($x_L$ and $x$) and the figures of merit ($y_L$ and $y$) as a function of $\phi$ (which is related to the magnetic flux given in Eq. (\ref{phi})) for both the voltage probe (blue) and voltage-temperature probe (red) heat engines. We represent the non-degenerate dot setup ($\epsilon_1\ne\epsilon_2=\epsilon_3$) with solid lines and the degenerate dot setup ($\epsilon_1=\epsilon_2=\epsilon_3$) with dash-dot lines across three different regimes of the $t/\gamma$ ratio. For the voltage probe heat engine, the degenerate dot setup exhibits higher asymmetry $x_L\ne1$ than the non-degenerate dot setup across all three regimes of the $t/\gamma$ ratio.
%The asymmetry is relatively small in the $t/\gamma=0.2$ regime, but increases with increasing the $t/\gamma$ ratio as shown in Fig. \ref{fig:xLyL}(a1-c1).
Unlike the asymmetry parameter, the generalized figure of merit $y_L$ is highest for the $t/\gamma<1$ regime and decreases as the $t/\gamma$ ratio increases, as illustrated in Figs. \ref{fig:xLyL}(a2-c2). In the $t/\gamma<1$ regime, $y_L$ is high for the degenerate dot setup, however, $y_L$ has higher values for the non-degenerate dot setup in the $t/\gamma=1$ and $t/\gamma>1$ regimes. \\
\indent
Unlike the voltage probe heat engine, in the presence of a non-vanishing magnetic field ($\phi\ne0$), TRS remains preserved ($x=1$) for the degenerate dot setup (red dashed line shown in Fig. \ref{fig:xLyL}(a1-c1)) across all three regimes of $t/\gamma$ ratio for the voltage-temperature probe heat engine. However, the symmetry is broken with $\phi$ for the non-degenerate dot setup, and the asymmetry parameter deviates from  1 as we increase the $t/\gamma$ ratio. Thus, to break TRS, in addition to a non-vanishing magnetic field, the system requires some anisotropy, e.g., $\epsilon_1\ne\epsilon_3$ for a voltage-temperature probe heat engine (also mentioned in Ref. \cite{Benenti2, Balachandran}). The generalized figure of merit $y$ shows a trend similar to that of the voltage probe, with maximum $y$ achieved in the $t/\gamma<1$ regime, and it decreases as the $t/\gamma$ ratio increases, as shown in Fig. \ref{fig:xLyL}(a2)-(c2). In contrast to the voltage-temperature probe heat engine, the voltage probe heat engine doesn't require additional dot energy anisotropy to break TRS. But, for comparison purposes, we consider the non-degenerate dot setup for both heat engines in all subsequent simulations. In addition, we observe that the thermoelectric performance is enhanced for the anisotropic case in the voltage probe heat engine.\\
\indent
We compare the performance of the voltage probe and the voltage-temperature probe heat engine in terms of maximum power, $\mathcal{P}_{max}$, and EMP, and analyze the role of broken TRS in enhancing the efficiency. Figure \ref{fig:vvsvt} illustrates maximum power and EMP as a function of $\phi$ for three different regimes of $t/\gamma$ ratio in the MNL regime for the voltage probe (blue) and voltage-temperature probe (red) heat engines. The voltage probe heat engine generates more power than the voltage-temperature probe heat engine in all three regimes of $t/\gamma$, with lower power in the $t/\gamma=0.2$ regime and highest power in the $t/\gamma=1$ regime. For $t/\gamma=0.2$, $\mathcal{P}_{max}$ is highest around $\phi=1.98\pi$ and it shows decreasing trend with $\phi$ values from $0$ to $\pi$ and then increases from $\pi$ to $2\pi$ as shown in Fig. \ref{fig:vvsvt}(a1). In Fig. \ref{fig:vvsvt}(b1) and \ref{fig:vvsvt}(c1), which correspond to the $t/\gamma=1$ and $t/\gamma=2$ regimes, respectively, the $\mathcal{P}_{max}$ curve exhibits two local maxima, forming hump-like structures that indicate the range of $\phi$ where the output power is higher compared to $\phi=0$ for both the voltage probe and voltage-temperature probe heat engines. It has been investigated in Refs. \cite{Benenti1, Benenti2, Balachandran, Brandner2, Brandner3,Zhang, Zahra} that broken TRS, with asymmetry parameter $x_L(x)>1$, can enhance the EMP as well as maximum efficiency. Figures \ref{fig:vvsvt}(a2), \ref{fig:vvsvt}(b2), and \ref{fig:vvsvt}(c2) demonstrate the EMP as a function of $\phi$ in the $t/\gamma=0.2$, $t/\gamma=1$, and $t/\gamma=2$ regime, respectively, for both the heat engines. The local maximum of the $\eta_{L}(\mathcal{P}_{max})/\eta_{CL}(\mathcal{P}_{max})$ curve identifies the range of $\phi$ where the EMP exceeds that at the symmetric point ($\phi=0$), indicating that breaking TRS can help enhance the EMP. For a voltage probe heat engine, the local maximum for EMP occurs within the range of $1.75\pi<\phi<2\pi$, $1.8\pi<\phi<2\pi$, and $1.88\pi<\phi<2\pi$ for the $t/\gamma=0.2$, $t/\gamma=1$, and $t/\gamma=2$ regimes, respectively. This behavior can be understood more clearly by comparing the EMP curve with Fig. \ref{fig:xLyL}. The EMP curve behavior depends on the $x_Ly_L$ behavior for the voltage probe heat engine. High asymmetries with $x_L>1$ and a high figure of merit $y_L$ give higher $x_Ly_L$ values, thus leading to an EMP that surpasses the symmetric point within this range under broken TRS. Maximum EMP reached at $\phi=1.85\pi$, $\phi=1.9\pi$, and $\phi=1.95\pi$ for the $t/\gamma=0.2$, $t/\gamma=1$, and $t/\gamma=2$ regimes, respectively.\\
\begin{figure*}[t!]
    \centering
    \includegraphics[scale=0.12]{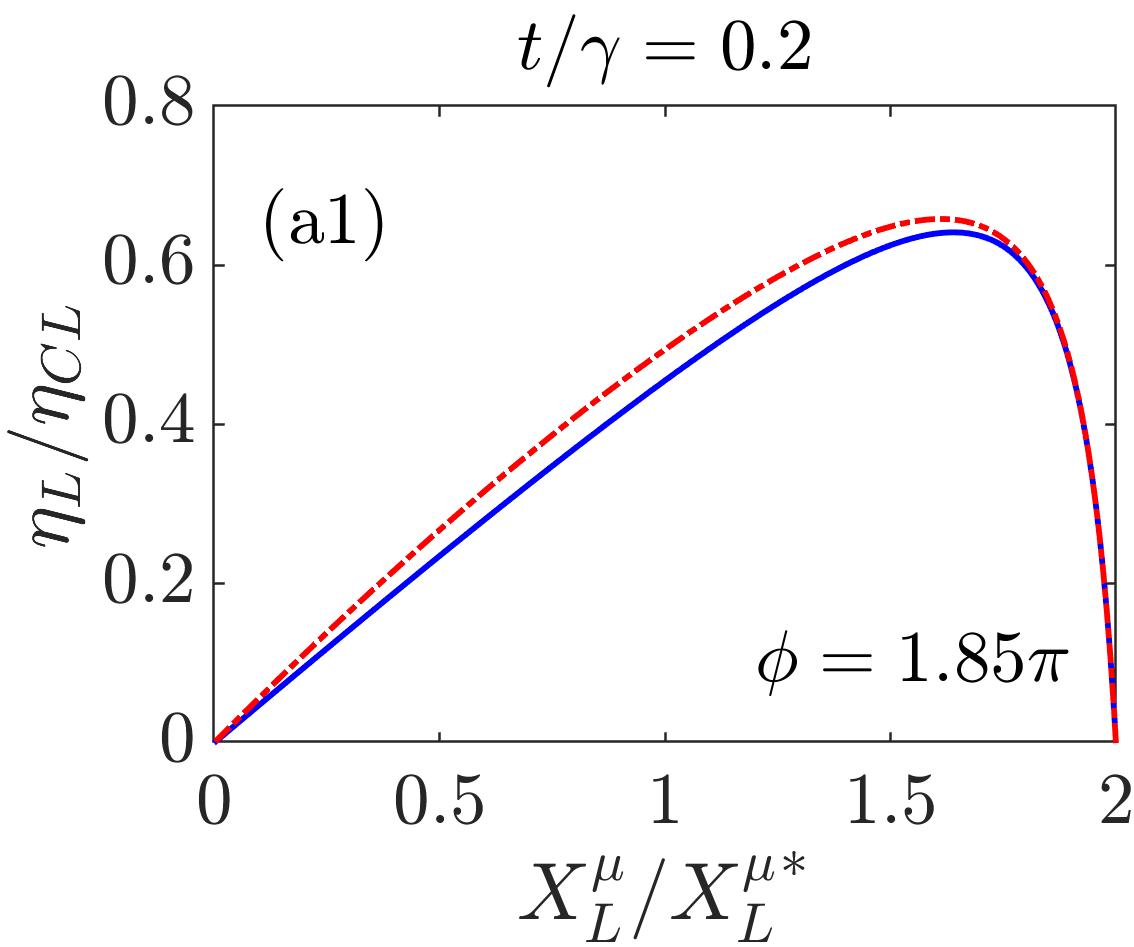}
    \includegraphics[scale=0.12]{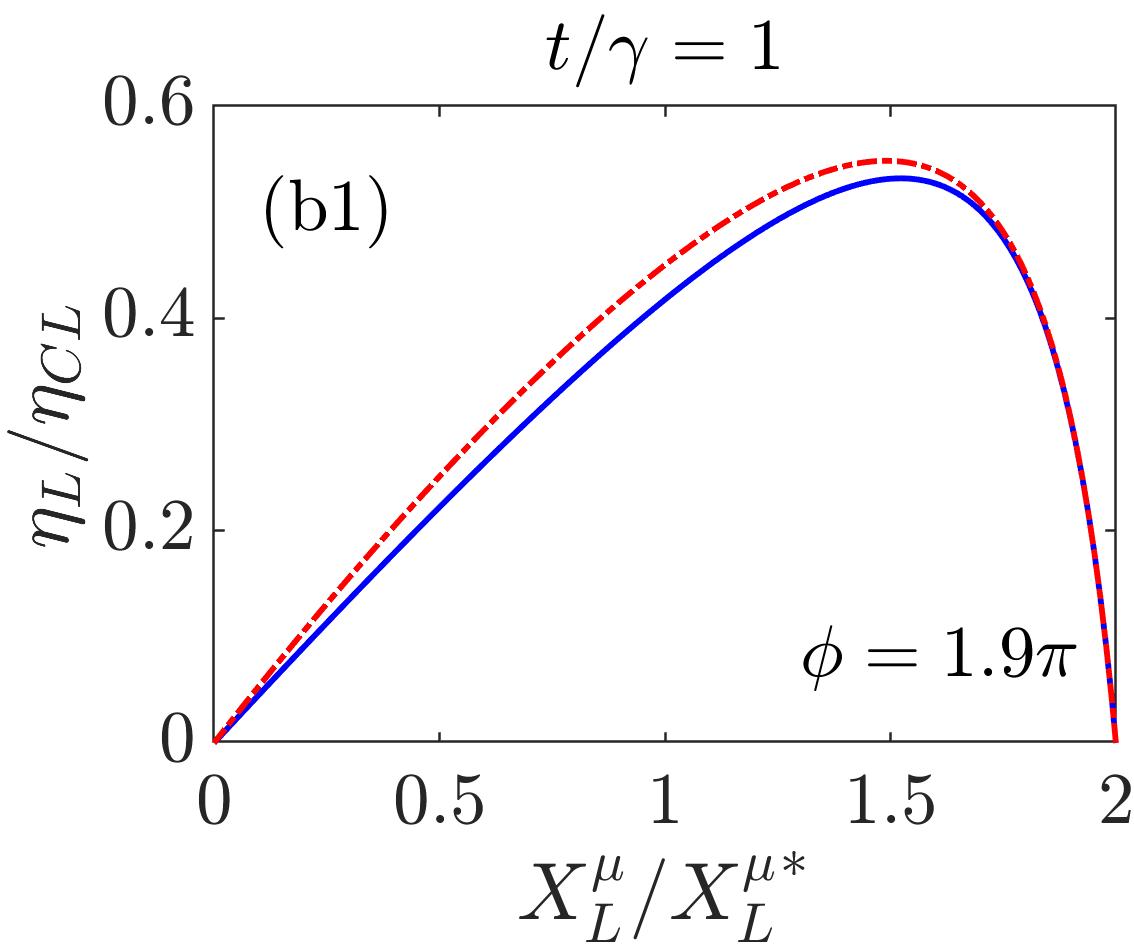}
    \includegraphics[scale=0.12]{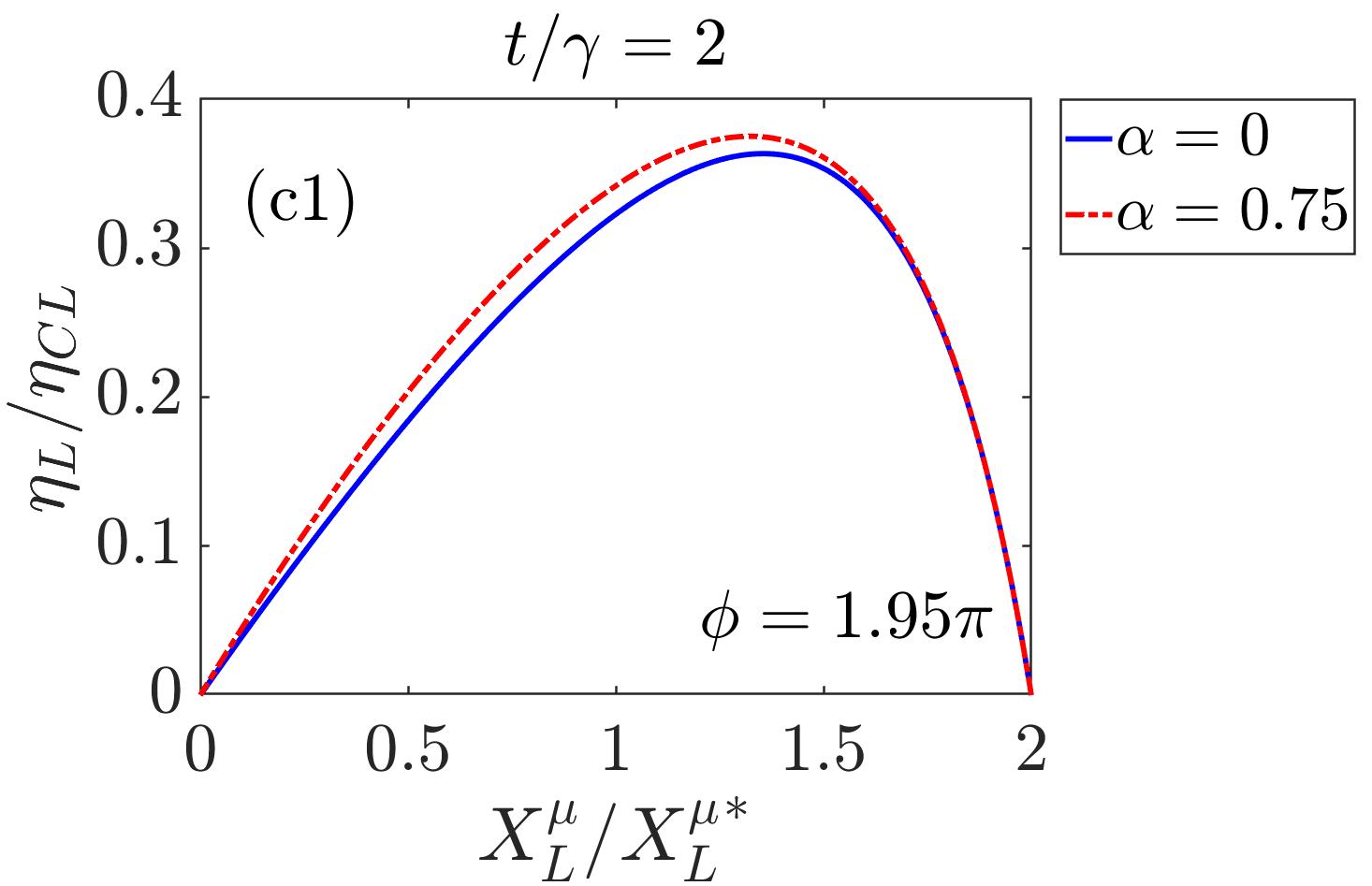}
    \includegraphics[scale=0.12]{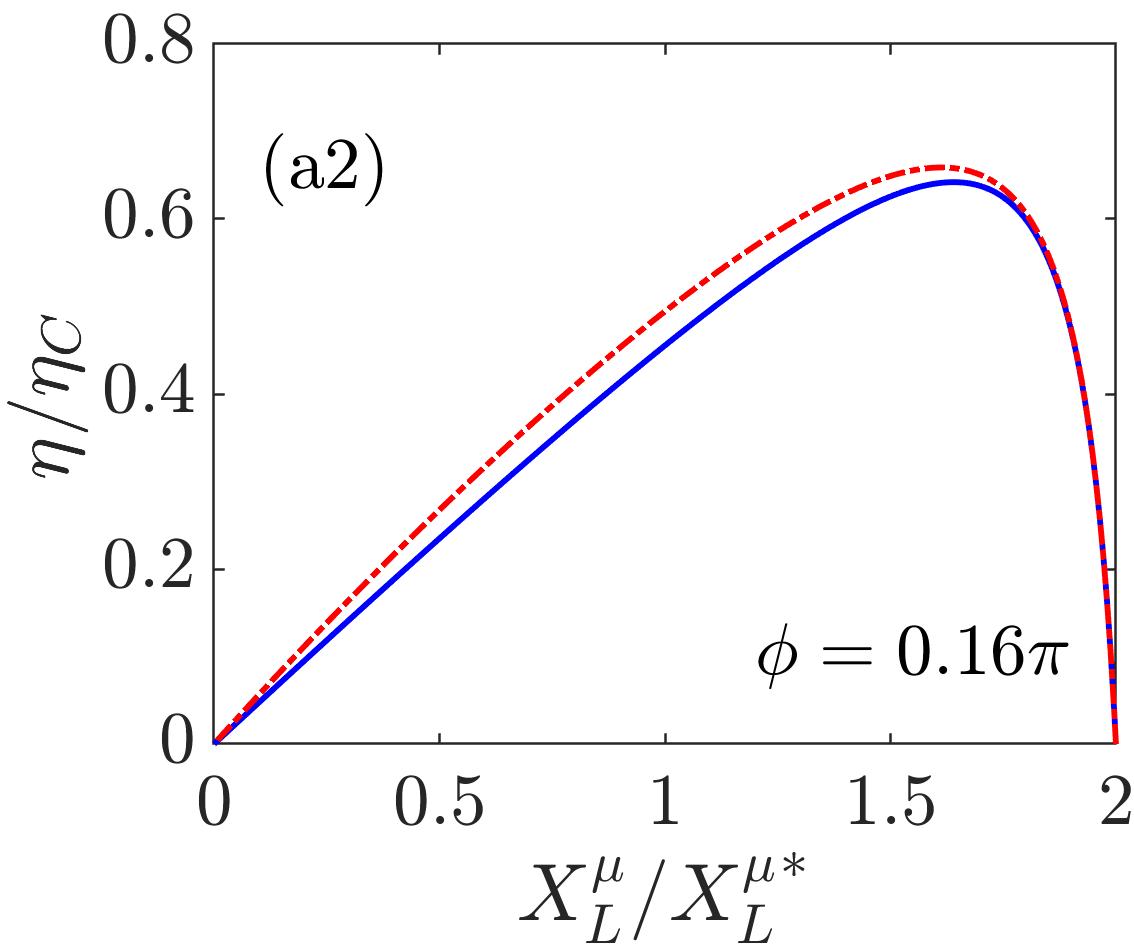}
    \includegraphics[scale=0.12]{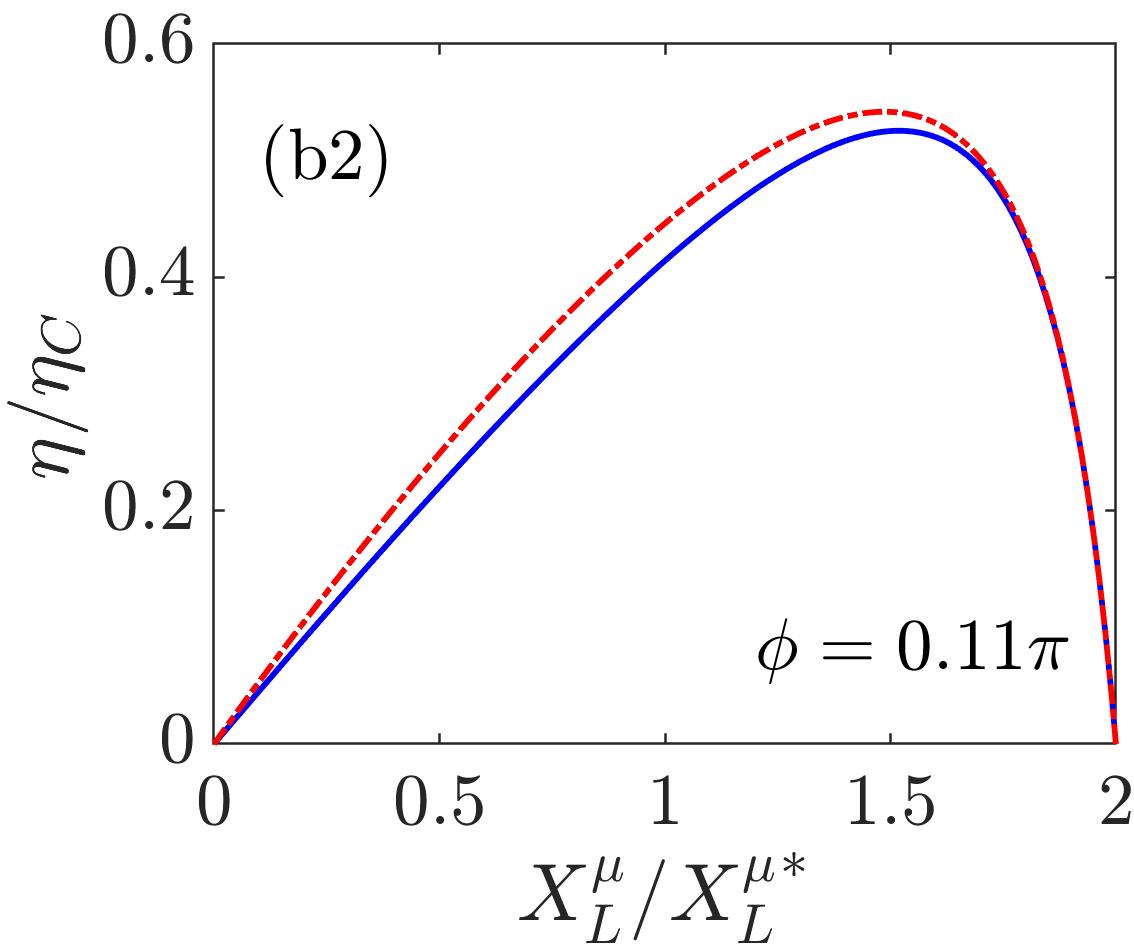}
    \includegraphics[scale=0.12]{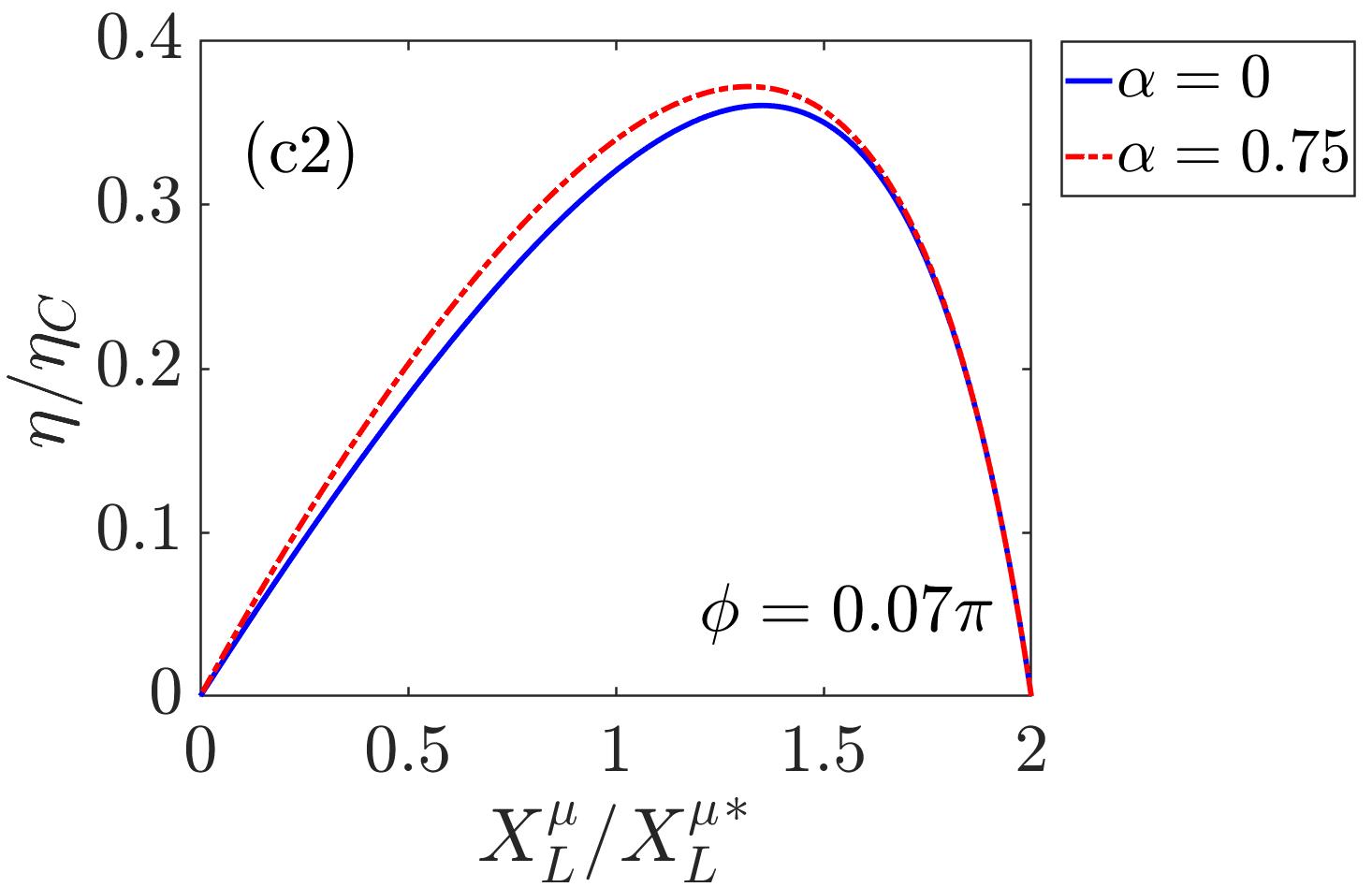}
    \caption{Efficiency at a given power, $\eta_L/\eta_{CL}$ and $\eta/\eta_{C}$ as a function of relative bias $X_L^{\mu}/X_L^{\mu^*}$ for the voltage probe (top panel) and voltage-temperature probe (bottom panel) heat engine in the MNL regime, respectively. Three regimes of the $t/\gamma$ ratio are demonstrated as (a1-a2) $t/\gamma=0.2$, (b1-b2) $t/\gamma=1$, and (c1-c2) $t/\gamma=2$. The chosen $\phi$ values correspond to the points of highest EMP under broken time-reversal symmetry for both heat engines. Parameters used are: $\epsilon_1=0.6$, $\epsilon_2=\epsilon_3=0.5$, $\gamma=\gamma_L=\gamma_R=\gamma_p=0.05$, $T_R=T=0.1$, $T_L=0.13$, $T_P=0.102$ (for the voltage probe), $\mu_R=\mu=0.3$.}
    \label{fig:nncL_mnl}
\end{figure*}
\indent
Likewise, EMP of the voltage-temperature probe heat engine exhibits a local maximum within the range of $0<\phi<0.31\pi$, $0<\phi<0.22\pi$, and $0<\phi<0.13\pi$ for the $t/\gamma=0.2$, $t/\gamma=1$, and $t/\gamma=2$ regimes, respectively. Within this range, we have a high asymmetry parameter with $x>1$ and a high figure of merit $y$ leading to a high $xy$ value, thus helping in enhancing the EMP as compared to the symmetric point ($\phi=0$). Under broken TRS, the EMP depends on $x_Ly_L$ and $xy$ for the voltage probe and voltage-temperature heat engines, respectively. Thus, asymmetry parameters with $x_L(x)>1$ and high figures of merit $y_L(y)$ are essential for enhancing the thermoelectric performance of these two heat engines. We observe that the voltage probe heat engine generates more power than the voltage-temperature probe heat, with the highest power in the $t/\gamma=1$ regime and lowest in the $t/\gamma=0.2$ regime. In contrast, the voltage-temperature probe heat engine is generally more efficient than the voltage probe heat engine, except for certain ranges of $\phi$, where $x_Ly_L$ is higher than $xy$, leading to higher EMP within that range for the voltage probe heat engine as illustrated in Figs. \ref{fig:vvsvt}(a2), \ref{fig:vvsvt}(b2), and \ref{fig:vvsvt}(a2). The introduction of minimal nonlinearity with a dissipation strength ratio $\alpha=0.75$ helps enhance the EMP of both heat engines, although power remains independent of the MNL term. For the weak probe coupling with $\gamma_P=0.005$ (not shown in the figure), we do not observe any significant difference between the performance of the voltage probe and voltage-temperature heat engine. Although the TRS is broken for weak probe coupling, it does not significantly enhance the EMP, as the asymmetry parameter doesn't deviate much from the symmetric point.\\
\begin{figure*}[t!]
    \centering
    \includegraphics[scale=0.12]{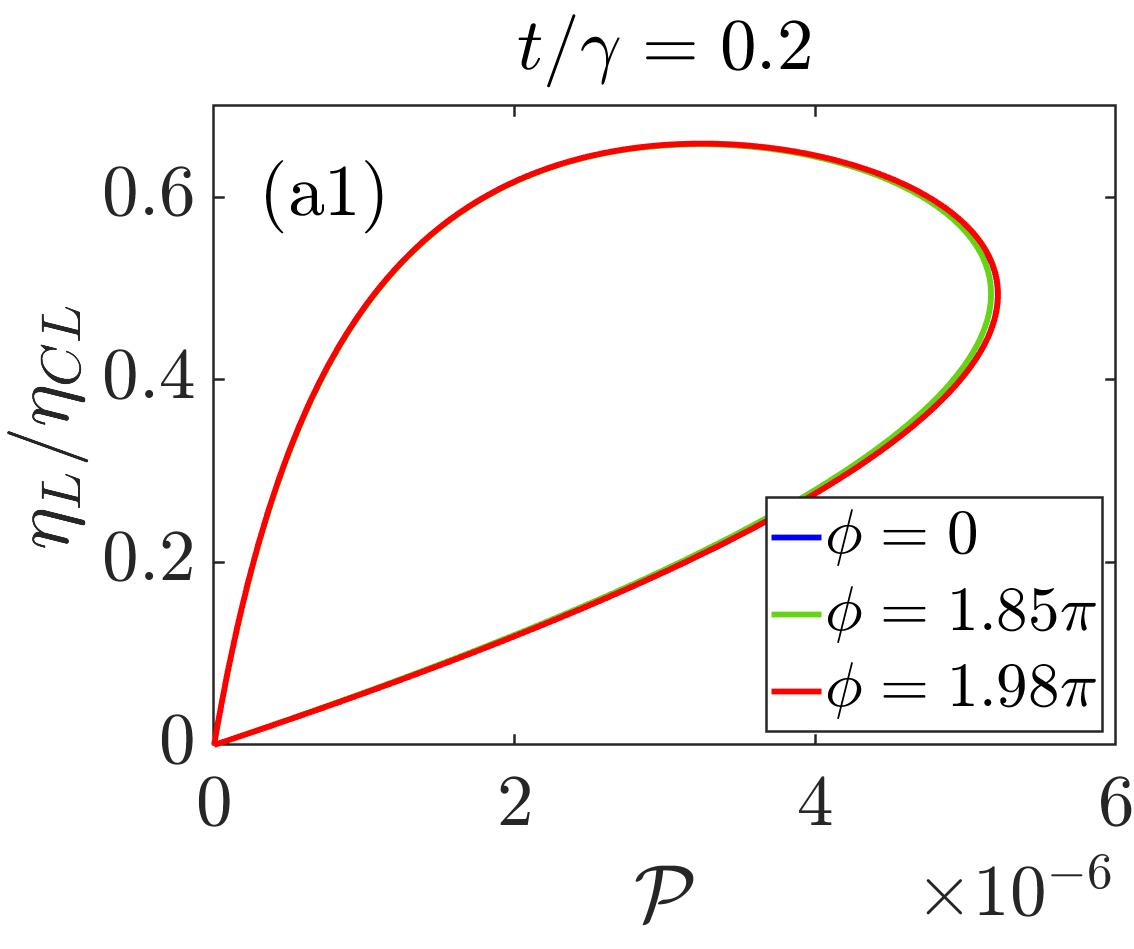}
    \includegraphics[scale=0.12]{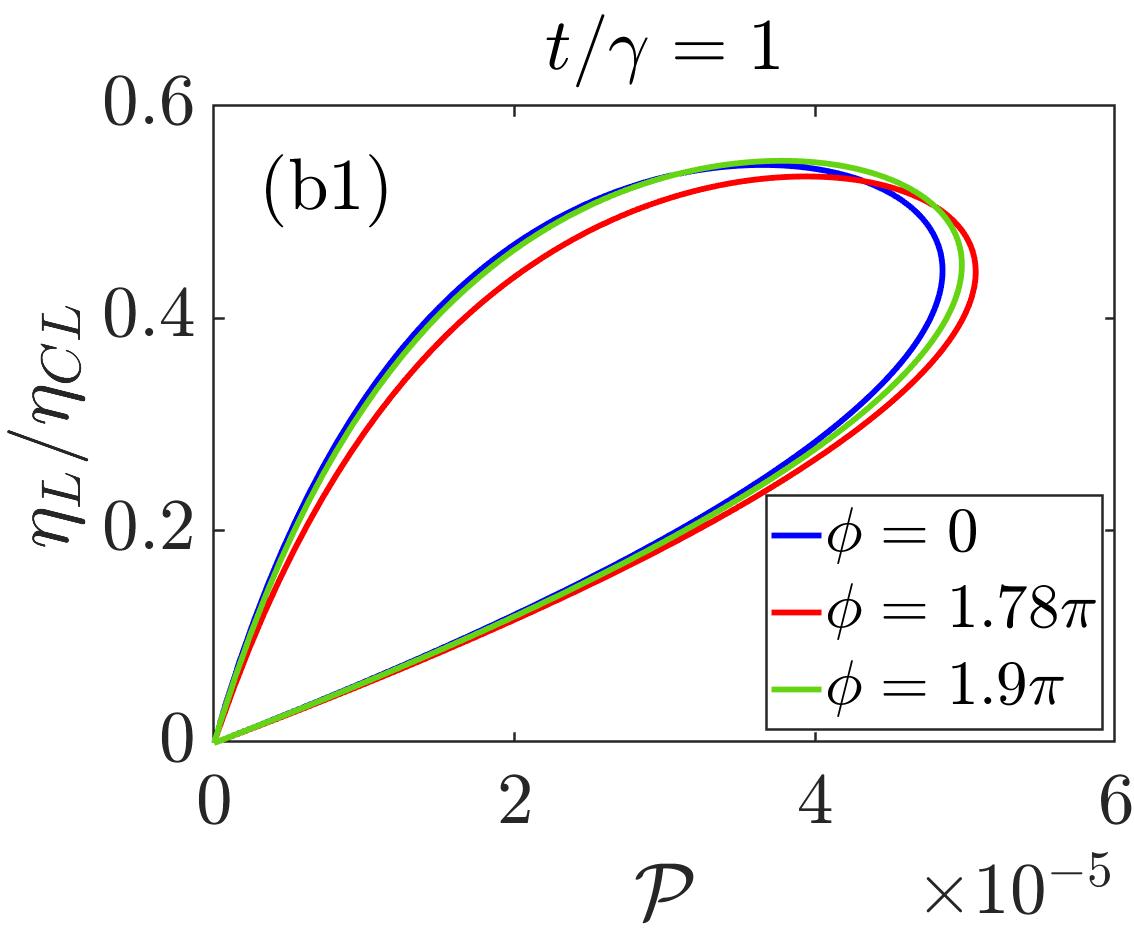}
    \includegraphics[scale=0.12]{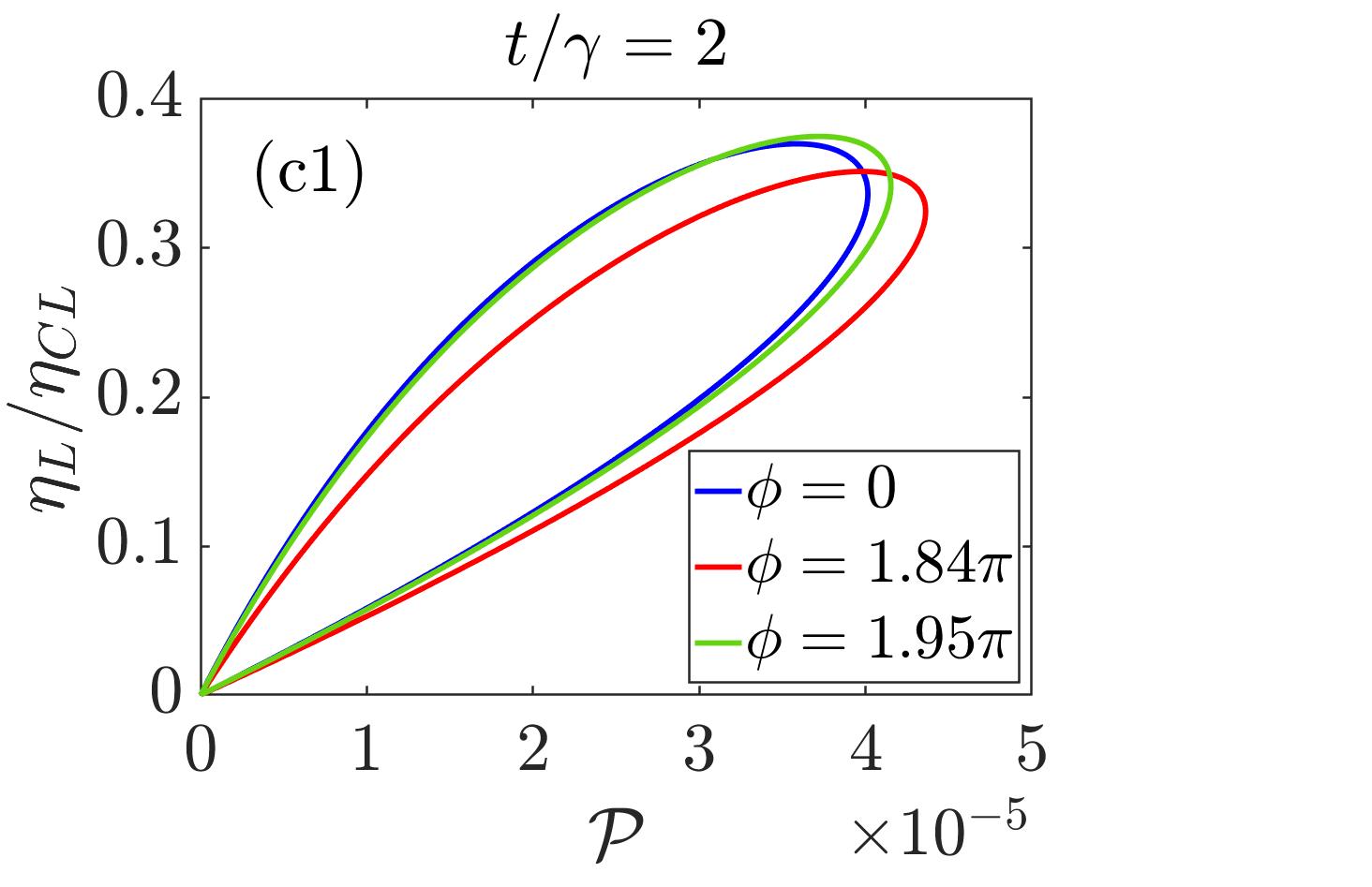}
    \includegraphics[scale=0.12]{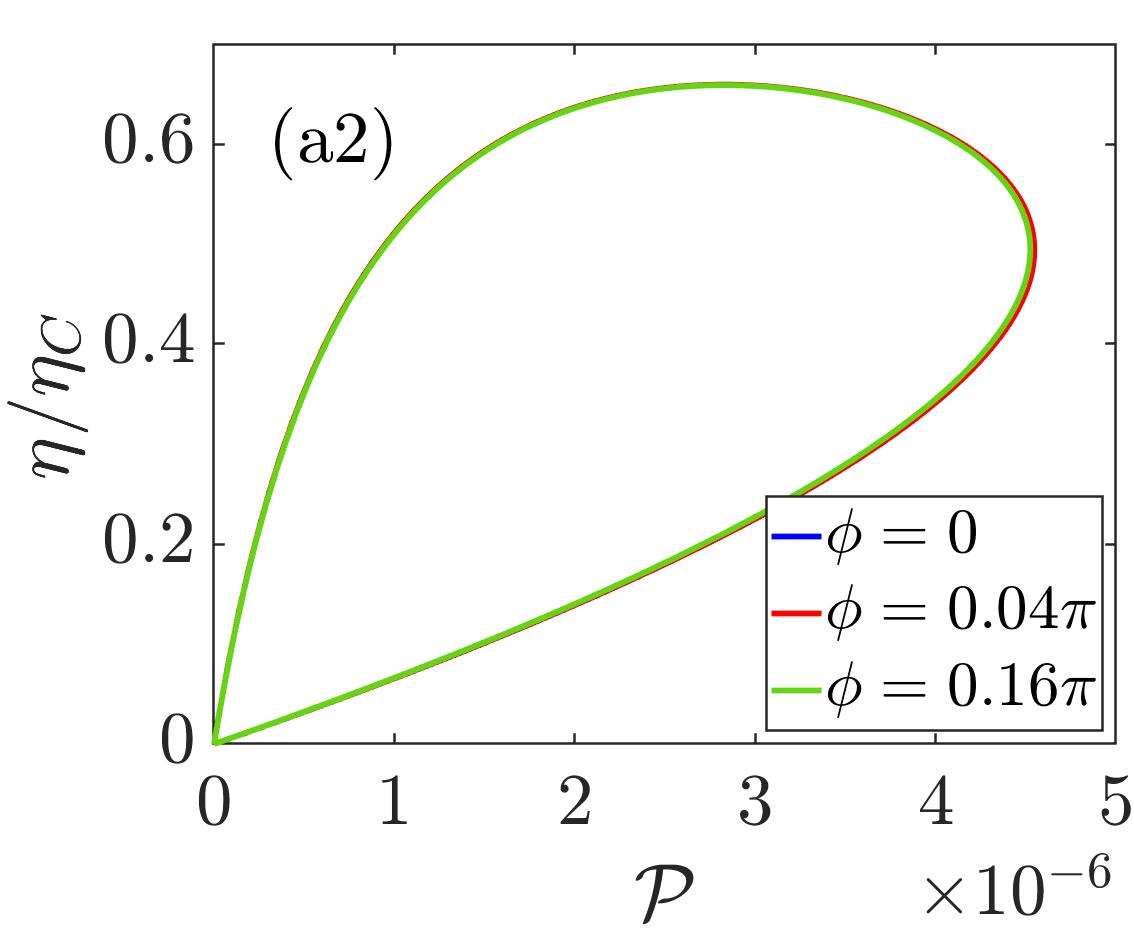}
    \includegraphics[scale=0.12]{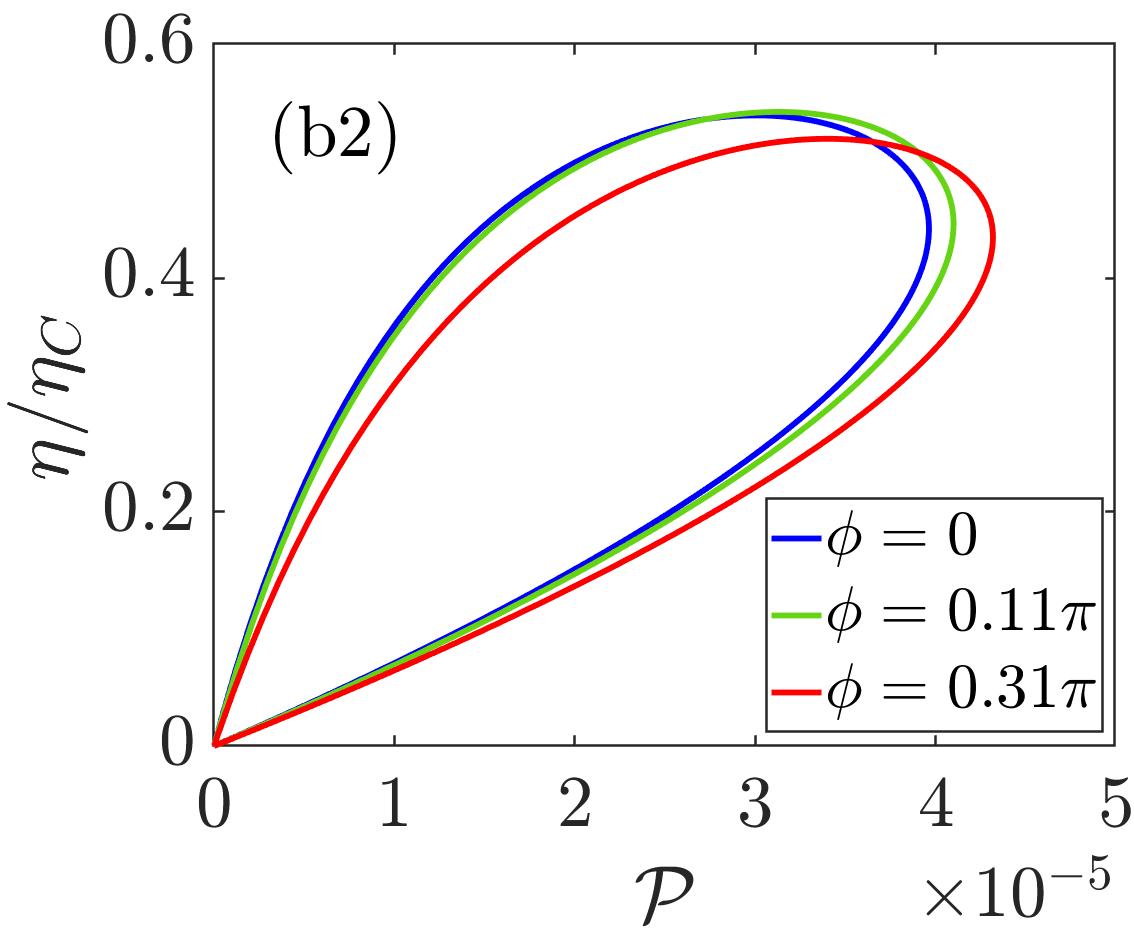}
    \includegraphics[scale=0.12]{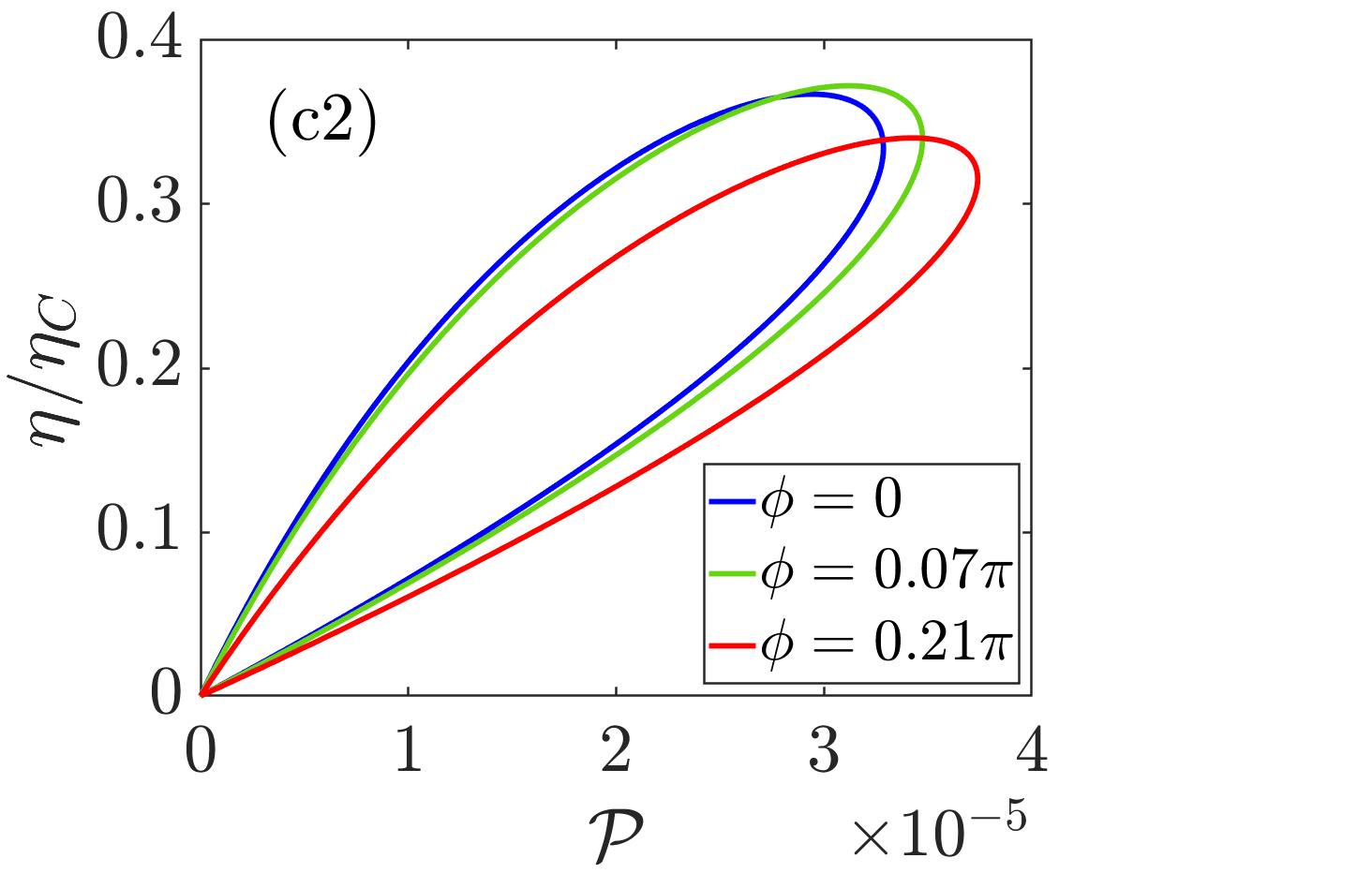}
    \includegraphics[scale=0.12]{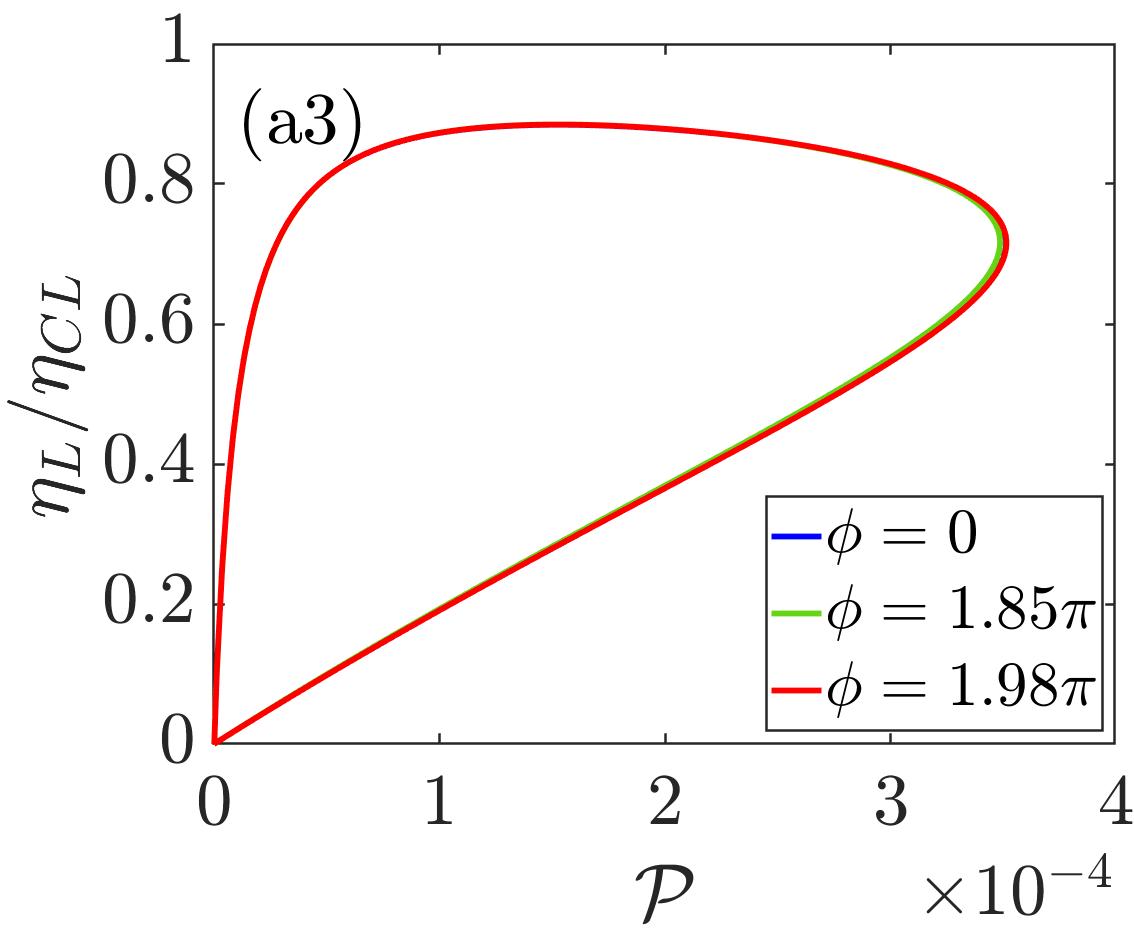}
    \includegraphics[scale=0.12]{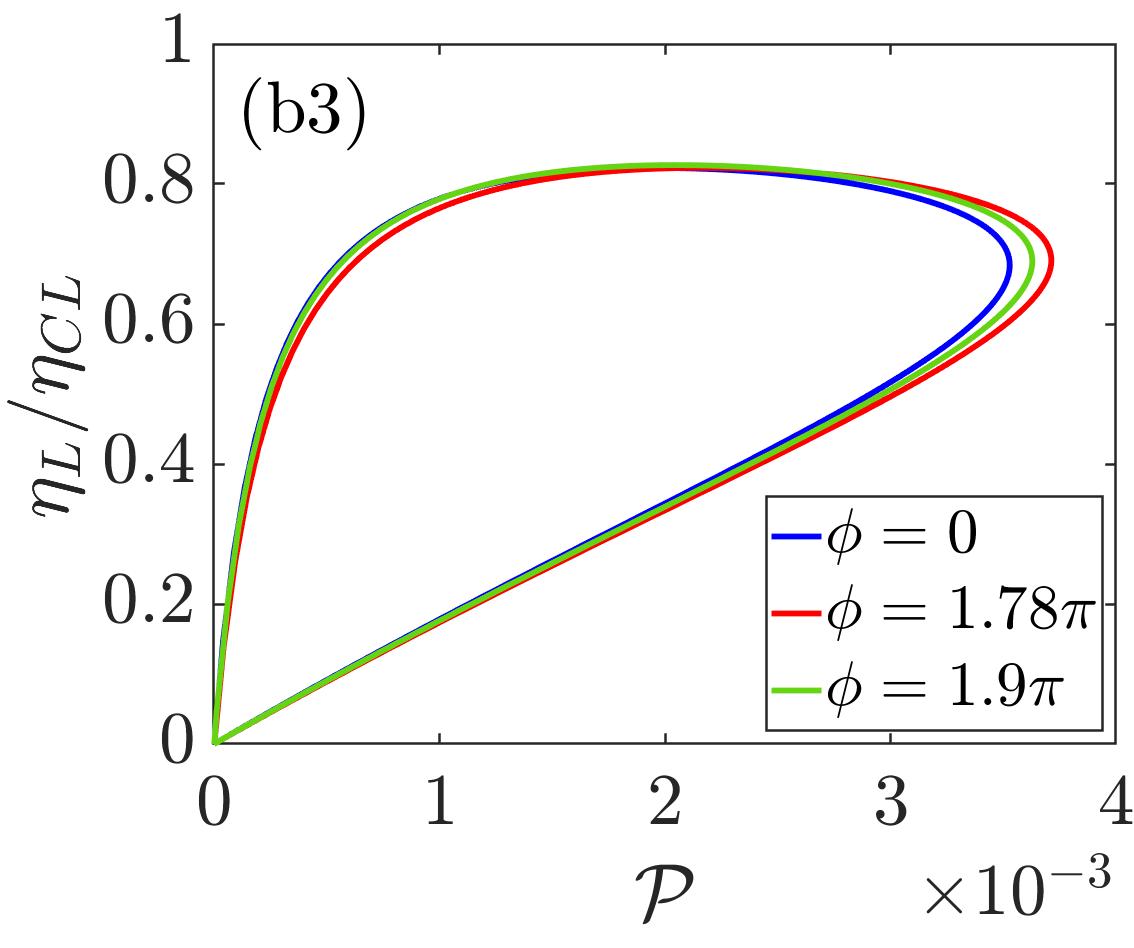}
    \includegraphics[scale=0.12]{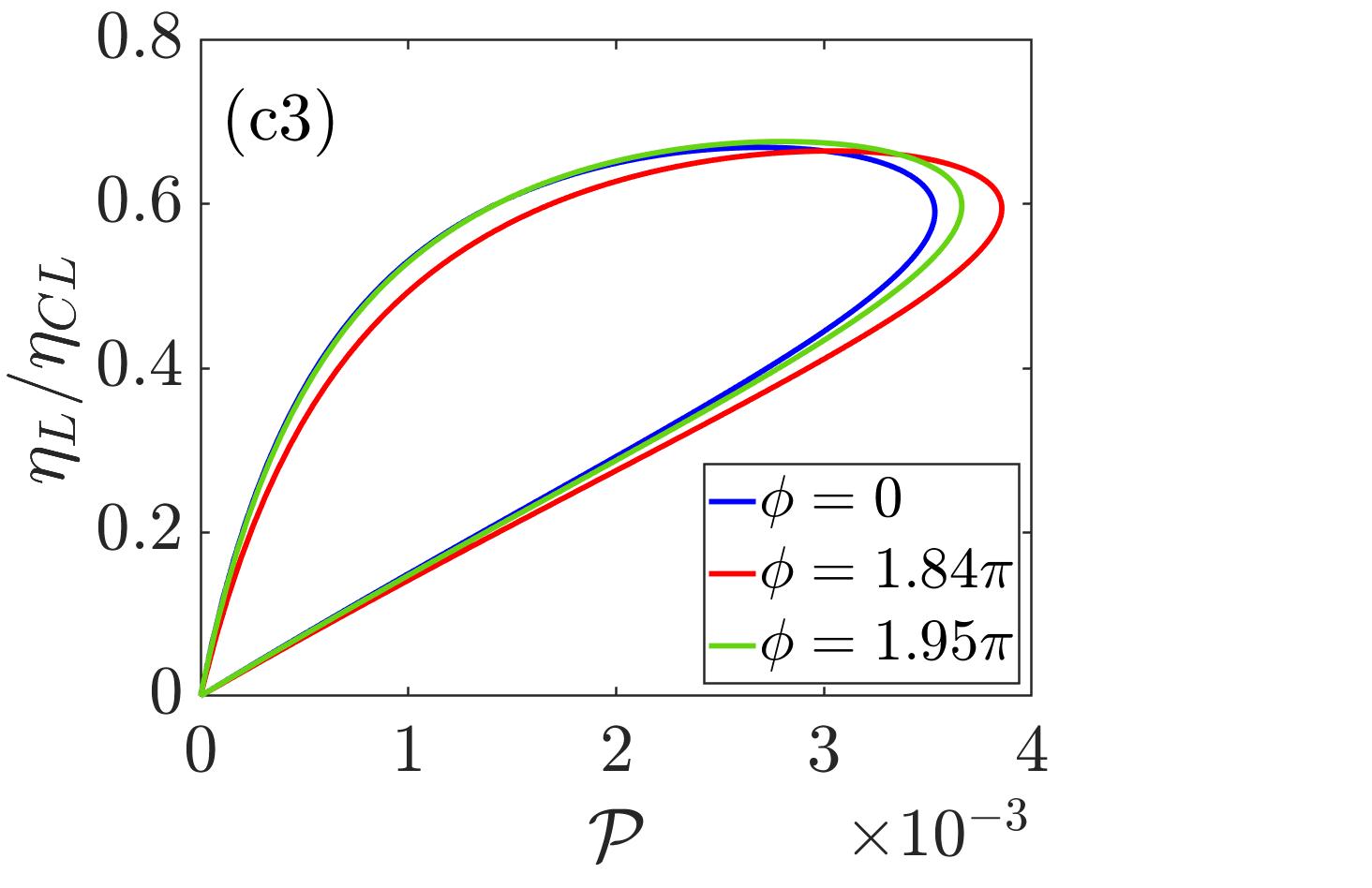}
    \caption{Power-efficiency plot for the voltage probe (top panel (a1-c1)) and voltage-temperature probe (middle panel (a2-c2)) heat engine in the MNL regime with $\alpha=0.75$ for three regimes of the $t/\gamma$ ratio: (a1-a2) $t/\gamma=0.2$, (b1-b2) $t/\gamma=1$, and (c1-c2) $t/\gamma=2$. Parameters: $\epsilon_1=0.6$, $\epsilon_2=\epsilon_3=0.5$, $\gamma=\gamma_L=\gamma_R=\gamma_p=0.05$, $T_R=T=0.1$, $T_L=0.13$, $T_P=0.102$ (for the voltage probe), $\mu_R=\mu=0.3$. The bottom panel (a3-c3) illustrates the power-efficiency plot for the voltage probe heat engine in the fully nonlinear regime with temperatures $T_R=T=0.1$, $T_L=0.4$, $T_P=0.13$. The chosen $\phi$ values correspond to the symmetric point (blue), the highest $\mathcal{P}_{max}$ (red), and the highest EMP (green).}
    \label{fig:power_nnc_mnl}
\end{figure*}
\indent
Breaking TRS introduces asymmetries and modifies quantum interference patterns, which can significantly affect transport properties and improve thermoelectric performance. Meanwhile, a high thermoelectric figure of merit indicates better thermoelectric performance, which enhances the efficiency by optimizing the interplay between the thermopower, electrical conductance, and thermal conductance. Thus, the combined effect of TRS breaking and high figures of merit is crucial for optimizing the thermoelectric performance of the heat engines. Figure. \ref{fig:scatter} shows a scatter plot of EMP, demonstrating the interplay between broken TRS, via its dependence on asymmetry parameters $x_L(x)$, and figures of merit $y_L(y)$ for both voltage probe (\ref{fig:scatter}(a1-c1)) and voltage-temperature probe (\ref{fig:scatter}(a2-c2)) heat engine in the MNL regime. The red-edged circles highlight the points where the EMP exceeds that of the symmetric point ($x_L(x)=1$). We observe that the TRS breaking enhances the EMP only when $x_L(x)>1$, accompanied by high figures of merit $y_L(y)$, resulting in larger $x_Ly_L(xy)$ values that boost the EMP beyond that at the symmetric point. We also find that large asymmetries do not necessarily imply large EMP or large efficiency at arbitrary power (as evident in Fig. \ref{fig:scatter}(c1)). It is the combined effect of TRS breaking and high figures of merit that plays a pivotal role in boosting the EMP beyond the symmetric point.\\
\indent
Figure \ref{fig:nncL_mnl} illustrates the efficiency at a given power as a function of relative voltage bias $X_L^{\mu}/{X_L^{\mu}}^*$, where ${X_L^{\mu}}^*$ represents the bias at which maximum power $\mathcal{P}_{max}$ is obtained, for the voltage probe (Fig. \ref{fig:nncL_mnl}(a1-c1)) and voltage-temperature probe (Fig. \ref{fig:nncL_mnl}(a2-c2)) heat engines in the MNL regime. The chosen $\phi$ values correspond to the point of highest EMP across all three regimes of the $t/\gamma$ ratio. The $t/\gamma<1$ regime is highly efficient but generates low power with almost one order of magnitude less than the $t/\gamma=1$ and $t/\gamma>1$ regimes (see Fig. \ref{fig:vvsvt} for power and EMP plot). Both power and EMP in the $t/\gamma>1$ regime are lower than in the $t/\gamma=1$ regime. Thus, the $t/\gamma=1$ regime is the optimal regime where it generates high power with significantly high EMP and maximum efficiency for both the heat engines. We observe that stronger coupling to the leads (i.e., $\gamma$) improves power output but reduces efficiency (results not shown in the figures), a trend also examined in Ref. \cite{Nakpathomkun}.  We define the region where $X_L^{\mu}/{X_L^{\mu}}^*>1$ as the favorable branch, since it is more efficient, with efficiency exceeding the EMP and reaching its maximum for a given power. Conversely, the region where $X_L^{\mu}/{X_L^{\mu}}^*<1$ is termed the unfavorable branch, as it exhibits lower efficiency for the same power output. Figure \ref{fig:nncL_mnl} also showcases the effect of nonlinearity on the efficiency by increasing the dissipation strength $\alpha$ for both the heat engines. The EMP and the efficiency at a given power are enhanced with increasing dissipation strength, although the power output remains unaffected for the heat engines. Thus, breaking time-reversal symmetry and introducing nonlinearity are advantageous for improving the efficiency of both heat engines.\\
\indent
We demonstrate the efficiency as a function of output power across all three regimes of $t/\gamma$ for the voltage probe and voltage-temperature probe heat engines in Fig. \ref{fig:power_nnc_mnl}(a1-c1) and Fig. \ref{fig:power_nnc_mnl}(a2-c2), respectively, in the MNL regime. The $\phi$ values correspond to the symmetric point, $\phi=0$ (blue), the highest $\mathcal{P}_{max}$ (red), and the highest EMP (green). We observe that the phase $\phi$ corresponding to the highest EMP (green curves) also yields the maximum efficiency at a power slightly lower than the maximum. The bottom panel, Fig. \ref{fig:power_nnc_mnl}(a3-c3), shows the thermoelectric performance of the voltage probe heat engine operating in the fully nonlinear regime. While the power remains unaffected by the introduction of the minimally nonlinear term to the heat current in the MNL heat engines, it is significantly enhanced when the voltage probe heat engine operates in the fully nonlinear regime. In the fully nonlinear regime, EMP and efficiency at a given power are also enhanced where the EMP overcomes the CA limit and almost reaches $0.71\eta_C$, $0.69\eta_C$, and $0.6\eta_C$ for $t/\gamma=0.2$, $t/\gamma=1$, and $t/\gamma=2$ regimes, respectively. The power-efficiency curves at different $\phi$ values that demonstrate maximum power and highest EMP are consistent with the MNL voltage probe heat engine. These results further support our analysis of the MNL heat engine, and we conclude that higher-order nonlinear effects, along with broken TRS, help enhance the thermoelectric performance of the voltage probe heat engine. We are encountering convergence issues while simulating the results for the voltage-temperature probe in the fully nonlinear regime. Consequently, we are unable to obtain results in this case. However, we assume the fully nonlinear results would remain consistent with those of the MNL voltage-temperature heat engine with broken TRS. \\
\begin{figure*}[t!]
    \centering
    \includegraphics[scale=0.12]{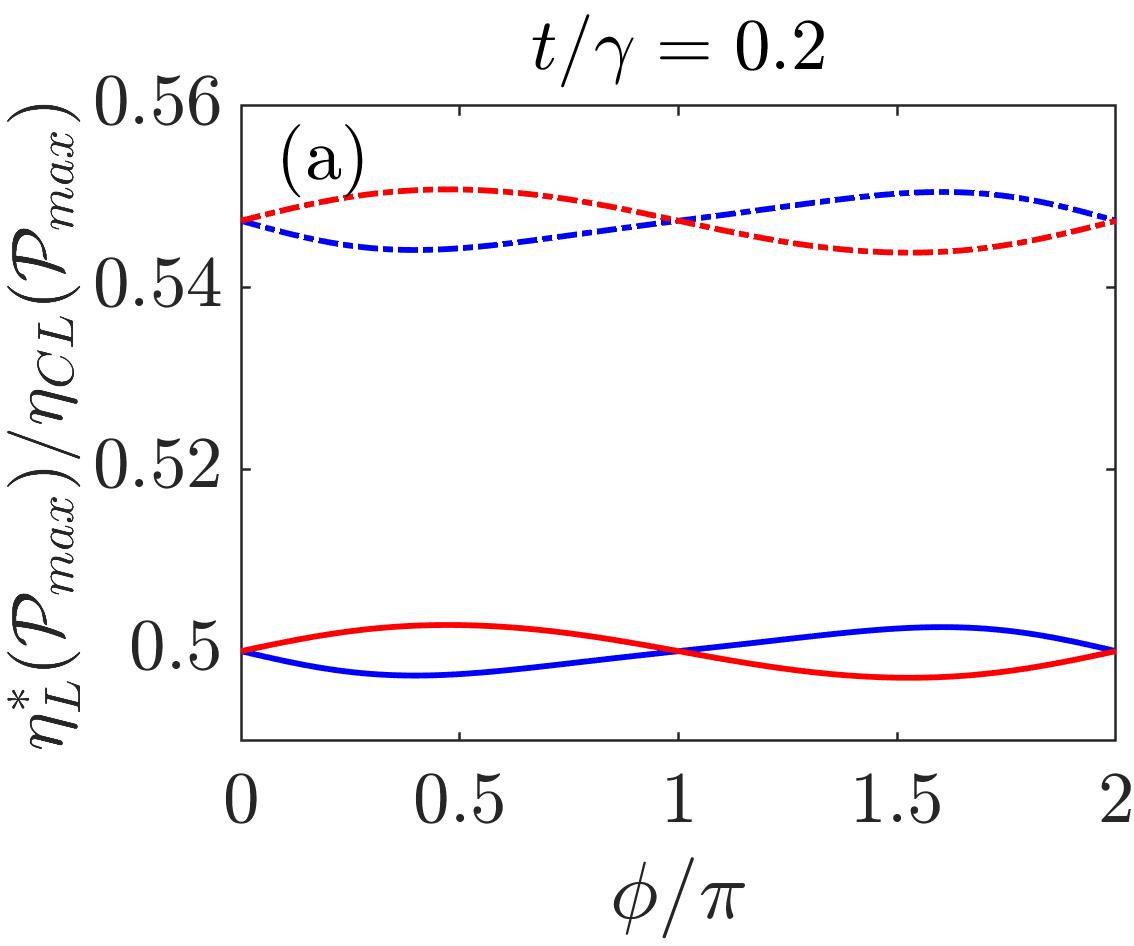}
    \includegraphics[scale=0.12]{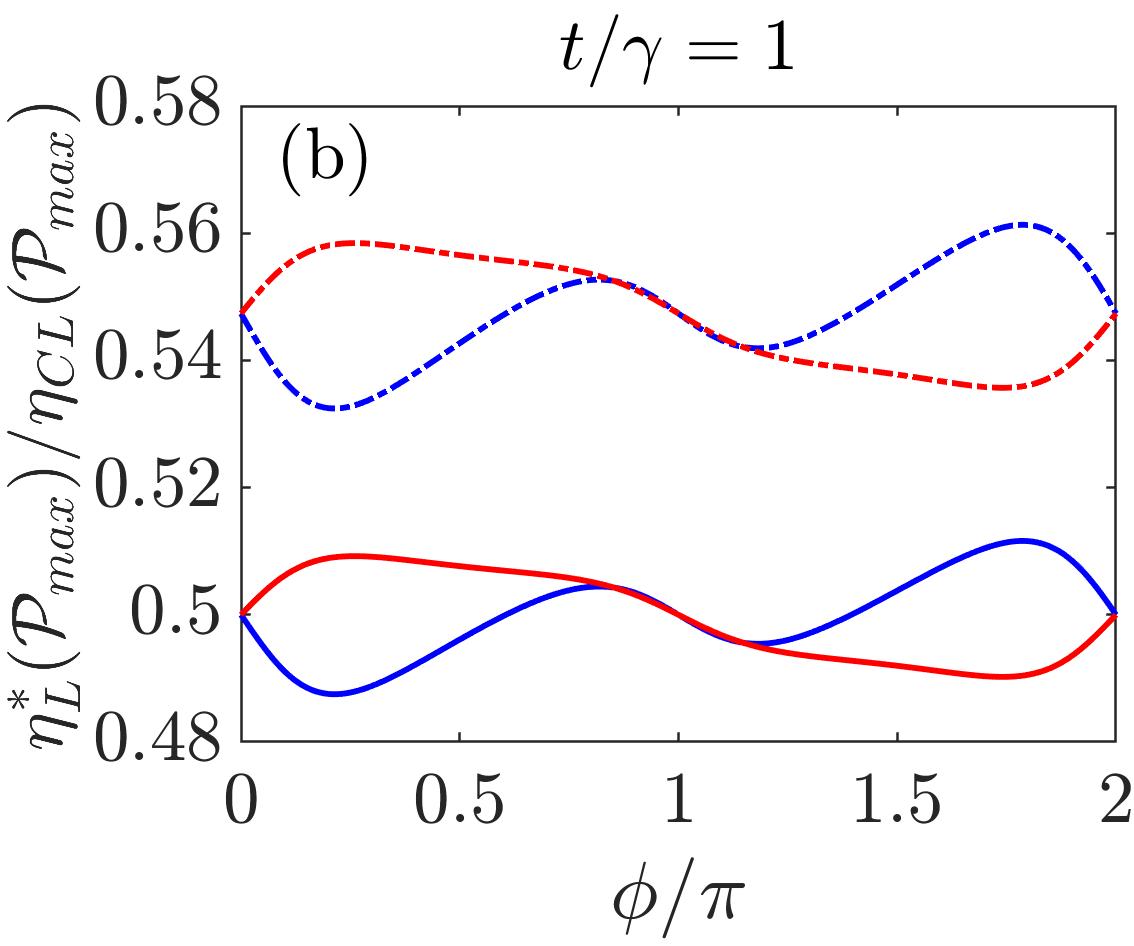}
    \includegraphics[scale=0.12]{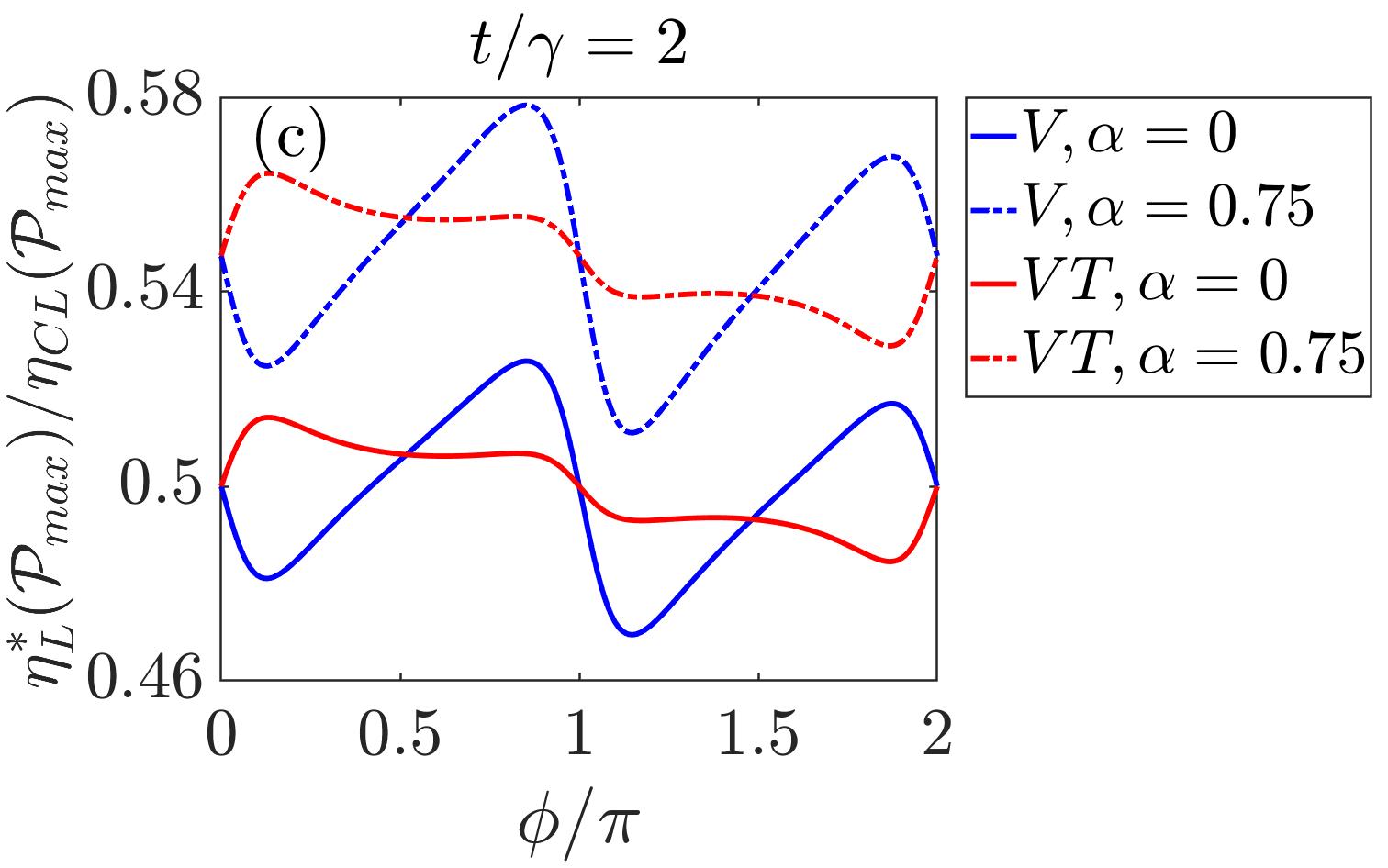}
    \caption{Upper bound on the EMP for the voltage probe (blue) and voltage-temperature probe (red) heat engine as a function $\phi$ in the MNL regime for three different regimes of $t/\gamma$ ratio: (a) $t/\gamma=0.2$, (b) $t/\gamma=1$, and (c) $t/\gamma=2$. Note that the upper bound on EMP for the voltage-temperature probe is denoted by $\eta^*(\mathcal{P}_{max})/\eta_C$ in the main text. The parameters used are $\epsilon_1=0.6$, $\epsilon_2=\epsilon_3=0.5$, $\gamma=\gamma_L=\gamma_R=\gamma_P=0.05$, $T_R=T=0.1$, $T_L=0.13$, $T_P=0.102$ (for the voltage probe), $\mu_R=\mu=0.3$.}
    \label{fig:bound}
\end{figure*}
\indent
The CA limit, $\eta_{CA}=\eta_C/2$ is no longer an upper bound on the EMP with broken TRS \cite{Benenti2, Balachandran, Brandner2, Brandner3, Zhang, Zahra} and it can be exceeded in the nonlinear regime \cite{Seifert, Esposito2, Esposito3, Izumida1, Izumida2, Izumida3, Izumida4, Long1, Long2, Iyyappan1, Zhang2, entropy_Liu}. With broken TRS, by setting the figures of merit $y_m=H_m$ and $y=H$, we obtain a new upper bound on the EMP in Eq. (\ref{npmaxb}) and Eq. (\ref{npmaxb_vtprobe}) for the MNL voltage probe and voltage-temperature probe heat engine, respectively. Although the structure looks similar, their respective Carnot efficiencies and asymmetry parameters characterize the difference in the upper bounds obtained for the MNL voltage probe and voltage-probe heat engine. Figure \ref{fig:bound} shows the plot for the upper bound on the EMP under broken TRS in the MNL regime for both the voltage probe and voltage-temperature probe heat engine. The CA limit is surpassed both under broken TRS and with the inclusion of nonlinearity. Increasing the dissipation strength $\alpha$ allows for a faster surpassing of the CA limit with broken TRS. \\
\begin{figure*}[t!]
    \centering
    \includegraphics[scale=0.12]{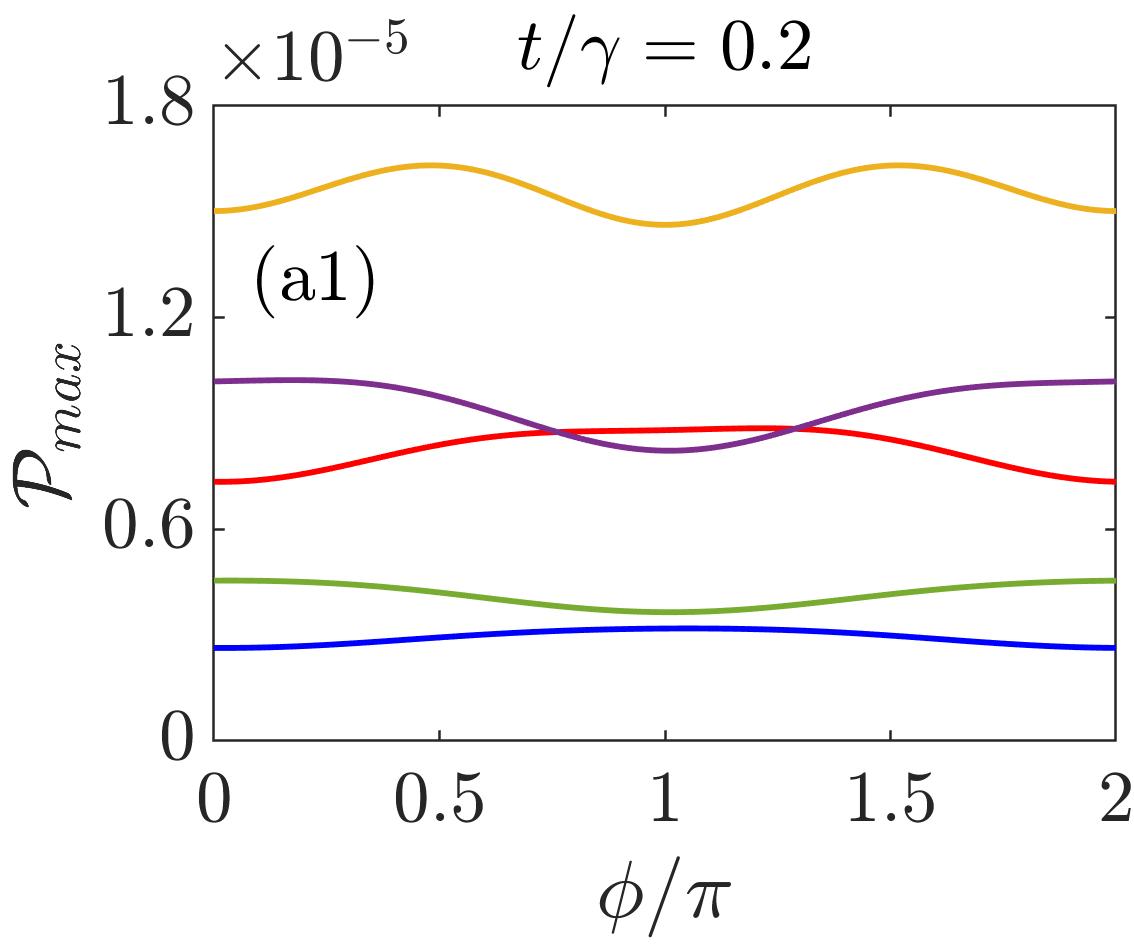}
    \includegraphics[scale=0.12]{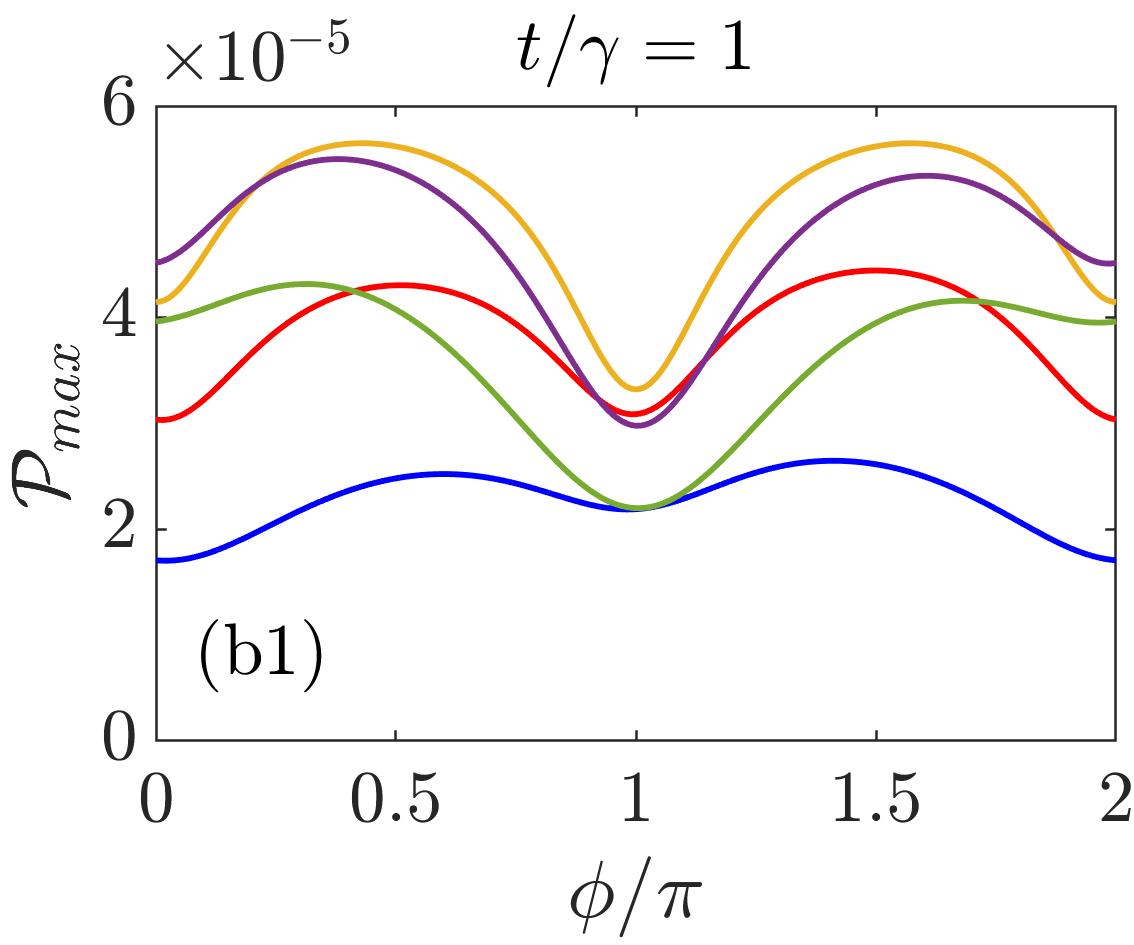}
    \includegraphics[scale=0.12]{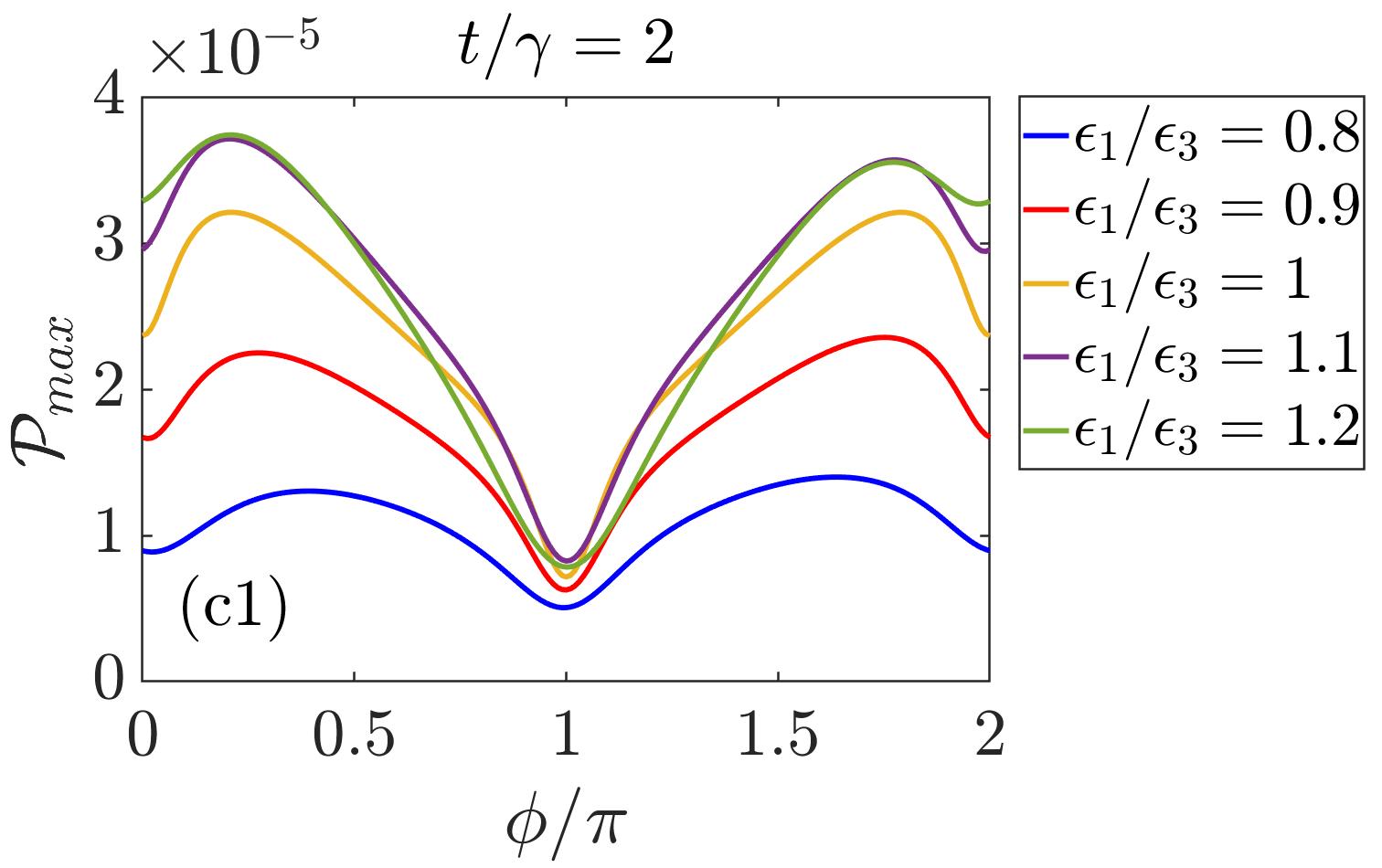}
    \includegraphics[scale=0.12]{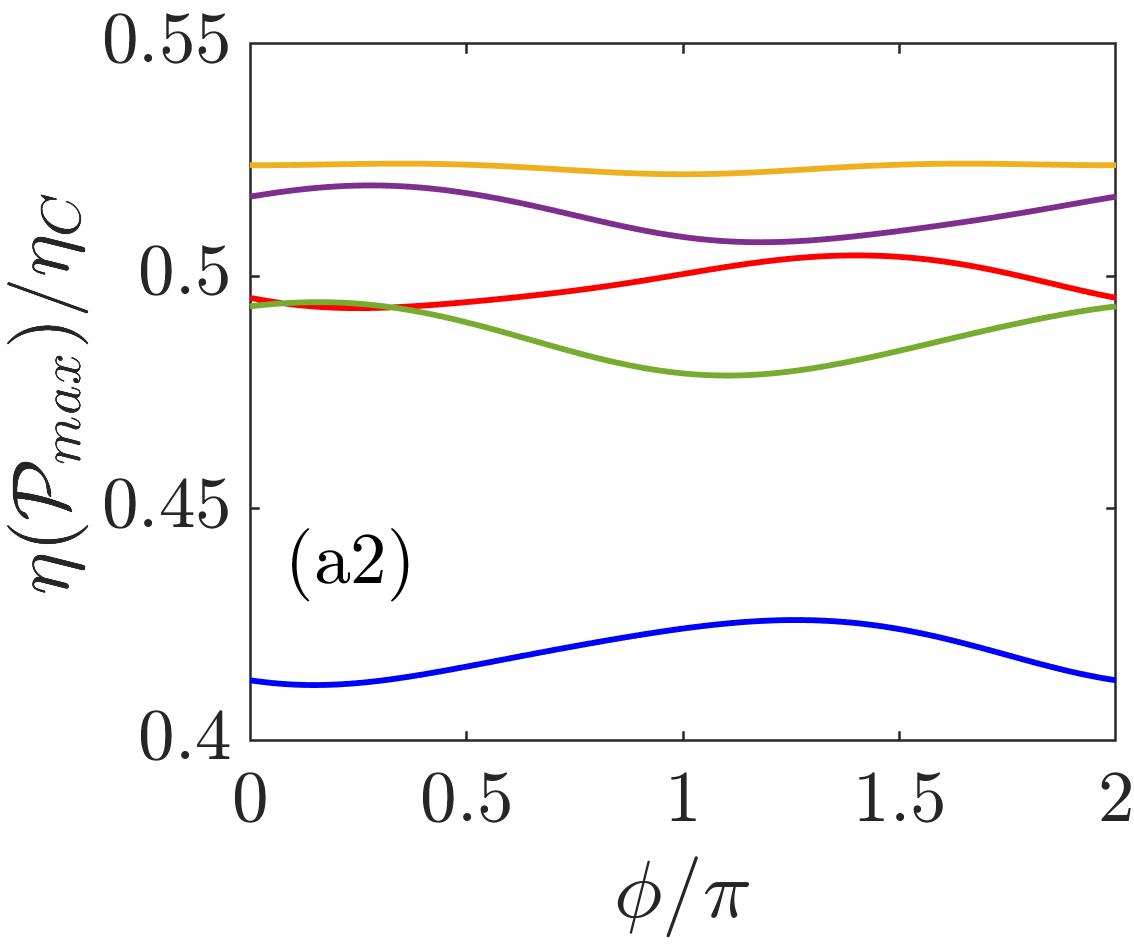}
    \includegraphics[scale=0.12]{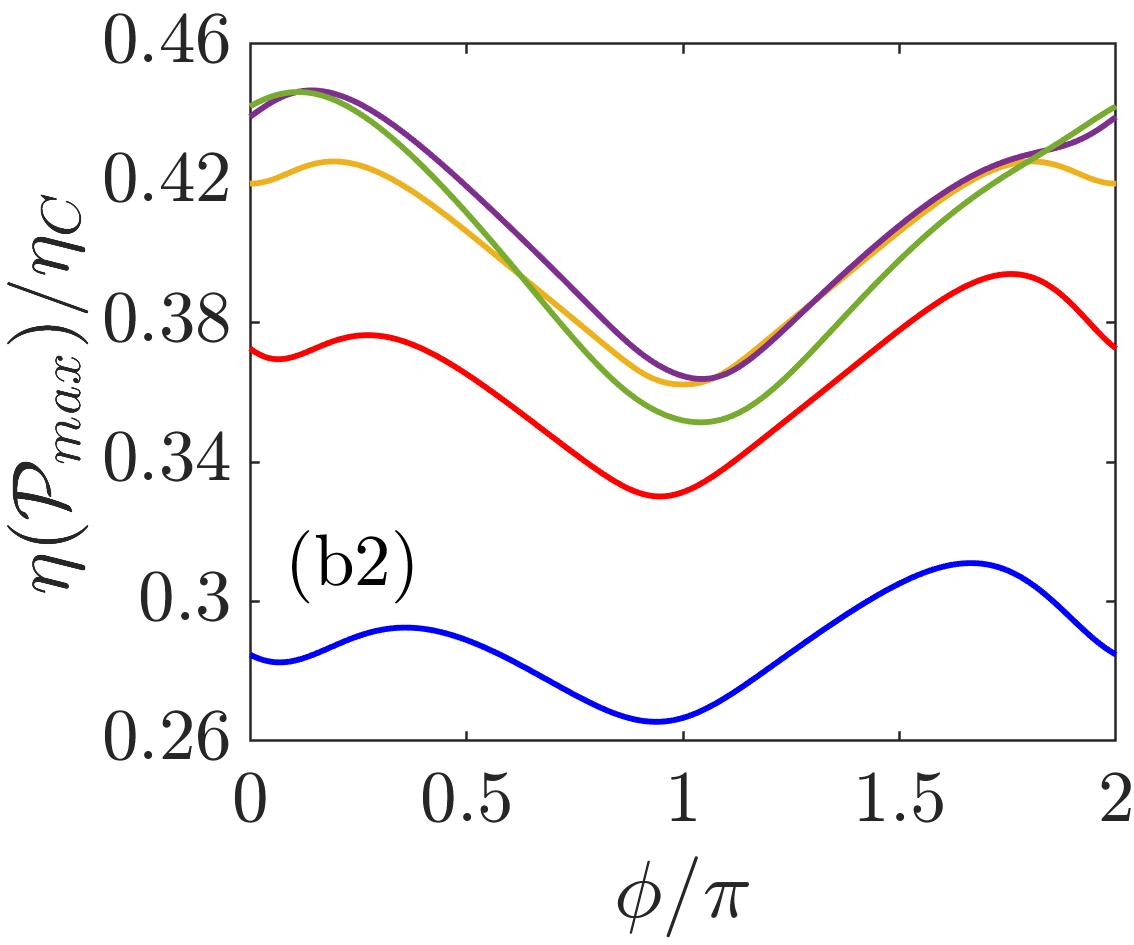}
    \includegraphics[scale=0.12]{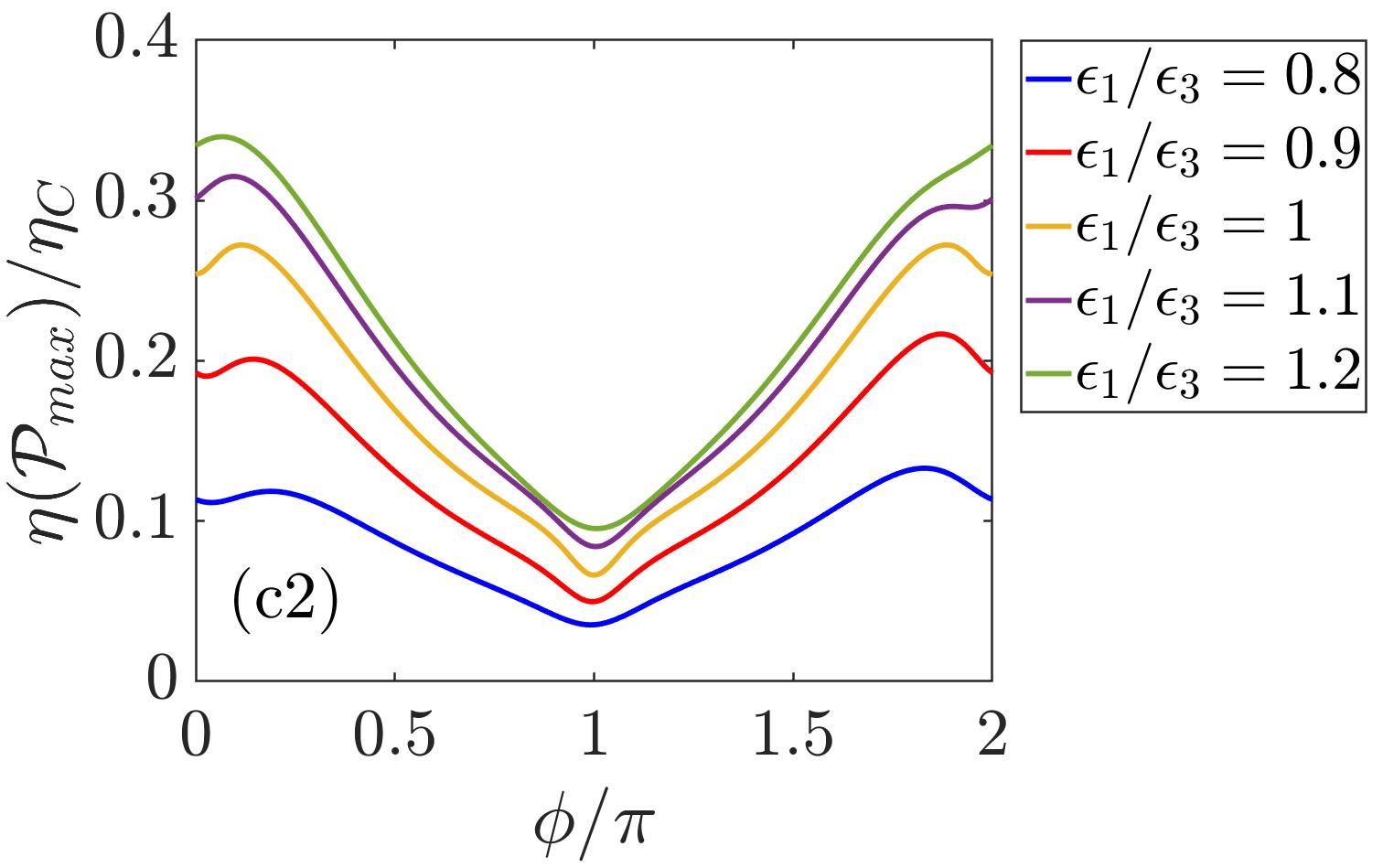}
    \includegraphics[scale=0.12]{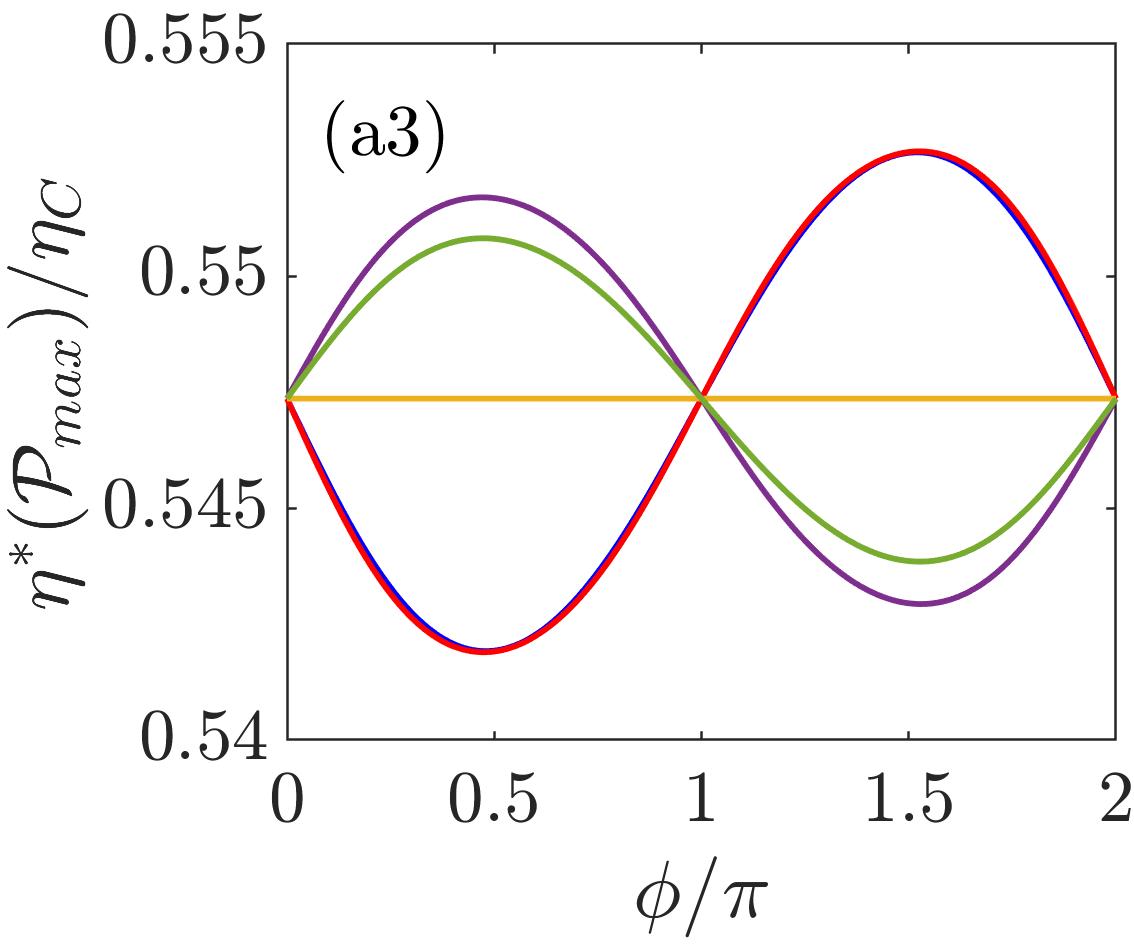}
    \includegraphics[scale=0.12]{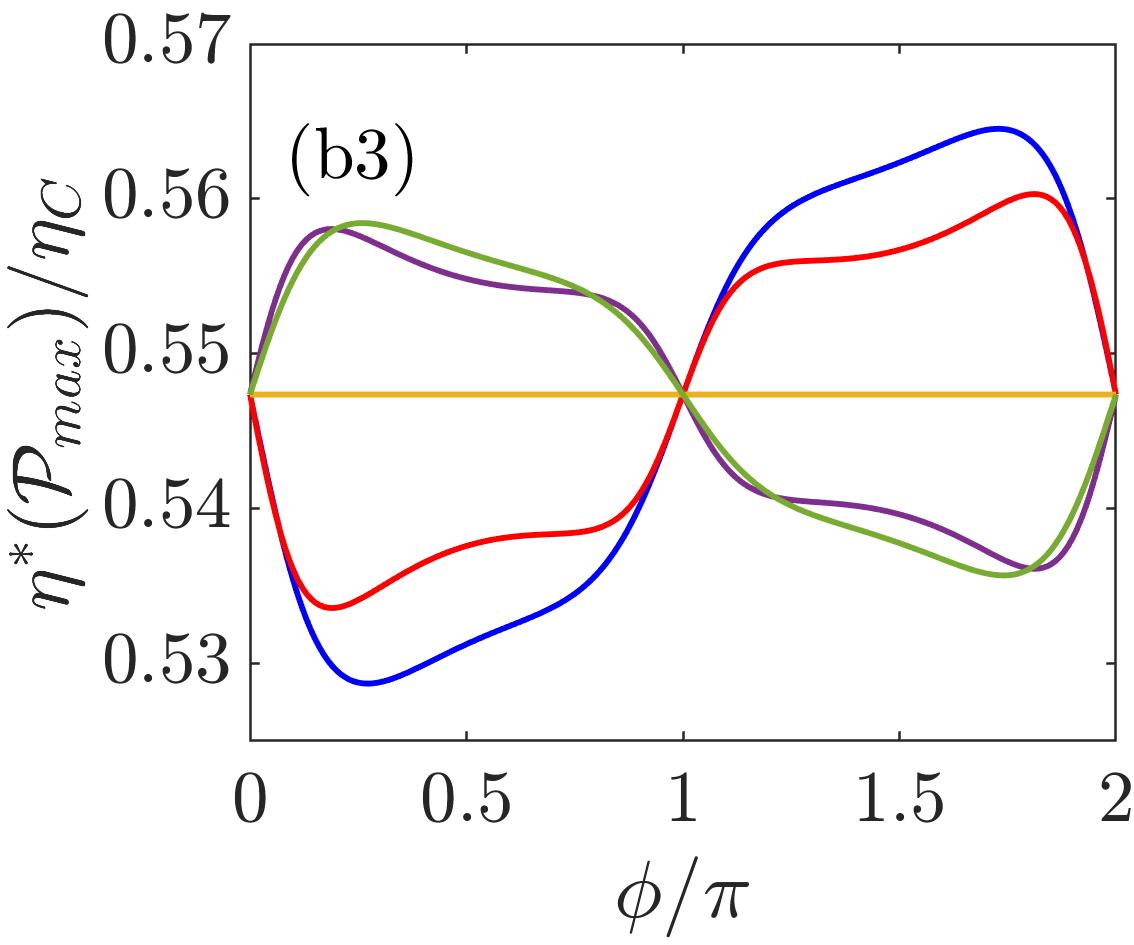}
    \includegraphics[scale=0.12]{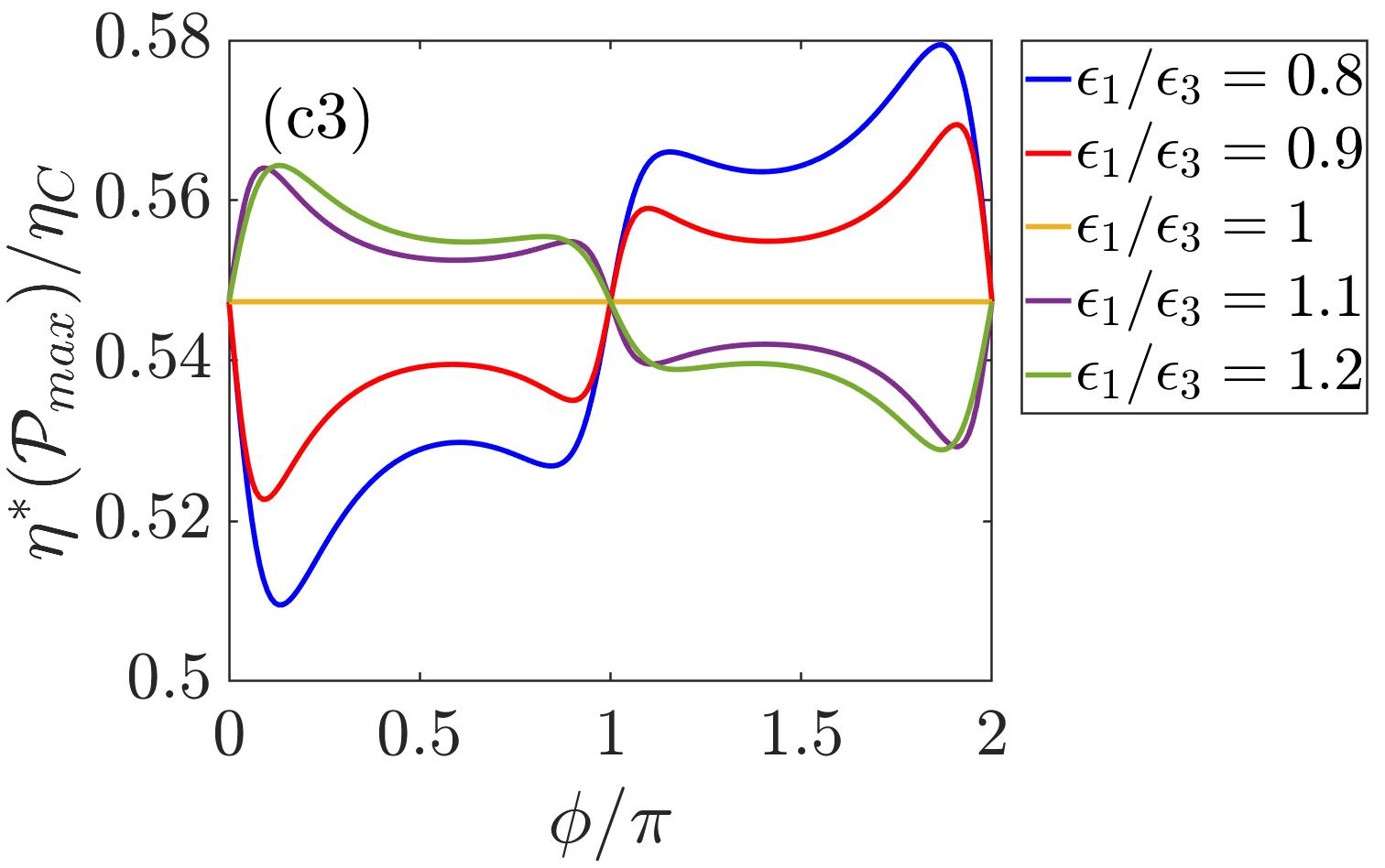}
    \caption{Maximum power ($\mathcal{P}_{max}$), EMP ($\eta(\mathcal{P}_{max})/\eta_C$) and upper bound on EMP ($\eta^*(\mathcal{P}_{max})/\eta_C$) as a function of $\phi$ for different anisotropic setup $\epsilon_1/\epsilon_3$ for an MNL voltage-temperature probe heat engine in three different regimes of $t/\gamma$ ratio: (a1-a2) $t/\gamma=0.2$, (b1-b2) $t/\gamma=1$, and $t/\gamma=2$. Parameters used are: $\gamma=\gamma_L=\gamma_R=\gamma_P=0.05$, $\epsilon_2=\epsilon_3=0.5$, $T_R=T=0.1$, $T_L=0.13$, $\mu_R=\mu=0.3$, $\alpha=0.75$.}
    \label{fig:anisotropy}
\end{figure*}
\indent Since anisotropy introduced by a non-degenerate quantum dots configuration, accompanied by a nonzero magnetic field, is a necessary condition for breaking the TRS ($x\ne1$) in a voltage-temperature probe heat engine, it is crucial to examine its impact on the engine's thermoelectric performance. Figure \ref{fig:anisotropy} depicts the effect of anisotropy, characterized by the ratio ($\epsilon_1/\epsilon_3$), on the thermoelectric performance of a voltage-temperature probe heat engine operating in the MNL regime. In the $t/\gamma<1$ regime, both the maximum power and EMP increase with the anisotropy ratio $\epsilon_1/\epsilon_3$, attain a maximum at the degenerate dot setup, $\epsilon_1/\epsilon_3=1$, and decrease thereafter as the ratio increases further, as depicted in Figs. \ref{fig:anisotropy}(a1) and \ref{fig:anisotropy}(a2). Figures \ref{fig:anisotropy}(b1) and \ref{fig:anisotropy}(b2) illustrate that in the $t/\gamma=1$ regime, although $\mathcal{P}_{max}$ is highest for the degenerate dot setup, the EMP goes beyond it when the anisotropic ratio $\epsilon_1/\epsilon_3>1$ attaining maximum at $\epsilon_1/\epsilon_3=1.1$. However, as the $t/\gamma$ ratio further increases, in the $t/\gamma>1$ regime,
the maximum in the thermoelectric performance shifts away from the degenerate dot setup. Instead, both maximum power and EMP continue to grow with increasing anisotropy, highlighting a shift in optimal configuration towards a more asymmetric configuration as shown in Figs. \ref{fig:anisotropy}(c1) and \ref{fig:anisotropy}(c2). Note that high thermoelectric performance is a consequence of the combined effect of high generalized figures of merit ($y$) and high asymmetries $x$, depicted in Fig. \ref{fig:aniso_xy} in appendix \ref{additional}. Figures \ref{fig:anisotropy}(a3-c3) illustrate the impact of anisotropy on the upper bound on EMP of a voltage-temperature probe heat engine in the MNL regime. In the linear-response regime, the CA limit remains the upper bound on the EMP as the TRS is not broken for the degenerate dot setup. However, the presence of the anisotropy $\epsilon_1/\epsilon_3\ne1$ breaks the TRS, and the CA limit is surpassed when $x>1$. Thus, TRS breaking with anisotropy and a nonzero magnetic field, as well as added nonlinearity, helps enhance the upper bound on EMP to exceed the CA limit. \\
\indent To summarize, we compare the thermoelectric performance of the voltage probe and voltage-temperature probe heat engine in the MNL regime under broken TRS. The symmetry is broken with a nonzero magnetic field, i.e., $\phi\ne0$ for a voltage heat engine. However, for a voltage-temperature probe heat engine, in addition to the nonzero magnetic field, some anisotropy (non-degenerate dot setup, $\epsilon_1\ne\epsilon_2=\epsilon_3$) is required in the system to break the TRS. We have demonstrated the effect of broken TRS with asymmetry parameters $x_L(x)>1$ in enhancing EMP and efficiency at given power in all three regimes of the $t/\gamma$ ratio. Although the two heat engines perform equivalently for the weak probe coupling, a significant difference in the power and efficiency is observed when the probe coupling is equal to the left-right reservoir coupling (i.e., $\gamma_L=\gamma_R=\gamma_P=\gamma$), with the voltage probe generating higher power output and high EMP in the regions with high $x_Ly_L$ values compared to the voltage-temperature probe. Note that high $x_Ly_L(xy)$ values correspond to high EMP for both the heat engines. We also observed that the $t/\gamma=1$ regime is the optimal operating regime for both heat engines. The nonlinear effect, combined with TRS, has been shown to enhance the efficiency of both heat engines. The performance of the voltage probe and voltage-temperature probe in the fully nonlinear regime is discussed in Appendix \ref{fully_nonlinear}. We have demonstrated the operation of the voltage probe heat engine in the fully nonlinear regime, achieving high power output and efficiency with EMP reaching $0.71\eta_C$, $0.69\eta_C$, and $0.6\eta_C$ for $t/\gamma=0.2$, $t/\gamma=1$, and $t/\gamma=2$ regimes, respectively. The effect of broken TRS remains consistent with the MNL regime. These results further support our analysis that higher nonlinearity and broken TRS enhance thermoelectric performance. Furthermore, we demonstrate the upper bounds on the EMP for the MNL heat engines, which overcome the CA limit under broken TRS and the introduction of nonlinearity in terms of the dissipation term.
\section{Conclusion}\label{sec:conclusion}
To conclude, we demonstrate the thermoelectric performance of an MNL voltage probe and voltage-temperature probe heat engine for a general setup with broken TRS. The voltage probe and voltage-temperature probe successfully introduce dissipative and non-dissipative inelastic scattering mechanisms into the system, respectively. The probe parameters are adjusted to block the net particle current and only allow the heat current with the voltage probe, whereas both the particle and heat currents are blocked with the voltage-temperature probe. Our calculations are based on extended Onsager relations where a nonlinear term ($-\gamma_h{J_L^N}^2$), meaning power dissipation, is introduced in the linear Onsager relation of the heat current. The nonlinear term can be seen as the inevitable power loss due to the finite time motion of the heat engine, and this can be treated as a natural extension of linear irreversible thermodynamics. We derive the analytical expression for the EMP and efficiency at a given power in terms of asymmetric parameters ($x_m$, $x$), generalized figures of merit ($y_m$, $y$), and $\alpha$ for the voltage probe and voltage temperature probe heat engine. We define a dimensionless parameter $\alpha$ with $0\le\alpha\le1$, denoting the dissipation strength ratio which accounts for the nonlinearity in the system. Two new universal upper bounds on the EMP are obtained for the voltage probe and voltage-temperature probe heat engines, which exceed the CA limit with nonlinearity and broken time-reversal symmetry.\\
\indent
To explicitly illustrate our results for the generalized setup, we investigate a three-terminal triple-dot AB heat engine using numerical simulation. We observe that in the presence of a nonzero magnetic field, unlike the voltage probe, the voltage-temperature probe heat engine requires anisotropy in the system, e.g., non-degenerate dot setup ($\epsilon_1\ne\epsilon_3$) to break the TRS. We demonstrate the tunability of the system with the $t/\gamma$ ratio. We obtain high efficiency but low output power in the $t/\gamma<1$ regime for both the voltage probe and voltage-temperature probe heat engine. The $t/\gamma=1$ regime is the optimal operating regime, generating the highest power and significantly higher efficiency than the $t/\gamma>1$ regime. The controllability of the $t/\gamma$ ratio in a triple-dot AB interferometer is investigated experimentally in Ref. \cite{Rogge, Uditendu, Chen, Noiri, exptunits}. Additionally, the voltage probe heat engine generates higher power than the voltage-temperature probe heat engine. However, the voltage-temperature probe heat engine is generally more efficient, except for certain ranges of $\phi$, where the voltage probe heat engine with high $x_Ly_L$ values outperforms it. We have demonstrated that the interplay of TRS breaking with the asymmetry parameters, $x_L(x)>1$ and high figures of merit, $y_L(y)$, leading to high values of $x_Ly_L(xy)$, are crucial for enhancing both the EMP and efficiency at a given power. For the MNL heat engines, the efficiency at a given power and the EMP are enhanced by adding nonlinearity in terms of dissipation, although the output power remains unaffected. We obtain two new upper bounds on the EMP for the MNL voltage probe and voltage-temperature probe heat engines with TRS. The difference in the upper bounds obtained for the voltage probe and the voltage-temperature heat engines is set by their respective Carnot efficiencies and asymmetry parameters. The CA limit on the EMP is exceeded with broken TRS, and the introduction of nonlinearity makes it faster to overcome the CA limit even in the time-reversal symmetry case. Our analysis is further validated by demonstrating the voltage probe heat engine operating in the fully nonlinear regime, where the EMP attains $0.71\eta_C$, $0.69\eta_C$, and $0.6\eta_C$ for $t/\gamma=0.2$, $t/\gamma=1$, and $t/\gamma=2$ regimes, respectively.\\
\indent
Our analysis reveals that the broken time-reversal symmetry can widen the parameter space to optimize the performance of heat engines (red-edged circular points in Fig. 5). We also illustrate the regimes (Fig. 8) where the CA limit can be exceeded in the MNL regime for a realistic model with broken TRS. Thus, our results provide useful guidelines in the fundamental understanding and optimization of real thermoelectric devices. Furthermore, our results will be of experimental relevance since the tuning of parameters such as $x_m$ and $y_m$, or $x$ and $y$, is indeed experimentally feasible given the recent progress in the phase-coherent thermal manipulation in the solid-state nanocircuits \cite{dubirmp,fornierinature,federico}.
\section*{Acknowledgement}
J.B. acknowledges the financial support received from
IIT Bhubaneswar in the form of an Institute Research Fellowship. M.B. acknowledges support from the Anusandhan National Research Foundation (ANRF), India, under the MATRICS scheme Grant No. MTR/2021/000566. B.K.A. acknowledges CRG Grant No. CRG/2023/003377 from ANRF, Government of India. B.K.A. would like to acknowledge funding from the National Mission on Interdisciplinary Cyber-Physical Systems (NM-ICPS) of the Department of Science and Technology, Govt. of India, through the I-HUB Quantum Technology Foundation, Pune, India.

\appendix
\section{Derivation of Onsager coefficients using Landauer-B\"{u}ttiker formalism}\label{coefficients}
For a noninteracting system, the particle current $J_{L}^N$ and the heat current $J_{L}^Q$ flowing from the left reservoir $L$ to the central system are defined using Landauer-B\"uttiker formalism as follows \cite{datta1997electronic, Sivan, Butcher_1990, Bergfield, Galperin1, Galperin2, Galperin3, Galperin4, Topp_2015}:
\begin{equation}\label{Landau_jln}
    \begin{split}
        J_{L}^N=\int_{-\infty}^{\infty} d\omega \Big[T_{LR}(\omega,\phi)f_{L}(\omega)-T_{RL}(\omega,\phi)f_{R}(\omega)\\
        +T_{LP}(\omega,\phi)f_{L}(\omega)-T_{PL}(\omega,\phi)f_{P}(\omega)\Big].
        \end{split}
\end{equation}
\begin{equation}\label{Landau_jlq}
    \begin{split}
        J_{L}^Q=\int_{-\infty}^{\infty} d\omega (\omega-\mu_L) \Big[T_{LR}(\omega,\phi)f_{L}(\omega)-T_{RL}(\omega,\phi)f_{R}(\omega)\\
        +T_{LP}(\omega,\phi)f_{L}(\omega)-T_{PL}(\omega,\phi)f_{P}(\omega)\Big].
        \end{split}
\end{equation}
\indent Similarly, the particle current $J_{P}^N$ and the heat current $J_{P}^Q$  flowing from the probe $P$ to the central system are defined as follows \cite{datta1997electronic, Sivan, Butcher_1990, Bergfield, Galperin1, Galperin2, Galperin3, Galperin4, Topp_2015}:
\begin{equation}\label{Landau_jpn}
    \begin{split}
        J_{P}^N=\int_{-\infty}^{\infty} d\omega \Big[T_{PL}(\omega,\phi)f_{P}(\omega)-T_{LP}(\omega,\phi)f_{L}(\omega)\\
        +T_{PR}(\omega,\phi)f_{P}(\omega)-T_{RP}(\omega,\phi)f_{R}(\omega)\Big].
        \end{split}
\end{equation}
\begin{equation}\label{Landau_jpq}
    \begin{split}
        J_{P}^Q=\int_{-\infty}^{\infty} d\omega (\omega-\mu_P)\Big[T_{PL}(\omega,\phi)f_{P}(\omega)-T_{LP}(\omega,\phi)f_{L}(\omega)\\
        +T_{PR}(\omega,\phi)f_{P}(\omega)-T_{RP}(\omega,\phi)f_{R}(\omega)\Big].
        \end{split}
\end{equation}
where $f_{\nu}(\omega,\mu_{\nu},T_{\nu})=[\mathrm{exp}(\frac{\omega-\mu_{\nu}}{T_{\nu}})+1]^{-1}$ is the Fermi-Dirac distribution function for $\nu=L,P,R$ . \\
\indent In the linear-response regime, we can expand the Fermi-Dirac distribution function around the reference point ($\mu,T$) as
\begin{equation}\label{fermiexpand}
    \begin{split}
        f_{\nu}(\omega,\mu_{\nu},T_{\nu})=f_a(\omega,\mu,T)+(\mu_{\nu}-\mu)\bigg(-\frac{\partial f_a}{\partial\omega}\bigg)\\
        +(T_{\nu}-T)\bigg(\frac{\omega-\mu}{T}\bigg)\bigg(-\frac{\partial f_a}{\partial\omega}\bigg),
    \end{split}
\end{equation}
where $f_a(\omega,\mu,T)=[\mathrm{exp}(\frac{\omega-\mu}{T})+1]^{-1}$ is equilibrium Fermi distribution at the reference point $(\mu,T)$. We consider the right reservoir as the reference point by setting $(\mu_R,T_R)=(\mu,T)$ and we can write $f_a(\omega,\mu,T)=f_R(\omega,\mu_R,T_R)$.\\
\indent By substituting Eq. (\ref{fermiexpand}) in Eq. (\ref{Landau_jln}), we can write the particle current $J_L^N$ as follows:
\begin{widetext}
    \begin{equation}\label{eq:jlnexpand1}
    \begin{split}
        J_L^N=\int_{-\infty}^{\infty} d\omega (T_{LR}+T_{LP})\bigg[f_a(\omega,\mu,T)+(\mu_{L}-\mu)(-\partial_{\omega}f_a)
        +(T_{L}-T)\Big(\frac{\omega-\mu}{T}\Big)(-\partial_{\omega} f_a)\bigg]\\
        -\int_{-\infty}^{\infty} d\omega \bigg[T_{PL}\Big[f_a(\omega,\mu,T)+(\mu_{P}-\mu)(-\partial_{\omega}f_a)
        +(T_{P}-T)\Big(\frac{\omega-\mu}{T}\Big)(-\partial_{\omega} f_a)\Big]
        +T_{RL}f_{a}\bigg].
    \end{split}
\end{equation}
We use the total probability conservation, $\sum_{\nu\ne\lambda}T_{\nu\lambda}=\sum_{\nu\ne\lambda}T_{\lambda\nu}$ for $\nu,\lambda=L,P,R$ to further simplify Eq. (\ref{eq:jlnexpand1}) as
\begin{equation}\label{eq:jlnexpand2}
    \begin{split}
        J_L^N=(\mu_{L}-\mu)\int_{-\infty}^{\infty} d\omega (T_{LR}+T_{LP})(-\partial_{\omega}f_a)
        +\bigg(\frac{T_{L}-T}{T}\bigg)\int_{-\infty}^{\infty} d\omega (T_{LR}+T_{LP})(\omega-\mu)(-\partial_{\omega} f_a)\\
        +(\mu_{P}-\mu)\int_{-\infty}^{\infty} d\omega (-T_{PL})(-\partial_{\omega}f_a)
        +\bigg(\frac{T_{P}-T}{T}\bigg)\int_{-\infty}^{\infty} d\omega (-T_{PL})(\omega-\mu)(-\partial_{\omega} f_a).
    \end{split}
\end{equation}
\end{widetext}
Equation (\ref{eq:jlnexpand2}) implies the thermodynamic flux and forces relation that can be written in terms of the Onsager coefficients as
\begin{equation}
    J_L^N=\mathcal{L}_{11}X_L^{\mu}+\mathcal{L}_{12}X_L^T
    +\mathcal{L}_{13}X_P^{\mu}+\mathcal{L}_{14}X_P^T,
\end{equation}
where $X_{\nu}^{\mu}=(\mu_{\nu}-\mu)/T$, $X_{\nu}^{T}=(T_{\nu}-T)/T^2$. Likewise, substituting Eq. (\ref{fermiexpand}) in Eqs. (\ref{Landau_jlq}), (\ref{Landau_jpn}), and (\ref{Landau_jpq}), we can obtain $J_L^Q$, $J_P^N$, and $J_P^Q$, respectively in the form of Onsager relations as follows:
\begin{equation}
    J_L^Q=\mathcal{L}_{21}X_L^{\mu}+\mathcal{L}_{22}X_L^T
    +\mathcal{L}_{23}X_P^{\mu}+\mathcal{L}_{24}X_P^T,
\end{equation}
\begin{equation}
    J_P^N=\mathcal{L}_{31}X_L^{\mu}+\mathcal{L}_{32}X_L^T
    +\mathcal{L}_{33}X_P^{\mu}+\mathcal{L}_{34}X_P^T,
\end{equation}
\begin{equation}
    J_P^Q=\mathcal{L}_{41}X_L^{\mu}+\mathcal{L}_{42}X_L^T
    +\mathcal{L}_{43}X_P^{\mu}+\mathcal{L}_{44}X_P^T,
\end{equation}
where the Onsager coefficients $\mathcal{L}_{ij}$ are defined as follows:
\begin{widetext}
\begin{equation}\label{onsager44}
\begin{alignedat}{2}
\mathcal{L}_{11} &=T\int_{-\infty}^{\infty}d\omega(-\partial_{\omega}f_a)(T_{LR}+T_{LP}), \quad &\qquad \mathcal{L}_{12} &=T\int_{-\infty}^{\infty}d\omega(-\partial_{\omega}f_a)(\omega-\mu)(T_{LR}+T_{LP})=\mathcal{L}_{21}, \\
\mathcal{L}_{13} &=T\int_{-\infty}^{\infty}d\omega(-\partial_{\omega}f_a)(-T_{PL}), \quad &\qquad \mathcal{L}_{14} &=T\int_{-\infty}^{\infty}d\omega(-\partial_{\omega}f_a)(\omega-\mu)(-T_{PL})=\mathcal{L}_{23}, \\
\mathcal{L}_{22} &=T\int_{-\infty}^{\infty}d\omega(-\partial_{\omega}f_a)(\omega-\mu)^2(T_{LR}+T_{LP}), \quad &\qquad \mathcal{L}_{24} &=T\int_{-\infty}^{\infty}d\omega(-\partial_{\omega}f_a)(\omega-\mu)^2(-T_{PL}), \\
\mathcal{L}_{31} &=T\int_{-\infty}^{\infty}d\omega(-\partial_{\omega}f_a)(-T_{LP}), \quad &\qquad \mathcal{L}_{32} &=T\int_{-\infty}^{\infty}d\omega(-\partial_{\omega}f_a)(\omega-\mu)(-T_{LP})=\mathcal{L}_{41}, \\
\mathcal{L}_{33} &=T\int_{-\infty}^{\infty}d\omega(-\partial_{\omega}f_a)(T_{PL}+T_{PR}), \quad &\qquad \mathcal{L}_{34} &=T\int_{-\infty}^{\infty}d\omega(-\partial_{\omega}f_a)(\omega-\mu)(T_{PL}+T_{PR})=\mathcal{L}_{43}, \\
\mathcal{L}_{42} &=T\int_{-\infty}^{\infty}d\omega(-\partial_{\omega}f_a)(\omega-\mu)^2(-T_{LP}), \quad &\qquad \mathcal{L}_{44} &=T\int_{-\infty}^{\infty}d\omega(-\partial_{\omega}f_a)(\omega-\mu)^2(T_{PL}+T_{PR}).
\end{alignedat}
\end{equation}
\end{widetext}
Here ($\partial_{\omega}f_a=\frac{\partial f_a}{\partial\omega}=-[4T\cosh^2{(\frac{\omega-\mu}{2T})}]^{-1}$) is the first-order derivative of the equilibrium Fermi distribution function with energy.
\section{Inelastic effects in the linear-response regime}\label{sec:inelastic}
Inelastic heat dissipative effects are introduced into the system with a voltage probe by enforcing the condition of zero net particle current to the probe ($J_P^N=0$) while permitting heat dissipation at the probe ($ J_P^Q\ne0$). The voltage probe condition demanding $J_P^N=0$ can be written using Eq. (\ref{Landau_jpn}) as follows:
\begin{equation}\label{eq:jpn0}
    \begin{split}
        \int_{-\infty}^{\infty} d\omega \big[(T_{PL}+T_{PR})f_{P}(\omega)-T_{LP}f_{L}(\omega)\\
        -T_{RP}f_{R}(\omega)\big]=0.
    \end{split}
\end{equation}
\begin{widetext}
In the linear-response regime, Eq. (\ref{eq:jpn0}) can further be expanded using Eq. (\ref{fermiexpand}) as follows:
\begin{equation}\label{eq:jpn0lin}
    \begin{split}
        \int_{-\infty}^{\infty} d\omega (T_{PL}+T_{PR})\Big[f_a(\omega,\mu,T)+(\mu_{P}-\mu)(-\partial_{\omega}f_a)
        +(T_{P}-T)\Big(\frac{\omega-\mu}{T}\Big)(-\partial_{\omega} f_a)\Big]\\
        -\int_{-\infty}^{\infty} d\omega \bigg[T_{LP}\Big[f_a(\omega,\mu,T)+(\mu_{L}-\mu)(-\partial_{\omega}f_a)
        +(T_{L}-T)\Big(\frac{\omega-\mu}{T}\Big)(-\partial_{\omega} f_a)\Big]
        +T_{RP}f_{a}\bigg]=0.
    \end{split}
\end{equation}
By using the total probability conservation, $T_{PL}+T_{PR}=T_{LP}+T_{RP}$ and solving the above equation, one can obtain the voltage probe chemical potential expressed as follows:
\begin{equation}\label{eq:muP_v}
    (\mu_P-\mu)=\frac{\begin{aligned} \bigg[(\mu_L-\mu)\int_{-\infty}^{\infty} d\omega T_{LP}(-\partial_{\omega}f_a)+\bigg(\frac{T_L-T}{T}\bigg)\int_{-\infty}^{\infty} d\omega T_{LP}(\omega-\mu)(-\partial_{\omega}f_a)\\
    -\bigg(\frac{T_P-T}{T}\bigg)\int_{-\infty}^{\infty} d\omega (T_{PL}+T_{PR})(\omega-\mu)(-\partial_{\omega}f_a)\bigg]\end{aligned}}
    {\int_{-\infty}^{\infty} d\omega (T_{PL}+T_{PR})(-\partial_{\omega}f_a)}.
\end{equation}
\end{widetext}
Further simplifying Eq. (\ref{eq:muP_v}), we can write the voltage probe chemical potential in terms of the thermodynamic forces and Onsager coefficients as follows:
\begin{equation}\label{eq:xpv}
    X_P^{\mu}=-\frac{(\mathcal{L}_{31}X_L^{\mu}+\mathcal{L}_{32}X_L^T+\mathcal{L}_{34}X_P^T)}{\mathcal{L}_{33}},
\end{equation}
where $X_{\nu}^{\mu}=(\mu_{\nu}-\mu)/T$ and $X_{\nu}^{T}=(T_{\nu}-T)/T^2$.\\
\indent It can be seen that Eq. (\ref{eq:xpv}) is the same as Eq. (\ref{xpv}) defined in Sec. \ref{sec:linear}. Thereafter, we follow the formalism as discussed in Sec. \ref{sec:linear} to evaluate the reduced Onsager matrix for the voltage probe defined in Eq. (\ref{vprobe}) and extend this formalism for the voltage-temperature probe that blocks both the net particle current ($J_P^N=0$) and heat current ($J_P^Q=0$) flow into the probe to obtain a ($2\times2$) reduced Onsager matrix as defined in Eq. (\ref{vtprobe}). From this point onward, we introduce the minimally nonlinear dissipative term to the heat current and follow the formalism outlined in Sec. \ref{sec:mnl} for the MNL heat engines.

\section{Theoretical Framework for Minimally Nonlinear Irreversible Heat Engines}\label{app:entropy_gauge}
To establish the theoretical framework of the MNL irreversible heat engine, it is essential to begin with an understanding of the total entropy production in the system. The theoretical formulation of an MNL irreversible heat engine in two-terminal systems with time-reversal symmetry has been previously discussed in Refs \cite{Izumida3, Izumida4}. For a steady-state heat engine, the internal state of the engine remains constant, allowing the application of the local equilibrium hypothesis \cite{callen1998thermodynamics}. For a three-terminal system, the total entropy production rate $\dot S_{tot}$ is written as the sum of the entropy changes in each heat reservoirs, $\dot S_{tot}=\sum_{\nu}\dot S_{\nu}=-\sum_{\nu}J_{\nu}^Q/T_{\nu}$, where $\nu=L, R, P$. In a voltage-temperature probe heat engine, he probe terminal is defined by the condition of vanishing net particle and heat currents, i.e., ($J_P^N=0$) and ($J_P^Q=0$). Under this condition, the total entropy production rate is expressed as follows:
\begin{equation}\label{entropy_VT}
    \dot S_{tot}=\dot S_L+\dot S_R=-\frac{J_L^Q}{T_L}-\frac{J_R^Q}{T_R}.
\end{equation}
By using the law of conservation of energy, $J_R^Q=\mathcal{P}-J_L^Q$, with output power $\mathcal{P}=-T_RX_L^{\mu}J_L^N$, Eq. (\ref{entropy_VT}) can be written as
\begin{equation}\label{entropy_JX}
    \dot S_{tot}=J_L^NX_L^{\mu}+J_L^QX_L^T,
\end{equation}
where $X_L^{\mu}=\frac{\mu_L-\mu_R}{T_R}$ and $X_L^T=\frac{1}{T_R}-\frac{1}{T_L}$ are the thermodynamic forces and $J_L^N$ and $J_L^Q$ are the thermodynamic fluxes (particle and heat current, respectively).\\
\indent For a general three-terminal system, we can perform the Taylor series expansion of the thermodynamic flux $J_i$ in terms of the thermodynamic force $X_i$ around the equilibrium point $X_i=0$, for $i=1,2,3,4$, as
\begin{equation}\label{Ji_expand}
    \begin{split}
        J_i &=\sum_{j=1}^4\mathcal{L}_{ij}X_j+\sum_{j,k=1}^4\mathcal{M}_{ijk}X_jX_k\\
        &\quad +\sum_{j,k,m=1}^4\mathcal{N}_{ijkm}X_jX_kX_m+\cdots,
    \end{split}
\end{equation}
where $\mathcal{L}_{ij}$, $\mathcal{M}_{ijk}$, and $\mathcal{N}_{ijkm}$ are the linear, quadratic, and cubic expansion coefficients, respectively. Note that in our model, the fluxes and forces are defined as $(J_1, J_2, J_3, J_3)\equiv(J_L^N, J_L^Q, J_P^N, J_P^Q)$ and $(X_1, X_2, X_3, X_4)\equiv(X_L^{\mu}, X_L^T, X_P^{\mu}, X_P^T)$. This expansion provides a complete framework for describing the entropy production rate of a heat engine operating in the far-from-equilibrium regime.\\
\indent The linear irreversible thermodynamics model assumes systems in local equilibrium, i.e., $X_i\to 0$, and defines the thermodynamic fluxes and forces through the linear Onsager relations between them. As discussed in Sec. \ref{sec:linear}, in the linear response regime, the thermodynamic fluxes and forces of the voltage-temperature probe heat engine are governed by the linear Onsager relations as follows:
\begin{equation}\label{entropy_jln}
    J_L^N=\mathcal{L}^{\prime\prime}_{11}X_L^{\mu}
    +\mathcal{L}^{\prime\prime}_{12}X_L^T,
\end{equation}
\begin{equation}\label{entropy_jlq}
    J_L^Q=\mathcal{L}^{\prime\prime}_{21}X_L^{\mu}
    +\mathcal{L}^{\prime\prime}_{22}X_L^T,
\end{equation}
where the reduced Onsager coefficients $\mathcal{L}_{ij}^{\prime\prime}$ are defined in Eq. (\ref{eq:coeff_VT}) of Sec. \ref{sec:linear}. The unitarity of the scattering matrix imposes much stronger bounds on these reduced Onsager coefficients than the positivity of the entropy production rate \cite{Brandner2} as discussed in Eq. (\ref{boundvt}), which can further be simplified as
\begin{equation}\label{entropy_boundvt}
    \begin{aligned}
        \mathcal{L}^{\prime\prime}_{11}\ge0,\,\,\ \mathcal{L}^{\prime\prime}_{22}\ge0,\\
        \mathcal{L}^{\prime\prime}_{11}\mathcal{L}^{\prime\prime}_{22}- \mathcal{L}^{\prime\prime}_{12}\mathcal{L}^{\prime\prime}_{21}\geq({\mathcal{L}^{\prime\prime}_{12}}-{\mathcal{L}^{\prime\prime}_{21}})^2.
    \end{aligned}
\end{equation}
The third inequality, which arises from the unitarity of the scattering matrix, bounds the degree of asymmetry quantified by the difference (${\mathcal{L}^{\prime\prime}_{12}}-{\mathcal{L}^{\prime\prime}_{21}}$). This inequality captures a fundamental trade-off between asymmetry and irreversibility: any enhancement in asymmetry due to broken time-reversal symmetry must be accompanied by increased entropy production, reflecting greater thermodynamic irreversibility.\\
\indent
Using Eq. (\ref{entropy_jln}) to change variables from $X_L^{\mu}$ to $J_L^N$, we can rewrite the Onsager relation in Eq. (\ref{entropy_jlq}), and the expression $J_R^Q=\mathcal{P}-J_L^Q$ in terms of $J_L^N$ as
\begin{equation}\label{entropy_jlqjln}
    J_L^Q=\frac{\mathcal{L}_{21}^{\prime\prime}}{\mathcal{L}_{11}^{\prime\prime}}J_L^N+\bigg(\frac{\mathcal{L}_{11}^{\prime\prime}\mathcal{L}_{22}^{\prime\prime}-\mathcal{L}_{12}^{\prime\prime}\mathcal{L}_{21}^{\prime\prime}}{\mathcal{L}_{11}^{\prime\prime}}\bigg)X_L^T,
\end{equation}
\begin{equation}\label{entropy_jrqjln}
\begin{split}
    J_R^Q &=\bigg(\frac{\mathcal{L}_{12}^{\prime\prime}T_RX_L^T-\mathcal{L}_{21}^{\prime\prime}}{\mathcal{L}_{11}^{\prime\prime}}\bigg)J_L^N-\bigg(\frac{\mathcal{L}_{11}^{\prime\prime}\mathcal{L}_{22}^{\prime\prime}-\mathcal{L}_{12}^{\prime\prime}\mathcal{L}_{21}^{\prime\prime}}{\mathcal{L}_{11}^{\prime\prime}}\bigg)X_L^T\\
    &\quad -\frac{T_R}{\mathcal{L}_{11}^{\prime\prime}}{J_L^N}^2.
\end{split}
\end{equation}
Using Eq. (\ref{entropy_jlqjln}) and Eq. (\ref{entropy_jrqjln}), we can express the total entropy production rate for the voltage-temperature probe heat engine as
\begin{equation}
\begin{split}
    \dot S_{tot} &=\bigg(\frac{\mathcal{L}_{21}^{\prime\prime}-\mathcal{L}_{12}^{\prime\prime}}{\mathcal{L}_{11}^{\prime\prime}}\bigg)X_L^TJ_L^N+\bigg(\frac{\mathcal{L}_{11}^{\prime\prime}\mathcal{L}_{22}^{\prime\prime}-\mathcal{L}_{12}^{\prime\prime}\mathcal{L}_{21}^{\prime\prime}}{\mathcal{L}_{11}^{\prime\prime}}\bigg){X_L^T}^2\\
    &\quad +\frac{{J_L^N}^2}{\mathcal{L}_{11}^{\prime\prime}}.
\end{split}
\end{equation}
Notably, the nonlinear term $\frac{T_R}{\mathcal{L}_{11}^{\prime\prime}}{J_L^N}^2$ appears exclusively in $J_R^Q$, in an asymmetric form, and serves to account for dissipation effect, thereby contributing to the overall entropy production rate. \\
\indent In our formulation of the MNL irreversible heat engine, we adopt extended Onsager relations that incorporate dissipation terms symmetrically into both sides of the heat fluxes described as follows:
\begin{equation}\label{entropy_jln_mnl}
    J_L^N=\mathcal{L}^{\prime\prime}_{11}X_L^{\mu}
    +\mathcal{L}^{\prime\prime}_{12}X_L^T,
\end{equation}
\begin{equation}\label{entropy_jlq_mnl}
    J_L^Q=\mathcal{L}^{\prime\prime}_{21}X_L^{\mu}
    +\mathcal{L}^{\prime\prime}_{22}X_L^T-\gamma_h{J_L^N}^2.
\end{equation}
Motivated by the framework of the low-dissipation Carnot cycle \cite{Esposito}, the nonlinear term $-\gamma_h{J_L^N}^2$ in $J_L^Q$ captures the irreversible dissipation into the hot reservoir, with $\gamma_h>0$ quantifying the dissipation strength. We assume that no other higher-order terms arise in Eq. (\ref{entropy_jln_mnl}) and Eq. (\ref{entropy_jlq_mnl}) and the bounds on the Onsager coefficients $\mathcal{L}_{ij}^{\prime\prime}$ defined in Eq. (\ref{entropy_boundvt}) still hold. By using $J_L^N$ as the control parameter instead of $X_L^{\mu}$, we can write
\begin{equation}\label{entropy_jlq_mnl_jnl}
\begin{split}
    J_L^Q &=\frac{\mathcal{L}_{21}^{\prime\prime}}{\mathcal{L}_{11}^{\prime\prime}}J_L^N+\bigg(\frac{\mathcal{L}_{11}^{\prime\prime}\mathcal{L}_{22}^{\prime\prime}-\mathcal{L}_{12}^{\prime\prime}\mathcal{L}_{21}^{\prime\prime}}{\mathcal{L}_{11}^{\prime\prime}}\bigg)X_L^T\\
    &\quad -\gamma_h{J_L^N}^2,
\end{split}
\end{equation}
\begin{equation}\label{entropy_jrq_mnl_jnl}
\begin{split}
    J_R^Q &=\bigg(\frac{\mathcal{L}_{12}^{\prime\prime}T_RX_L^T-\mathcal{L}_{21}^{\prime\prime}}{\mathcal{L}_{11}^{\prime\prime}}\bigg)J_L^N-\bigg(\frac{\mathcal{L}_{11}^{\prime\prime}\mathcal{L}_{22}^{\prime\prime}-\mathcal{L}_{12}^{\prime\prime}\mathcal{L}_{21}^{\prime\prime}}{\mathcal{L}_{11}^{\prime\prime}}\bigg)X_L^T\\
    &\quad -\gamma_c{J_L^N}^2,
\end{split}
\end{equation}
where $\gamma_c=T_R/\mathcal{L}_{11}^{\prime\prime}-\gamma_h>0$ is the strength of dissipation to the cold reservoir. The entropy production rate for the MNL voltage-temperature probe heat engine is expressed as:
\begin{equation}\label{entropy_mnl}
\begin{split}
    \dot S_{tot} &=\bigg(\frac{\mathcal{L}_{21}^{\prime\prime}-\mathcal{L}_{12}^{\prime\prime}}{\mathcal{L}_{11}^{\prime\prime}}\bigg)X_L^TJ_L^N+\bigg(\frac{\mathcal{L}_{11}^{\prime\prime}\mathcal{L}_{22}^{\prime\prime}-\mathcal{L}_{12}^{\prime\prime}\mathcal{L}_{21}^{\prime\prime}}{\mathcal{L}_{11}^{\prime\prime}}\bigg){X_L^T}^2\\
    &\quad +\frac{\gamma_h}{T_L}{J_L^N}^2+\frac{\gamma_c}{T_R}{J_L^N}^2.
\end{split}
\end{equation}
The positivity of the entropy production rate is assured from the bounds on Onsager coefficients imposed by the unitarity in Eq. (\ref{entropy_boundvt}) and from the non-negativity of $\gamma_h$ and $\gamma_c$. We can understand the contribution of each term in the entropy production rate, which reflects different physical mechanisms of dissipation. The first term, $(\mathcal{L}_{21}^{\prime\prime}-\mathcal{L}_{12}^{\prime\prime})X_L^TJ_L^N/\mathcal{L}_{11}^{\prime\prime}$, arises due to the asymmetry between the cross-coefficients $\mathcal{L}_{12}^{\prime\prime}$ and $\mathcal{L}_{21}^{\prime\prime}$ due broken TRS. This term can be either positive or negative, depending on the sign of $\mathcal{L}_{21}^{\prime\prime}-\mathcal{L}_{12}^{\prime\prime}$, and thus can either enhance or suppress entropy production, reflecting the impact of nonreciprocal transport. However, a large negative contribution from this term is constrained by the unitarity condition, as discussed in Eq. (\ref{entropy_boundvt}). The second term in Eq. (\ref{entropy_mnl}) captures the net dissipation due to heat transport, taking into account the coupling between particle and heat currents. Finally, the third and fourth terms, $\gamma_h{J_L^N}^2/T_L$ and $\gamma_c{J_L^N}^2/T_R$, respectively, represent the intrinsic dissipation arising from finite-time operation. These terms are associated purely with particle transport, analogous to Joule heating, and are always positive. Thus, the MNL model captures the leading-order effects of dissipation that are neglected in linear response. In our formulation of the MNL voltage-temperature probe heat engine, the nonlinear correction to the entropy production rate captures realistic irreversible losses and illustrates how Onsager asymmetry and nonlinear effects jointly influence entropy generation and the overall performance of the heat engine. We extend the similar formalism to the MNL voltage probe heat engine under broken TRS.
\subsection{Gauge-invariance of Minimally Nonlinear Irreversible Thermodynamics (MNLIT)}
Gauge invariance in the context of MNLIT refers to the invariance of physical quantities, such as the entropy production, under redefinitions of the thermodynamic fluxes and forces, provided these transformations preserve the entropy production rate as defined in Eq. (\ref{entropy_JX}). If we redefine the thermodynamic forces and the fluxes as follows:
\begin{equation}\label{GT_force}
    \tilde{X}_L^{\mu}=X_L^{\mu}+\lambda X_L^T, \quad \tilde{X}_L^T=X_L^T,
\end{equation}
\begin{equation}\label{GT_flux}
    \tilde{J}_L^N=J_L^N, \quad \tilde{J}_L^Q=J_L^Q-\lambda J_L^N.
\end{equation}
Then one can show that
\begin{equation}
    \tilde{J}_L^N\tilde{X}_L^{\mu}+\tilde{J}_L^Q\tilde{X}_L^T=J_L^NX_L^{\mu}+J_L^QX_L^T.
\end{equation}
This is a gauge transformation of the thermodynamic fluxes and forces that keeps the entropy production rate invariant. The question now is whether the nonlinear term in the MNLIT equations respects this invariance. Applying the transformation given in Eq. (\ref{GT_force}) and Eq. (\ref{GT_flux}) to the expression of $J_L^N$ and $J_L^Q$ defined in Eq. (\ref{entropy_jln_mnl}) and Eq. (\ref{entropy_jlq_mnl}), respectively, we obtain
\begin{equation}\label{GT_jln}
    J_L^N=\mathcal{L}_{11}^{\prime\prime}(\tilde{X}_L^{\mu}-\lambda \tilde{X}_L^T)+\mathcal{L}_{12}^{\prime\prime}\tilde{X}_L^T,
\end{equation}
\begin{equation}\label{GT_jlq}
    J_L^Q=\mathcal{L}_{21}^{\prime\prime}(\tilde{X}_L^{\mu}-\lambda \tilde{X}_L^T)+\mathcal{L}_{22}^{\prime\prime}\tilde{X}_L^T-\gamma_h \big({\tilde{J}_{L}^N}\big)^{2}.
\end{equation}
Now, using Eq. (\ref{GT_jln}) and Eq. (\ref{GT_jlq}), we can express $\tilde{J}_L^Q=J_L^Q-\lambda J_L^N$ as
\begin{equation}
\begin{split}
    \tilde{J}_L^Q &=\mathcal{L}_{21}^{\prime\prime}(\tilde{X}_L^{\mu}-\lambda \tilde{X}_L^T)+\mathcal{L}_{22}^{\prime\prime}\tilde{X}_L^T-\gamma_h \big({\tilde{J}_{L}^N}\big)^{2}\\
    &\quad -\lambda \big[\mathcal{L}_{11}^{\prime\prime}(\tilde{X}_L^{\mu}-\lambda \tilde{X}_L^T)+\mathcal{L}_{12}^{\prime\prime}\tilde{X}_L^T\big],
\end{split}
\end{equation}
which can further be simplified as
\begin{equation}
\begin{split}
    \tilde{J}_L^Q &=(\mathcal{L}_{21}^{\prime\prime}-\lambda\mathcal{L}_{11}^{\prime\prime})\tilde{X}_L^{\mu}\\
    &\quad +\big[\mathcal{L}_{22}^{\prime\prime}-\lambda(\mathcal{L}_{12}^{\prime\prime}+\mathcal{L}_{21}^{\prime\prime})+\lambda^2\mathcal{L}_{11}^{\prime\prime}\big]\tilde{X}_L^T
    -\gamma_h \big({\tilde{J}_{L}^N}\big)^{2}.
\end{split}
\end{equation}
Thus, the gauge transformation of the heat current for the MNLIT case is given as
\begin{equation}
    \tilde{J}_L^Q=L_{21}^{\prime\prime}\tilde{X}_L^{\mu}+L_{22}^{\prime\prime}\tilde{X}_L^T
    -\gamma_h \big({\tilde{J}_{L}^N}\big)^{2},
\end{equation}
where $L_{21}^{\prime\prime}=\mathcal{L}_{21}^{\prime\prime}-\lambda\mathcal{L}_{11}^{\prime\prime}$ and $L_{22}^{\prime\prime}=\mathcal{L}_{22}^{\prime\prime}-\lambda(\mathcal{L}_{12}^{\prime\prime}+\mathcal{L}_{21}^{\prime\prime})+\lambda^2\mathcal{L}_{11}^{\prime\prime}$ are the new linear coefficients which depends on $\lambda$. However, the heat current $J_L^Q$ remains invariant under this transformation. This indicates that the MNLIT model is gauge invariant under the transformation of thermodynamic forces and fluxes that leave the entropy production rate invariant. The overall structure of the heat current, including the nonlinear term, is preserved under the transformation with the redefined linear coefficients. Thus, the minimal nonlinearity introduced into the linear Onsager relation respects the gauge symmetry of the nonequilibrium thermodynamics.
\section{Carnot efficiency for a three-terminal heat engine}\label{Carnot}
Carnot efficiency is the upper bound of the efficiency of a thermoelectric heat engine, which can be derived from the zero entropy production rate condition \cite{Sadi}. A voltage probe heat engine operating between temperatures $T_R<T_P<T_L$, the Carnot efficiency is defined as follows \cite{Mazza}:\\
If $J_L^Q$ is only positive
\begin{equation}\label{eq:etaCL}
    \eta_{C,L}=1-\frac{T_R}{T_L}+\frac{J_P^Q}{J_L^Q}\Big(1-\frac{T_R}{T_P}\Big),
\end{equation}
and if both $J_L^Q$ and $J_P^Q$ are positive
\begin{equation}\label{eq:etaCLP}
    \eta_{C,LP}=1-\frac{T_R}{T_L}\Bigg(1+\frac{\frac{T_L}{T_P}-1}{1+\frac{J_L^Q}{J_P^Q}}\Bigg).
\end{equation}
For a voltage-temperature probe with $J_P^Q=0$, the Carnot efficiency is defined as
\begin{equation}\label{eq:eta}
    \eta_C=1-\frac{T_R}{T_L}.
\end{equation}
One can observe that the Carnot efficiency of the voltage-temperature probe heat engine depends only on the temperatures of the left (hot) and the right (cold) reservoirs, like the two-terminal heat engine.
However, the Carnot efficiency is not constant for a voltage probe heat engine. It depends on the temperatures of all three reservoirs and is a function of the heat currents from the left reservoir and the probe.
\begin{figure*}[t!]
    \centering
    \includegraphics[scale=0.12]{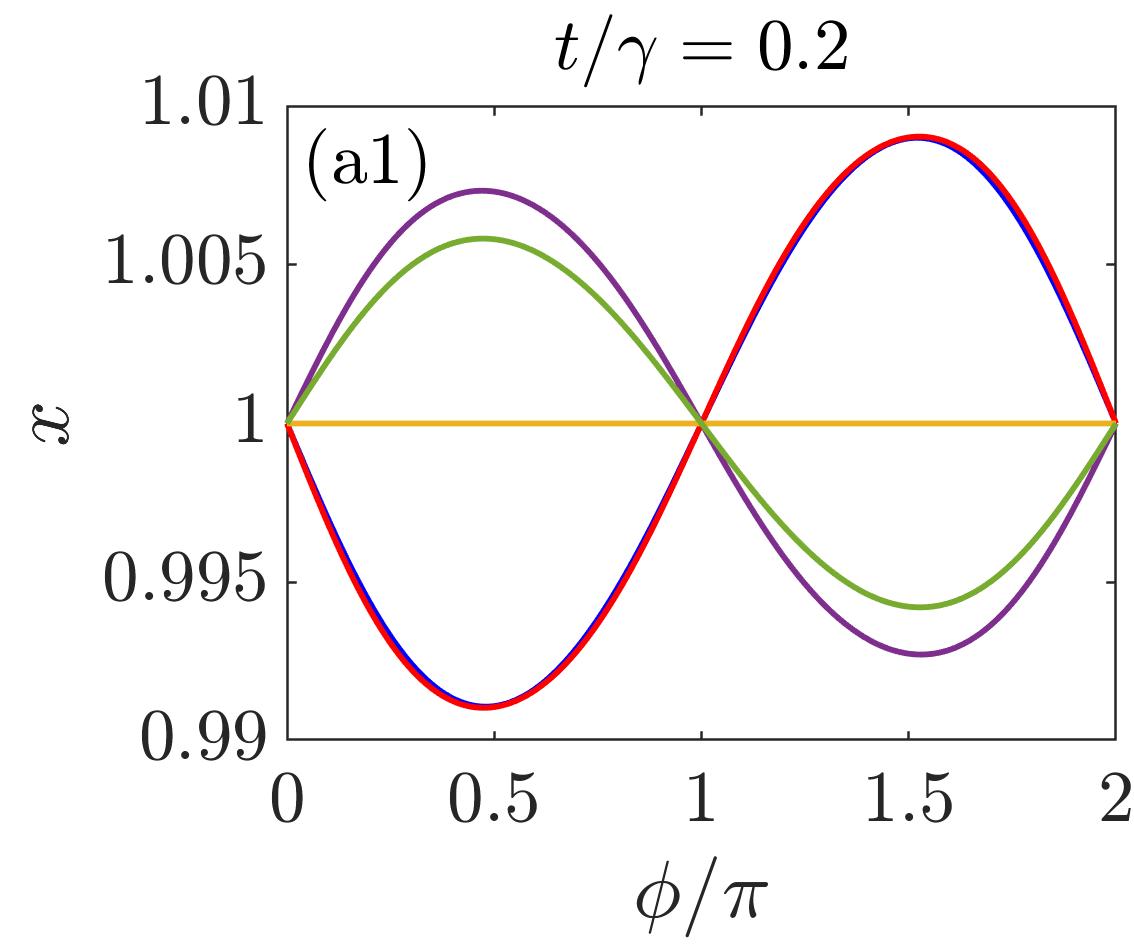}
    \includegraphics[scale=0.12]{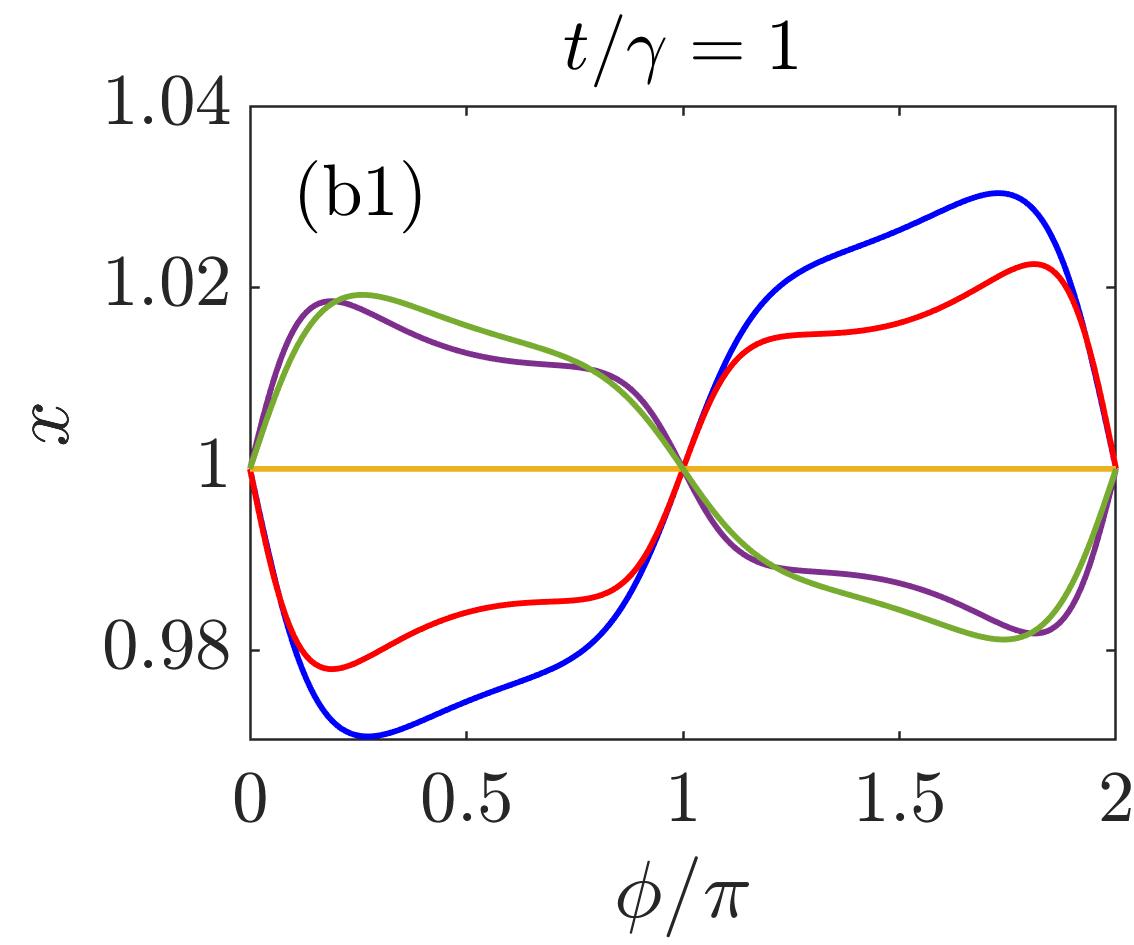}
    \includegraphics[scale=0.12]{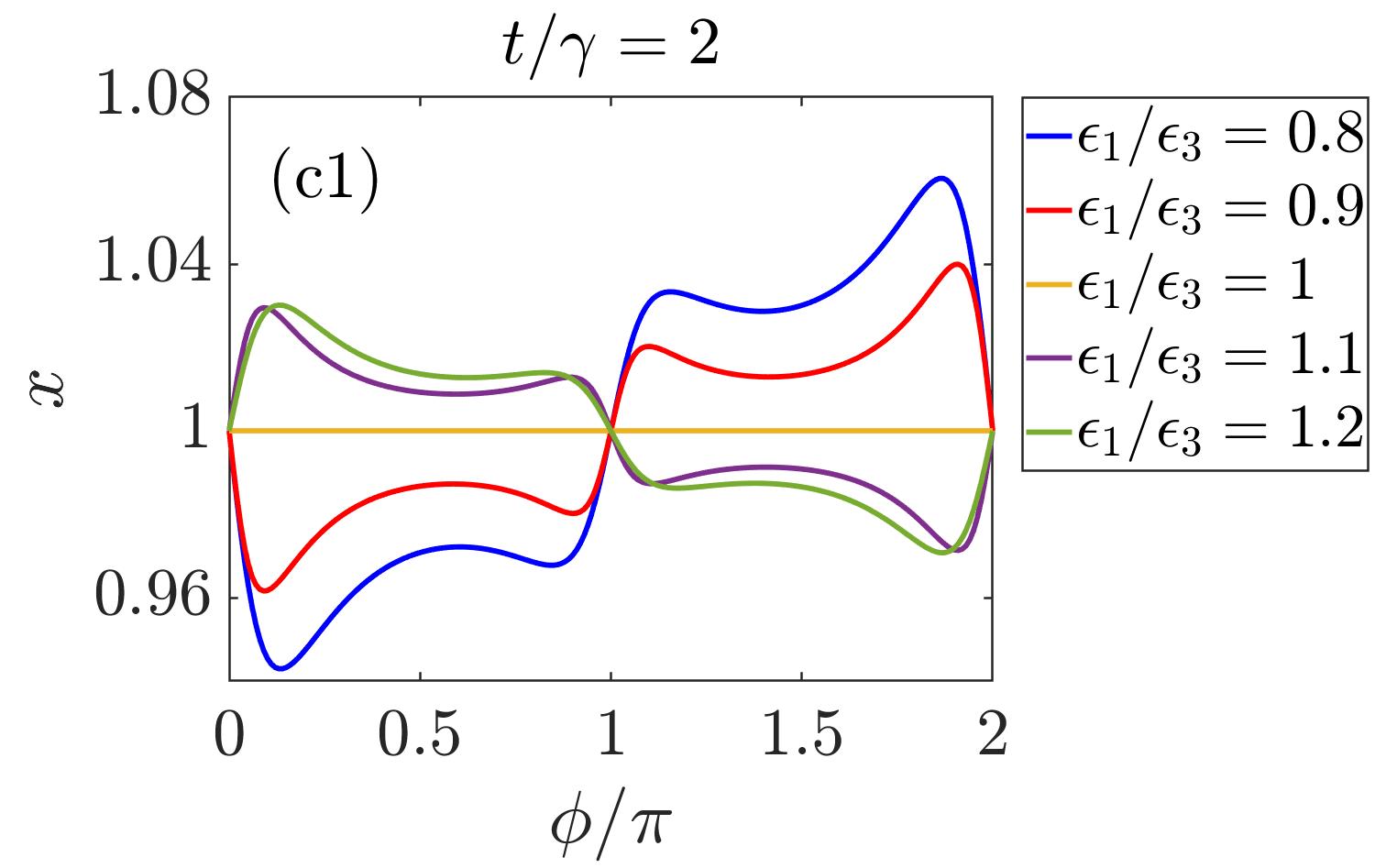}
    \includegraphics[scale=0.12]{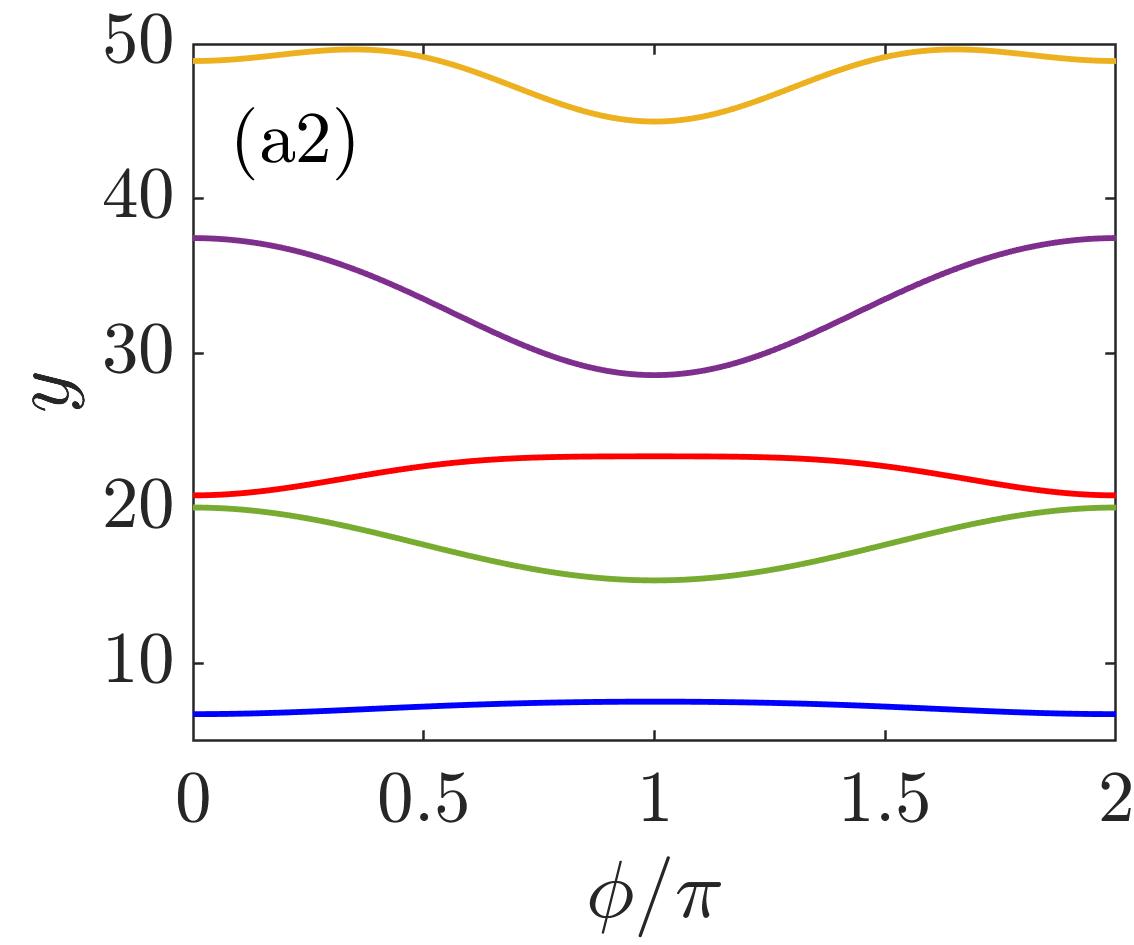}
    \includegraphics[scale=0.12]{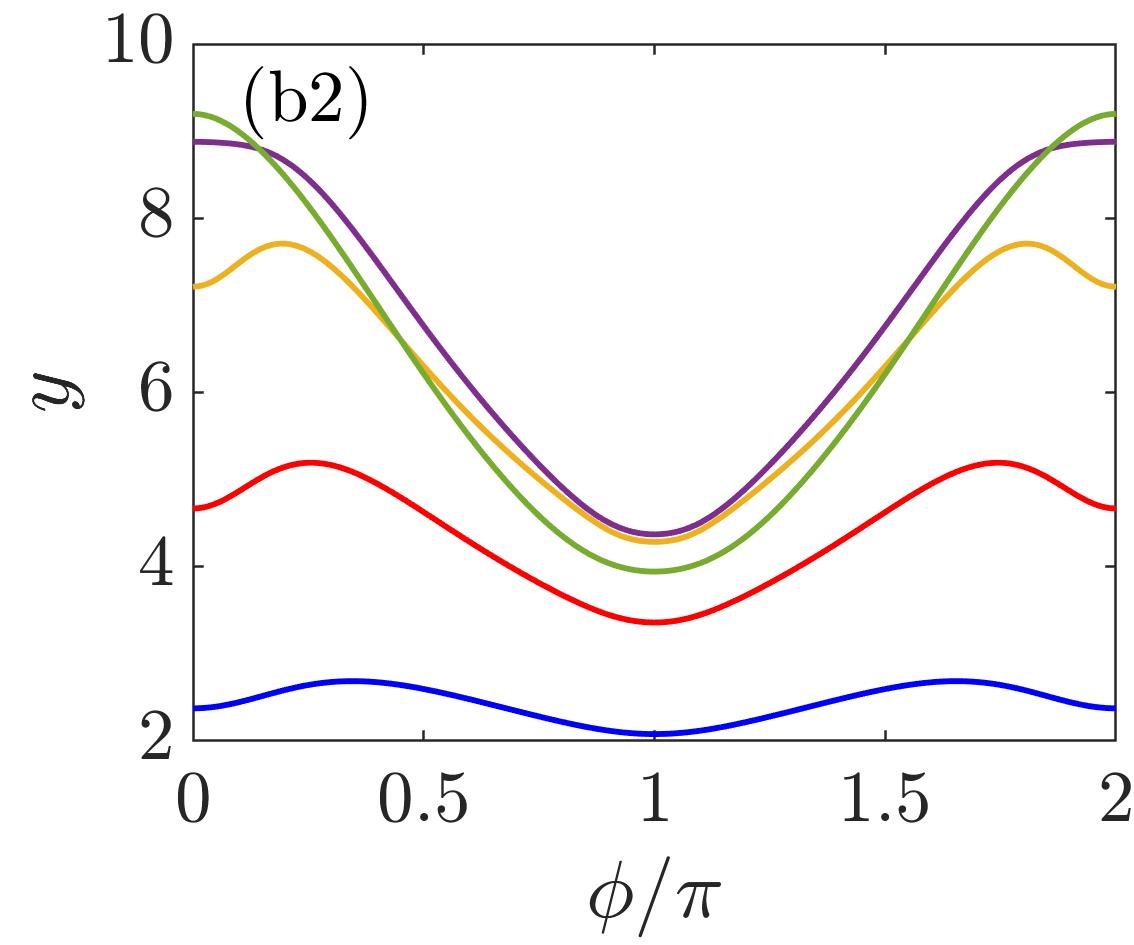}
    \includegraphics[scale=0.12]{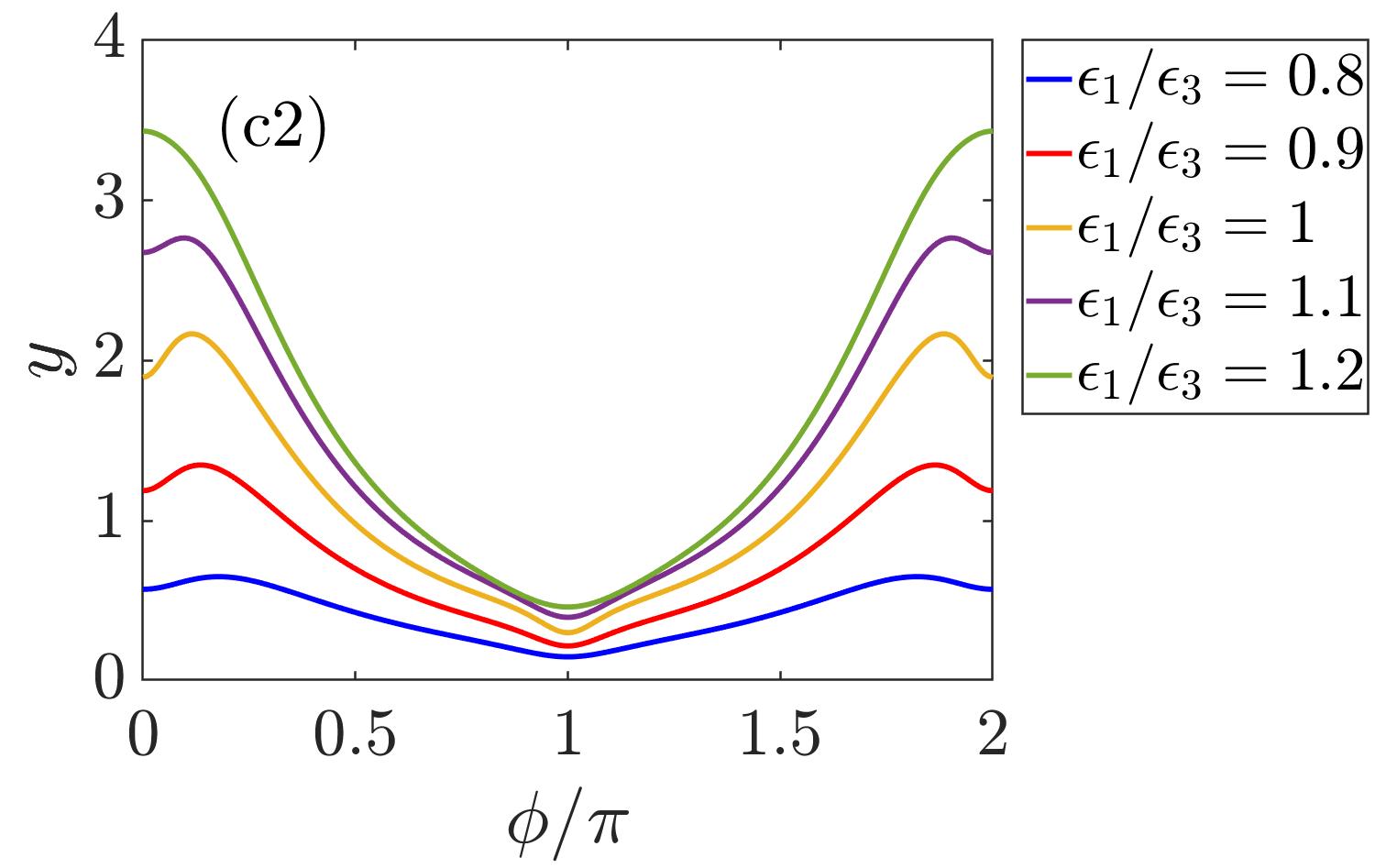}
    \caption{Asymmetry parameter ($x$) and figure of merit ($y$) as a function of $\phi$ for different anisotropic setups $\epsilon_1/\epsilon_3$ for a voltage-temperature probe heat engine in three different regimes of $t/\gamma$ ratio: (a1-a2) $t/\gamma=0.2$, (b1-b2) $t/\gamma=1$, and $t/\gamma=2$. Parameters used are: $\gamma=\gamma_L=\gamma_R=\gamma_P=0.05$, $\epsilon_2=\epsilon_3=0.5$, $T_R=T=0.1$, $T_L=0.13$, $\mu_R=\mu=0.3$, $\alpha=0.75$.}
    \label{fig:aniso_xy}
\end{figure*}
\section{Derivation of heat dissipation to the cold reservoir}\label{additional}
In this section, we derive the heat dissipation to the cold reservoir by applying the energy conservation law to both the voltage probe and voltage-temperature probe heat engine in the MNL regime.\\
\subsection{Voltage probe}
\indent For a voltage probe heat engine, the particle current defined in Eq. (\ref{eq:jlnv}) shows that $J_L^N$ and $X_L^{\mu}$ are uniquely related at a fixed value of $X_L^T$ and $X_P^T$. Thus, we can choose $J_L^N$ as our control parameter instead of $X_L^{\mu}$ by the relation
\begin{equation}\label{xlmu}
    X_L^{\mu}=\frac{J_L^N-(\mathcal{L}_{12}^{\prime}+\mathcal{L}_{13}^{\prime}\xi)X_L^T}{\mathcal{L}_{11}^{\prime}}.
\end{equation}
From the energy conservation law, the output power of the heat engine is equal to the sum of all the heat currents exchanged between the system and the three reservoirs i.e., $\mathcal{P}=J_L^Q+J_P^Q+J_R^Q$, where $J_R^Q$ is the heat current from the right reservoir. Thus, applying the energy conservation law and using Eq. (\ref{xlmu}) in Eqs. (\ref{eq:jlqv}), (\ref{eq:jpqv}), and (\ref{power}), we can express all the heat currents in terms of $J_L^N$ as follows:
\begin{widetext}
    \begin{equation}
        J_L^Q =\bigg[\bigg(\frac{\mathcal{L}_{11}^{\prime}\mathcal{L}_{22}^{\prime}-\mathcal{L}_{21}^{\prime}\mathcal{L}_{12}^{\prime}}{\mathcal{L}_{11}^{\prime}}\bigg)+\bigg(\frac{\mathcal{L}_{11}^{\prime}\mathcal{L}_{23}^{\prime}-\mathcal{L}_{21}^{\prime}\mathcal{L}_{13}^{\prime}}{\mathcal{L}_{11}^{\prime}}\bigg)\xi\bigg]X_L^T
        +\frac{\mathcal{L}_{21}^{\prime}}{\mathcal{L}_{11}^{\prime}}J_L^N-\gamma_h{J_L^N}^2,
    \end{equation}
    \begin{equation}
        J_P^Q =\bigg[\bigg(\frac{\mathcal{L}_{11}^{\prime}\mathcal{L}_{32}^{\prime}-\mathcal{L}_{31}^{\prime}\mathcal{L}_{12}^{\prime}}{\mathcal{L}_{11}^{\prime}}\bigg)+\bigg(\frac{\mathcal{L}_{11}^{\prime}\mathcal{L}_{33}^{\prime}-\mathcal{L}_{31}^{\prime}\mathcal{L}_{13}^{\prime}}{\mathcal{L}_{11}^{\prime}}\bigg)\xi\bigg]X_L^T
        +\frac{\mathcal{L}_{31}^{\prime}}{\mathcal{L}_{11}^{\prime}}J_L^N,
    \end{equation}
    \begin{equation}
        J_R^Q =\bigg[\frac{(\mathcal{L}_{12}^{\prime}+\mathcal{L}_{13}^{\prime}\xi)(\mathcal{L}_{21}^{\prime}+\mathcal{L}_{31}^{\prime})}{\mathcal{L}_{11}^{\prime}}-\big(\mathcal{L}_{22}^{\prime}+\mathcal{L}_{32}^{\prime}+(\mathcal{L}_{23}^{\prime}+\mathcal{L}_{33}^{\prime})\xi\big)\bigg]X_L^T
        +\bigg[\frac{T_R}{\mathcal{L}_{11}^{\prime}}(\mathcal{L}_{12}^{\prime}+\mathcal{L}_{13}^{\prime}\xi)X_L^T-\frac{(\mathcal{L}_{21}^{\prime}+\mathcal{L}_{31}^{\prime})}{\mathcal{L}_{11}^{\prime}}\bigg]J_L^N-\gamma_c{J_L^N}^2,
    \end{equation}
\end{widetext}
where $\gamma_c=\frac{T}{\mathcal{L}_{11}^{\prime}}-\gamma_h>0$ denotes the strength of dissipation to the cold (right) reservoir for the voltage probe heat engine.\\
\subsection{Voltage-temperature probe}
\indent For a voltage-temperature probe, at a fixed value of $X_L^T$, we can consider $J_L^N$ as our control parameter instead of $X_L^{\mu}$ using Eq. (\ref{eq:jlnvt}) as follows:
\begin{equation}\label{xlmuvt}
    X_L^{\mu}=\frac{J_L^N-\mathcal{L}^{\prime\prime}_{12}X_L^T}{\mathcal{L}^{\prime\prime}_{11}}.
\end{equation}
Following the energy conservation law, the heat current to the right (cold) reservoir is given by $J_R^Q=\mathcal{P}-J_L^Q$. Thus, using Eq. (\ref{xlmuvt}) in Eq. (\ref{eq:jlqvt}) and Eq. (\ref{powervt}), we can write the heat currents in terms of $J_L^N$ expressed as follows:
\begin{equation}
    J_L^Q =\bigg(\frac{\mathcal{L}^{\prime\prime}_{11}\mathcal{L}^{\prime\prime}_{22}-\mathcal{L}^{\prime\prime}_{12}\mathcal{L}^{\prime\prime}_{21}}{\mathcal{L}^{\prime\prime}_{11}}\bigg)X_L^T+\frac{\mathcal{L}^{\prime\prime}_{21}}{\mathcal{L}^{\prime\prime}_{11}}J_L^N-\gamma_h{J_L^N}^2,
\end{equation}
\begin{equation}
    \begin{split}
        J_R^Q =\bigg(\frac{\mathcal{L}^{\prime\prime}_{12}\mathcal{L}^{\prime\prime}_{21}-\mathcal{L}^{\prime\prime}_{11}\mathcal{L}^{\prime\prime}_{22}}{\mathcal{L}^{\prime\prime}_{11}}\bigg)X_L^T+\bigg(\frac{\mathcal{L}^{\prime\prime}_{12}\eta_C-\mathcal{L}^{\prime\prime}_{21}}{\mathcal{L}^{\prime\prime}_{11}}\bigg)J_L^N\\
        -\gamma_c{J_L^N}^2,
    \end{split}
\end{equation}
where $\gamma_c=\frac{T}{\mathcal{L}_{11}^{\prime\prime}}-\gamma_h>0$ represents the strength of dissipation to the cold (right) reservoir for the voltage-temperature probe heat engine.
\section{Inelastic effects in fully nonlinear regime}\label{fully_nonlinear}
\textit{Voltage probe}. In a fully nonlinear regime, the voltage probe condition $J_P^N=0$ provides a unique solution for probe chemical potential that can be obtained numerically using the Newton-Raphson method \cite{Bedkihal1}
\begin{equation}
    \mu_P^{k+1}=\mu_P^k-J_P^N(\mu_P^k)\Bigg[\frac{\partial J_P^N(\mu_P^k)}{\partial\mu_P}\Bigg]^{-1},
\end{equation}
where the particle current $J_P^N(\mu_P^k)$ and its derivative are calculated from Eq. (\ref{Landau_jpn}) using the Fermi distribution for the probe at $\mu_P^k$.\\
\indent
\textit{Voltage-temperature probe}. The voltage-temperature probe demands $J_P^N=0$ and $J_P^Q=0$ to introduce non-dissipative effects into the system. In the fully nonlinear regime, the probe chemical potential and temperature are evaluated self-consistently using the two-dimensional Newton-Raphson method \cite{Bedkihal1} as follows:
\begin{equation}
    \mu_P^{k+1}=\mu_P^k-D_{1,1}^{-1}J_P^N(\mu_P^k,T_P^k)-D_{1,2}^{-1}J_P^Q(\mu_P^k,T_P^k),
\end{equation}
\begin{equation}
    T_P^{k+1}=T_P^k-D_{2,1}^{-1}J_P^N(\mu_P^k,T_P^k)-D_{2,2}^{-1}J_P^Q(\mu_P^k,T_P^k),
\end{equation}
where $D_{i,j}^{-1}$ are the elements of the inverse of the Jacobian matrix $D$ that are re-evaluated at every iterations,
\begin{equation}
\renewcommand{\arraystretch}{2}
    D(\mu_P,T_P)=
    \begin{pmatrix}
        \frac{\partial J_P^N(\mu_P,T_P)}{\partial\mu_P} & \frac{\partial J_P^N(\mu_P,T_P)}{\partial T_P}\\
        \frac{\partial J_P^Q(\mu_P,T_P)}{\partial\mu_P} & \frac{\partial J_P^Q(\mu_P,T_P)}{\partial T_P}
    \end{pmatrix}.
\end{equation}
Note that the probe distribution function $f_P(\omega,\phi)$ is of the form of a Fermi distribution, which is now a function of magnetic flux for both the voltage probe and voltage-temperature probe.\\
\indent Figure \ref{fig:power_nnc_mnl}(a3-c3) illustrates the thermoelectric performance of the voltage probe heat engine operating in the fully nonlinear regime for three different regimes of $t/\gamma$, and the results are discussed in Sec. \ref{sec:results}. We conclude that higher-order nonlinear effects with broken time-reversal symmetry help enhance the thermoelectric performance of the voltage probe heat engine.
\begin{figure}[t!]
    \centering
    \includegraphics[scale=0.12]{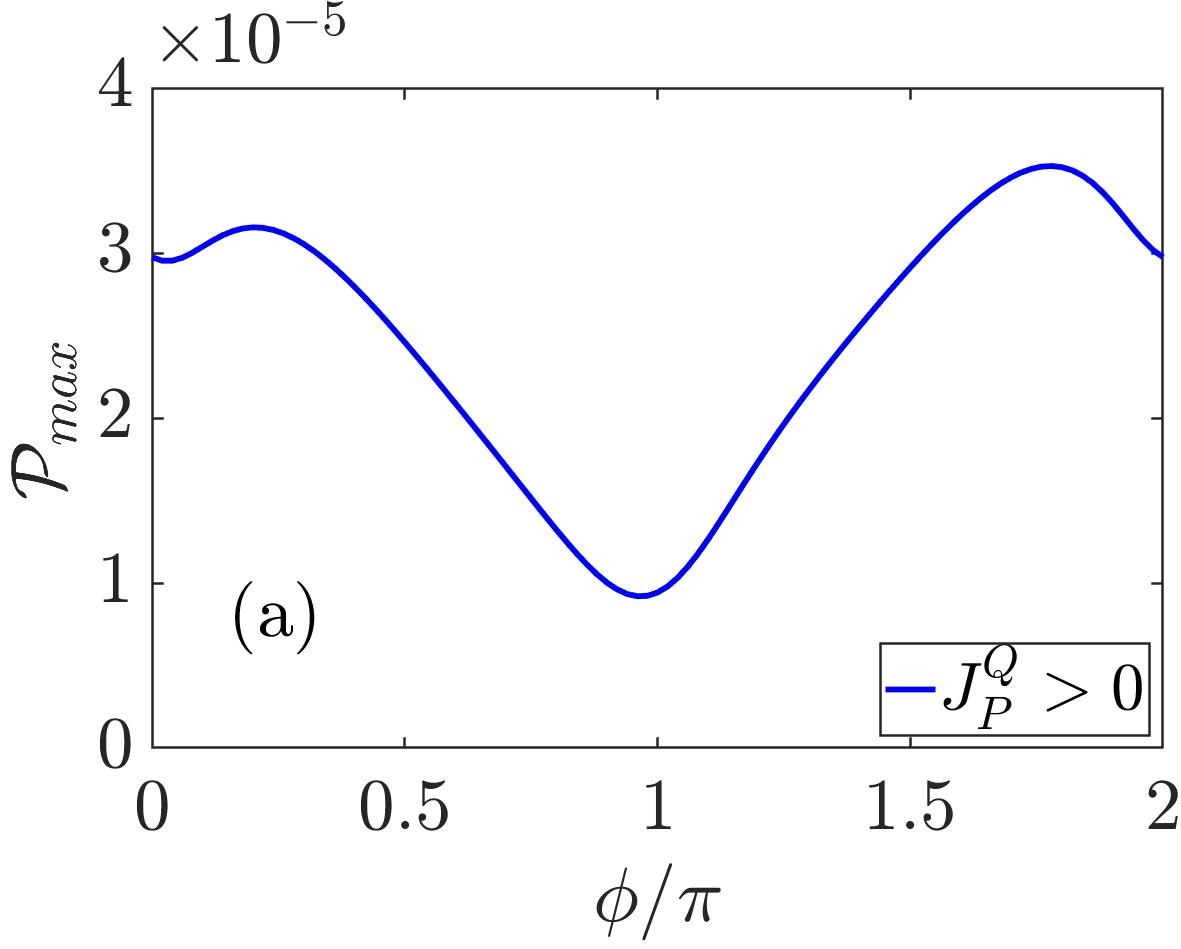}
    \includegraphics[scale=0.12]{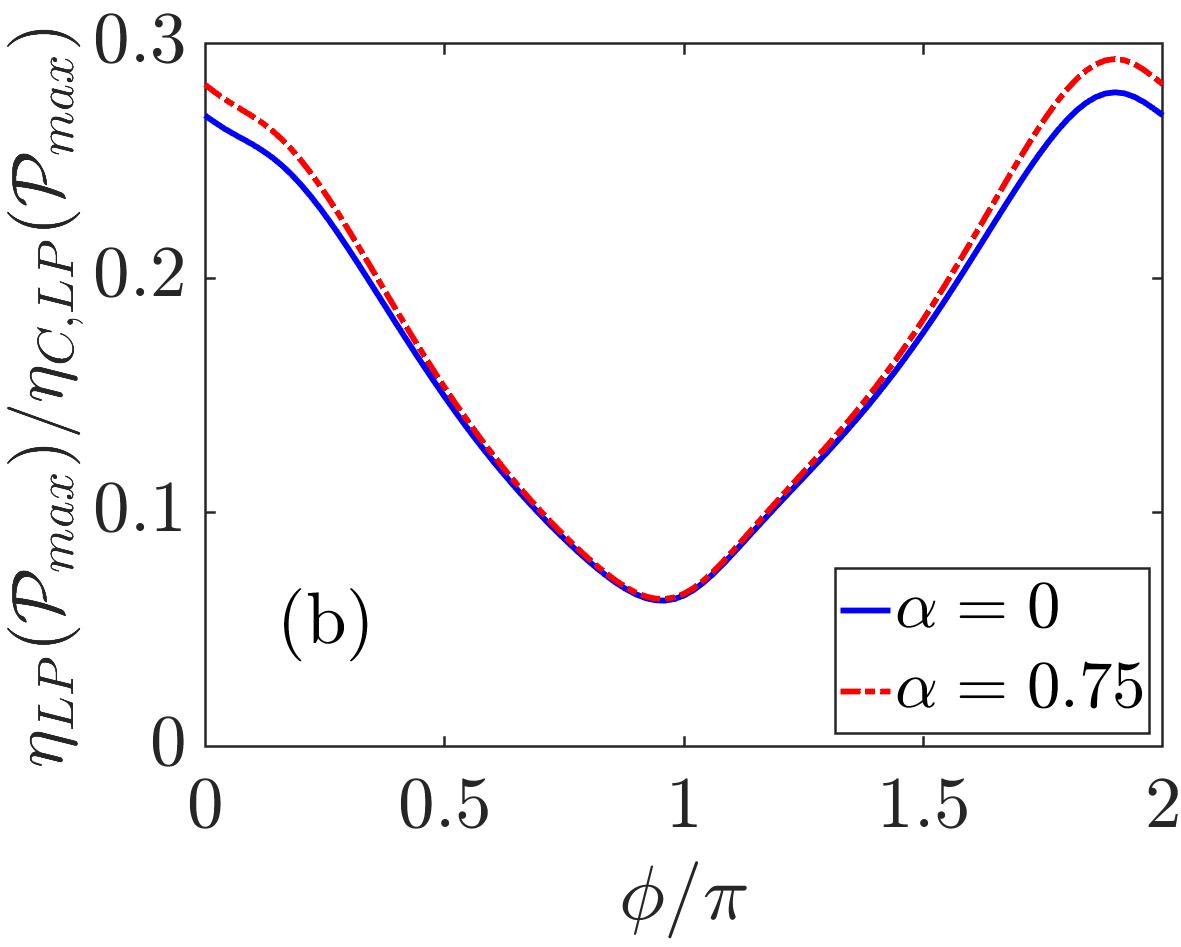}
    \caption{(a) Maximum power and (b) EMP as a function of $\phi$ for a voltage probe heat engine with $J_P^Q>0$ in the MNL regime for $t/\gamma=2$ regime. The parameters used are: $\gamma=\gamma_L=\gamma_R=\gamma_P=0.05$, $\epsilon_1=0.6$, $\epsilon_2=\epsilon_3=0.5$, $T_R=0.1$, $T_L=0.13$, $T_P=0.128$, $\mu_R=\mu=0.3$.}
    \label{fig:vnncLP}
\end{figure}
\section{Voltage probe heat engine with $J_P^Q>0$}\label{sec:LP}
In a voltage probe heat engine, the probe can either act as a heat sink and absorb heat from the system, $J_P^Q<0$, or act as a heat source to release heat into the system, $J_P^Q>0$. In a voltage probe heat engine, the probe can function either as a heat sink, absorbing heat from the system ($J_P^Q<0$), or as a heat source, releasing heat into the system ($J_P^Q>0$). In each case, the efficiency and Carnot efficiency are distinctly defined in Eqs. (\ref{etaL}), (\ref{etaLP}), (\ref{eq:etaCL}), and (\ref{eq:etaCLP}). Our previous discussions have focused exclusively on the case where $J_P^Q<0$. In this section, we discuss the case where $J_P^Q>0$. Figure \ref{fig:vnncLP} demonstrates the maximum power and EMP plot as a function of $\phi$ in the $t/\gamma=2$ regime. In this case, $\mathcal{P}_{max}$ and EMP exhibit a similar trend with $\phi$ as observed in the $J_P^Q<0$ case. The $\mathcal{P}_{max}$ curve exhibits local maxima within a certain range of $\phi$ where it surpasses the power at the symmetric point ($\phi=0$). Figure \ref{fig:vnncLP}(b) shows the EMP achieving a local maximum within the range $1.8\pi<\phi<2\pi$ with the highest value around $\phi=1.9\pi$ in the linear-response regime ($\alpha=0$). This local maximum is the consequence of broken time-reversal symmetry, with $x_{LP}>1$ and a higher figure of merit $y_{LP}$ (plots not shown). The effect of nonlinearity is also evident in the EMP plot, where it increases with growing dissipation strength $\alpha$. We observe that the $J_P^Q>0$ case is less efficient and produces lower power compared to the $J_P^Q<0$ case.
\bibliography{ref}
\end{document}